\documentclass[iop]{emulateapj}
\usepackage{graphicx}
\usepackage{amsmath}
\usepackage{atbegshi}
\def\hackaltaffiltext#1#2{\AtBeginShipoutNext{\footnotetext[#1]{#2}\stepcounter{footnote}}}

\begin{document}

\newcommand{\ocen}{$\omega$~Cen}

\title{Homogeneous analysis of globular clusters from the APOGEE
survey with the BACCHUS code. II. The Southern clusters and overview}

\author{
Szabolcs~M{\'e}sz{\'a}ros\altaffilmark{1,2,3}, 
Thomas~Masseron\altaffilmark{4,5},
D.~A.~Garc\'{\i}a-Hern{\'a}ndez\altaffilmark{4,5}, 
Carlos~Allende~Prieto\altaffilmark{4,5},
Timothy~C.~Beers\altaffilmark{6},
Dmitry~Bizyaev\altaffilmark{7}, 
Drew~Chojnowski\altaffilmark{8}, 
Roger~E.~Cohen\altaffilmark{9}, 
Katia~Cunha\altaffilmark{10,11}, 
Flavia~Dell'Agli\altaffilmark{4,5},
Garrett~Ebelke\altaffilmark{12}, 
Jos{\'e}~G.~Fern{\'a}ndez-Trincado\altaffilmark{13,14}, 
Peter~Frinchaboy\altaffilmark{15}, 
Doug~Geisler\altaffilmark{16,17,18}, 
Sten~Hasselquist\altaffilmark{19,20}, 
Fred~Hearty\altaffilmark{21}, 
Jon~Holtzman\altaffilmark{22}, 
Jennifer~Johnson\altaffilmark{23}, 
Richard~R.~Lane\altaffilmark{24},
Ivan~Lacerna\altaffilmark{13,25},
Penelop{\'e}~Longa-Pe{\~n}a\altaffilmark{26},
Steven~R.~Majewski\altaffilmark{12}, 
Sarah~L.~Martell\altaffilmark{27}, 
Dante~Minniti\altaffilmark{28,29,30}, 
David~Nataf\altaffilmark{31}, 
David~L.~Nidever\altaffilmark{32,35},
Kaike~Pan\altaffilmark{7}, 
Ricardo~P.~Schiavon\altaffilmark{33}, 
Matthew~Shetrone\altaffilmark{34}, 
Verne~V.~Smith\altaffilmark{35}, 
Jennifer~S.~Sobeck\altaffilmark{11}, 
Guy~S.~Stringfellow\altaffilmark{36},
L{\'a}szl{\'o}~Szigeti\altaffilmark{1,3}, 
Baitian~Tang\altaffilmark{37}, 
John~C.~Wilson\altaffilmark{12}, 
Olga~Zamora\altaffilmark{4,5}
}
\altaffiltext{1}{ELTE E\"otv\"os Lor\'and University, Gothard Astrophysical Observatory, 9700 Szombathely, Szent Imre H. st. 112, Hungary}
\altaffiltext{2}{Premium Postdoctoral Fellow of the Hungarian Academy of Sciences (meszi@gothard.hu)}
\altaffiltext{3}{MTA-ELTE Exoplanet Research Group, 9700 Szombathely, Szent Imre h. st. 112, Hungary}
\altaffiltext{4}{Instituto de Astrof{\'{\i}}sica de Canarias (IAC), E-38205 La Laguna, Tenerife, Spain}
\altaffiltext{5}{Universidad de La Laguna (ULL), Departamento de Astrof\'{\i}sica, 38206 La Laguna, Tenerife, Spain}
\altaffiltext{6}{Dept. of Physics and JINA Center for the Evolution of the Elements, University of Notre Dame, Notre Dame, IN 46556, USA}
\altaffiltext{7}{Apache Point Observatory, P.O. Box 59, Sunspot, NM 88349}
\altaffiltext{8}{Dept. of Astronomy, New Mexico State University, Las Cruces, NM 88003, USA}
\altaffiltext{9}{Space Telescope Science Institute, 3700 San Martin Drive, Baltimore, MD 21218, USA}
\altaffiltext{10}{Steward Observatory, University of Arizona, 933 North Cherry Avenue, Tucson, AZ 85721, USA} 
\altaffiltext{11}{Observat\'orio Nacional, S\~ao Crist\'ov\~ao, Rio de Janeiro, Brazil}
\altaffiltext{12}{Dept. of Astronomy, University of Virginia, Charlottesville, VA 22904-4325, USA}
\altaffiltext{13}{Instituto de Astronom\'ia y Ciencias Planetarias, Universidad de Atacama, Copayapu 485, Copiap\'o, Chile}
\altaffiltext{14}{Institut Utinam, CNRS-UMR 6213, Universit\'e Bourgogne-Franche-Compt\'e, OSU THETA Franche-Compt\'e, Observatoire de Besan\c{c}on, BP 1615, 251010 Besan\c{c}on Cedex, France}
\altaffiltext{15}{Department of Physics \& Astronomy, Texas Christian University, Fort Worth, TX 76129, USA}
\altaffiltext{16}{Departamento de Astronom\'{\i}a, Universidad de Concepci\'on, Casilla 160-C, Concepci\'on, Chile}
\hackaltaffiltext{17}{Instituto de Investigación Multidisciplinario en Ciencia y Tecnolog\'{\i}a, Universidad de La Serena. Avenida Raúl Bitrán 
S/N, La Serena, Chile}
\hackaltaffiltext{18}{Departamento de Astronom\'{\i}a, Facultad de Ciencias, Universidad de La Serena. Av.
Juan Cisternas 1200, La Serena, Chile}
\hackaltaffiltext{19}{Dept. of Physics \& Astronomy, University of Utah, Salt Lake City, UT, 84112, USA}
\hackaltaffiltext{20}{NSF Astronomy and Astrophysics Postdoctoral Fellow}
\hackaltaffiltext{21}{Dept. of Astronomy and Astrophysics, The Pennsylvania State University, University Park, PA 16802, USA}
\hackaltaffiltext{22}{New Mexico State University, Las Cruces, NM 88003, USA}
\hackaltaffiltext{23}{Dept. of Astronomy, The Ohio State University, 140 W. 18th Ave., Columbus, OH 43210, USA}
\hackaltaffiltext{24}{Instituto de Astro{\'{\i}}sica, Pontificia Universidad Catalica de Chile, Av. Vicuna Mackenna 4860, 782-0436 Macul, Santiago, Chile}
\hackaltaffiltext{25}{Instituto Milenio de Astrof\'isica, Av. Vicu\~na Mackenna 4860, Macul, Santiago, Chile}
\hackaltaffiltext{26}{Centro de Astronomia, Universidad de Antofagasta, Avenida Angamos 601, Antofagasta 1270300, Chile}
\hackaltaffiltext{27}{School of Physics, University of New South Wales, NSW 2052, Australia}
\hackaltaffiltext{28}{Departamento de Ciencias Fisicas, Facultad de Ciencias Exactas, Universidad Andres Bello, 
Av. Fernandez Concha 700, Las Condes, Santiago, Chile}
\hackaltaffiltext{29}{Millennium Institute of Astrophysics, Av. Vicuna Mackenna 4860, 782-0436, Santiago, Chile}
\hackaltaffiltext{30}{Vatican Observatory, V00120 Vatican City State, Italy}
\hackaltaffiltext{31}{Dept. of Physics and Astronomy, Johns Hopkins University, Baltimore, MD, 21218, USA}
\hackaltaffiltext{32}{Dept. of Physics, Montana State University, P.O. Box 173840, Bozeman, MT 59717-3840, USA ; National Optical Astronomy Observatory, 
950 North Cherry Avenue, Tucson, AZ 85719, USA}
\hackaltaffiltext{33}{Astrophysics Research Institute, IC2, Liverpool Science Park,
Liverpool John Moores University, 146 Brownlow Hill, Liverpool, L3 5RF, UK}
\hackaltaffiltext{34}{University of Texas at Austin, McDonald Observatory, Fort Davis, TX 79734, USA}
\hackaltaffiltext{35}{National Optical Astronomy Observatory, Tucson, AZ 85719, USA}
\hackaltaffiltext{36}{Center for Astrophysics and Space Astronomy, Dept. of Astrophysical and Planetary Sciences, 
University of Colorado, 389 UCB, Boulder, CO 80309-0389, USA}
\hackaltaffiltext{37}{School of Physics and Astronomy, Sun Yat-sen University, Zhuhai, China}

\begin{abstract}

We investigate the Fe, C, N, O, Mg, Al, Si, K, Ca, Ce and Nd abundances of 2283 red giant stars in 31 globular clusters 
from high-resolution spectra observed in both the northern and southern hemisphere by the SDSS-IV APOGEE-2 survey. This 
unprecedented homogeneous dataset, largest to date, allows us to discuss the intrinsic Fe spread, the shape and statistics 
of Al-Mg and N-C anticorrelations as a function of cluster mass, luminosity, age and metallicity for all 31 clusters. We 
find that the Fe spread does not depend on these parameters within our uncertainties including cluster metallicity, 
contradicting earlier observations. We do not confirm the metallicity variations previously observed in M22 and NGC 1851. 
Some clusters show a bimodal Al distribution, while others exhibit a continuous distribution as has been previously 
reported in the literature. We confirm more than 2 populations in \ocen \ and NGC 6752, and find new ones in M79. We 
discuss the scatter of Al by implementing a correction to the standard chemical evolution of Al in the Milky Way. 
After correction, its dependence on cluster mass is increased suggesting that the extent of Al enrichment as a function 
of mass was suppressed before the correction. We observe a turnover in the Mg-Al anticorrelation at very low Mg in 
\ocen, similar to the pattern previously reported in M15 and M92. \ocen \ may also have a weak K-Mg anticorrelation, 
and if confirmed, it would be only the third cluster known to show such a pattern.

\end{abstract}

\section{Introduction}

During most of the 20th century it was believed that globular clusters (GCs) exhibit only one generation of stars. However, 
detailed photometric and spectroscopic studies of Galactic globular clusters over the past thirty years have revealed great 
complexity in the elemental abundances of their stars, from the main sequence through to the asymptotic giant branch. Most light 
elements show star-to-star variations in almost all GCs and these large variations are generally interpreted as the result of 
chemical feedback from an earlier generation of stars \citep{gratton01, cohen02}, rather than inhomogeneities 
in the original stellar cloud from which these stars formed. Thus, the current scenario of GC evolution generally assumes that 
more than one population of stars were formed in each cluster, and the chemical makeup of stars that formed later is 
polluted by material produced by the first generation. 

The origin of the polluting material remains to be established and it has obvious bearings on the timescales for the 
formation of the cluster itself and its mass budget. Proposed candidate polluters include intermediate mass stars in
their asymptotic giant branch (AGB) phase \citep{ventura01}, fast rotating massive stars losing
mass during their main sequence phase \citep{decressin01}, novae \citep{maccarone01}, 
massive binaries \citep{demink01}, and supermassive stars \citep{deni01}. 
These potential contributions obviously operate on
different time scales and require a different amount of stellar mass in the first generation. In order to constrain these 
models and to gain an overall understanding of the multiple stellar populations in globular clusters we need 
comprehensive studies with a relatively complete and unbiased data set. This 
requires a focused effort by Galactic archaeology surveys to obtain and uniformly analyze spectra for 
large samples of globular cluster stars across a wide range of metallicity.

There are two main fronts in exploring multiple populations (MPs) in GCs: photometry and spectroscopy. 
Several larger photometric surveys have been conducted to explore MPs in almost all GCs 
\citep[e.g.,][]{piotto01, sara01, piotto02, milone02, soto01}, using the data from the 
Hubble Space Telescope achieving unprecedented photometric precision. Using high-resolution spectroscopy
the Lick-Texas group \citep[e.g.,][]{sneden01, sneden02, sneden03, sneden04, sneden05, kraft02, kraft03, ivans01}
conducted the first large survey of northern clusters using three different 
telescopes and spectrographs. Also using high-resolution spectroscopy 
\citet{carretta02, carretta03, carretta01} have 
carried out the first detailed survey of southern clusters with the VLT telescopes, exploring the Na-O and Al-Mg 
anticorrelations, which are the result of Ne-Na and Mg-Al cycles occurring in the H-burning shell of the first 
population stars whose nucleosynthetic products were later distributed through the cluster. We refer the 
reader to \citet{bastian03} for a complete overview on MPs in GCs.

With the appearance of high spectral resolution sky 
surveys some of these southern clusters were revisited by the Gaia-ESO survey 
\citep{gilmore01} focusing on the same two element pairs \citep{pancino01}. The first homogeneous exploration of 10 
northern clusters was carried out by \citet{meszaros03}, which was updated 
by \citet{masseron01}, both using data from the SDSS-III \citep{eis11} Apache Point Observatory Galactic Evolution Experiment 
(APOGEE) survey. Results for 
additional clusters observed by APOGEE were published by \citet{schiavon01}, \citet{tang01}, and \citet{trin01}. Its successor,  
SDSS-IV \citep{blanton01} APOGEE-2 \citep{majewski01} started in the summer of 2014 and ends in 2020, further expanding the number 
of observed GCs from the southern hemisphere. Comparison of northern and southern clusters 
was difficult previously because many observations were carried out with 
different telescopes and abundance determination techniques that may have systematic errors of their own. The APOGEE survey 
is the first spectroscopic survey that covers both the northern and southern sky by installing two twin spectrographs, 
identical in design, on the Sloan 2.5 meter telescope \citep{gunn01} at the Apache Point Observatory (APO) and the du Pont 
2.5m telescope at Las Campanas Observatory (LCO). In an effort to create the first truly systematic study of the chemical 
makeup of multiple populations in all GCs, \citet{masseron01} reanalyzed the 10 clusters observed from APO \citep{meszaros03} 
with an updated pipeline.

\begin{deluxetable*}{llrrrrrrrrrrr}[!ht]
\tabletypesize{\scriptsize}
\tablecaption{Properties of Clusters from the Literature}
\tablewidth{0pt}
\tablehead{
\colhead{ID} & 
\colhead{Name} & 
\colhead{N$_{\rm 1}$\tablenotemark{a}} &
\colhead{N$_{\rm 2}$\tablenotemark{b}} &
\colhead{[Fe/H]} & 
\colhead{E(B$-$V)} &
\colhead{R$_{\rm t}$} &
\colhead{V$_{\rm disp}$} &
\colhead{RA} &
\colhead{Dec} &
\colhead{V$_{\rm helio}$} &
\colhead{$\mu_{\alpha *}$} & 
\colhead{$\mu_{\delta}$} \\
\colhead{} & \colhead{} & 
\colhead{All} &
\colhead{S/N$>$70} &
\colhead{} & 
\colhead{} &
\colhead{\'} &
\colhead{km/s} &
\colhead{} &
\colhead{} &
\colhead{km/s} & 
\colhead{mas yr$^{-1}$} & 
\colhead{mas yr$^{-1}$} 
}
\startdata
NGC 104		& 47 Tuc 	&  186 	&  151 	& -0.72 &   0.04 &	42.9 	& 12.2  & 00 24 05.67 &  -72 04 52.6  & -17.2   &	5.25	&	-2.53	\\
NGC 288		& 			&   43 	&   40 	& -1.32 &   0.03 &	12.9 	&  3.3  & 00 52 45.24 &  -26 34 57.4  & -44.8   &	4.22	& -5.65		\\
NGC 362		& 			&   56 	&   40 	& -1.26 &   0.05 &	16.1 	&  8.8  & 01 03 14.26 &  -70 50 55.6  & 223.5   &	6.71	& -2.51	\\
NGC 1851	& 			&   43 	&   30 	& -1.18 &   0.02 &	11.7 	& 10.2  & 05 14 06.76 &  -40 02 47.6  & 320.2   &	2.12 	& -0.63	\\
NGC 1904	& M79		&   26 	&   25	& -1.60 &   0.01 &	8.3  	&  6.5  & 05 24 11.09 &  -24 31 29.0  & 205.6   &	2.47	& -1.59	\\
NGC 2808	& 			&   77 	&   71 	& -1.14 &   0.22 &	15.6 	& 14.4  & 09 12 03.10 &  -64 51 48.6  & 103.7   &	1.02	& 0.28	\\
NGC 3201	& 			&  179 	&  152 	& -1.59 &   0.24 &	28.5 	&  5.0  & 10 17 36.82 &  -46 24 44.9  & 494.3   &	8.35	& -2.00	\\
NGC 4147    &  			&	 3	&	 1	& -1.80 &   0.02 &	 6.3 	&  3.1	& 12 10 06.30 &  +18 32 33.5  & 179.1	& -1.71	& -2.10	\\
NGC 4590	& M68		&   37 	&   36	& -2.23 &   0.05 &	13.7 	&  3.7  & 12 39 27.98 &  -26 44 38.6  & -93.2   &	-2.75	& 1.78	\\
NGC 5024   	& M53		&   41 	&   39	& -2.10 &   0.02 &	30.3 	&  5.9  & 13 12 55.25 &  +18 10 05.4  & -63.1   &	-0.11	& -1.35	\\
NGC 5053	& 			&   17 	&   17 	& -2.27 &   0.01 &	11.8 	&  1.6  & 13 16 27.09 &  +17 42 00.9  & 42.5    &	-0.37	& -1.26	\\
NGC 5139	& \ocen 	&  898 	&  775 	& -1.53 &   0.12 &	57.0 	& 17.6  & 13 26 47.24 &  -47 28 46.5  & 232.7   &	-3.24	& -6.73	\\
NGC 5272    & M3		&  153 	&  148	& -1.50 &   0.01 &	38.2 	&  8.1  & 13 42 11.62 &  +28 22 38.2  & -147.2   &	-0.14	& -2.64	\\
NGC 5466   	&   		&   15	&    7 	& -1.98 &   0.00 &	34.2 	&  1.6  & 14 05 27.29 &  +28 32 04.0  & 106.9   &	-5.41	& -0.79	\\
NGC 5634	&			&	 2	&	 0	& -1.88	&	0.05 &	 8.4 	&  5.3	& 14 29 37.30 &	 -05 58 35.0  & -16.2	& -1.67	& -1.55	\\
NGC 5904	& M5 		&  207 	&  191 	& -1.29 &   0.03 &	28.4 	&  7.7  & 15 18 33.22 &  +02 04 51.7  &  53.8   &	4.06	& -9.89	\\
NGC 6121	& M4		&  158 	&  153 	& -1.16 &   0.35 &	32.5 	&  4.6  & 16 23 35.22 &  -26 31 32.7  &  71.0   &	-12.48	& -18.99	\\
NGC 6171  	& M107		&   66 	&   55 	& -1.02 &   0.33 &	17.4 	&  4.3  & 16 32 31.86 &  -13 03 13.6  & -34.7   &	-1.93	& -5.98	\\
NGC 6205   	& M13		&  127 	&  103 	& -1.53 &   0.02 &	25.2 	&  9.2  & 16 41 41.24 &  +36 27 35.5  & -244.4   &	-3.18	& -2.56	\\
NGC 6218	& M12		&   86 	&   54 	& -1.37 &   0.19 &	17.6 	&  4.5  & 16 47 14.18 &  -01 56 54.7  & -41.2   &	-0.15	& -6.77	\\
NGC 6229	&			&	 7	&	 5	& -1.47	&	0.01 &	10.0 	&  7.1	& 16 46 58.79 &  +47 31 39.9  & -138.3	& -1.19	& -0.46	\\
NGC 6254	& M10		&   87 	&   84	& -1.56 &   0.28 &	21.5 	&  6.2  & 16 57 09.05 &  -04 06 01.1  &  74.0   &	-4.72	& -6.54	\\
NGC 6316	&			&	 1	&	 1	& -0.45	&  	0.54 &	 5.9 	&  9.0	& 17 16 37.30 &  -28 08 24.4  &  99.1	& -4.97	& -4.61	\\
NGC 6341   	& M92		&   70 	&   67	& -2.31 &   0.02 &	15.2 	&  8.0  & 17 17 07.39 &  +43 08 09.4  & -120.7   &	-4.93	& -0.57	\\
NGC 6388	& 			&   26 	&    9 	& -0.55 &   0.37 &	 6.2 	& 18.2  & 17 36 17.23 &  -44 44 07.8  &  83.4   &	-1.33	& -2.68	\\
NGC 6397	& 			&  158 	&  141 	& -2.02 &   0.18 &	15.8 	&  5.2  & 17 40 42.09 &  -53 40 27.6  &  18.4   &	3.30	& -17.60		\\
NGC 6441	&			&	17	&	 5	& -0.46 &	0.47 &	 8.0 	& 18.8 	& 17 50 13.06 &  -37 03 05.2  &  17.1   & -2.51	& -5.32	\\
NGC 6522	& 			&    7 	&    5 	& -1.34 &   0.48 &	16.4 	&  8.2  & 18 03 34.02 &  -30 02 02.3  & -14.0   &	2.62	& -6.40	\\
NGC 6528	&			&	 2	&	 1	& -0.11 & 	0.54 &	16.6 	&  6.4  & 18 04 49.64 &  -30 03 22.6  & 211.0   & -2.17	& -5.52	\\
NGC 6539	&			&	 1	&	 1	& -0.63	&	1.02 &	21.5 	&  5.9	& 18 04 49.68 &  -07 35 09.1  &  35.6	& -6.82	& -3.48	\\
NGC 6544	&			&	 7 	&	 7	& -1.40	& 	0.76 &	2.05 	&  6.4  & 18 07 20.58 &  -24 59 50.4  & -36.4	&	-2.34	& -18.66	\\
NGC 6553	& 			&    8 	&    7 	& -0.18 &   0.63 &	 8.2 	&  8.5  & 18 09 17.60 &  -25 54 31.3  &   0.5   &	0.30	& -0.41	\\
NGC 6656	& M22		&	80	&	20	& -1.70 &   0.34 &	29.0 	&  8.4	& 18 36 23.94 &  -23 54 17.1  & -147.8   & 9.82	& -5.54	\\
NGC 6715	& M54		&	22	&	 7	& -1.49 &   0.15 &	10.0 	& 16.2 	& 18 55 03.33 &  -30 28 47.5  &  142.3   & -2.73	& -1.38	\\
NGC 6752	& 			&  153 	&  138 	& -1.54 &   0.04 &	55.3 	&  8.3  & 19 10 52.11 &  -59 59 04.4  & -26.2   &	-3.17	& -4.01	\\
NGC 6760	&			&	 3	&	 3	& -0.40	& 	0.77 &	 7.2 	& \nodata & 19 11 12.01 & +01 01 49.7 &  -1.6	& -1.11	& -3.59	\\	
NGC 6809	& M55 		&   96 	&   92	& -1.94 &   0.08 &	16.3 	&  4.8  & 19 39 59.71 &  -30 57 53.1  & 174.8   &	-3.41	& -9.27	\\
NGC 6838   	& M71		&   39 	&   35 	& -0.78 &   0.25 &	 9.0 	&  3.3  & 19 53 46.49 &  +18 46 45.1  & -22.5   &	-3.41	& -2.61	\\
NGC 7078   	& M15		&  133 	&  104 	& -2.37 &   0.10 &	21.5 	& 12.9  & 21 29 58.33 &  +12 10 01.2  & -106.5  &	-0.63	& -3.80	\\
NGC 7089    & M2		&   26 	&   24	& -1.65 &   0.06 &	21.5 	& 10.6  & 21 33 27.02 &  -00 49 23.7  &  -3.6   &	3.51	& -2.16	\\
Pal 5		& 			&    5 	&    5 	& -1.41 &   0.03 &	16.3 	&  0.6  & 15 16 05.25 &  -00 06 41.8  & -58.4   &	-2.77	& -2.67	\\
Pal 6    	&   		&    5 	&    4 	& -0.91 &   1.46 &	 8.4 	& \nodata & 17 43 42.20 &  -26 13 21.0 &181.0   & -9.17	& -5.26	\\
Terzan 5	&			&	 7	&	 7	& -0.23 &   2.28 &	13.3	&  19.0	& 17 48 04.80 &  -24 46 45.0  &  -82.3   & -1.71	& -4.64  \\
Terzan 12	&			&	 1	&	 1	& -0.50	& 	2.06 &	\nodata	& \nodata & 18 12 15.80 & -22 44 31.0 &  94.1	 & -6.07	& -2.63	 
\enddata
\tablecomments{Average metallicities, reddenings, tidal radii and coordinates were taken from \citet{harris01}. 
Radial and dispersion velocities are from \citet{baum01}. Proper motions were taken from \citet{baum02}.}
\tablenotetext{a}{The number of all stars in our sample.}
\tablenotetext{b}{The number of stars with S/N$>$70.}
\end{deluxetable*}

\begin{deluxetable*}{llrrrrrrrrrrr}
\tabletypesize{\scriptsize}
\tablewidth{0pt}
\tablecaption{Atmospheric Parameters and Abundances of Individual Stars}
\tablehead{
\colhead{2MASS ID} & \colhead{Cluster} & 
\colhead{Status} & \colhead{T$_{\rm eff}$} &
\colhead{log g} & \colhead{[Fe/H]} & 
\colhead{$\sigma_{\rm [Fe/H]}$} & \colhead{[C/Fe]} & 
\colhead{limit\tablenotemark{a}} & \colhead{$\sigma_{\rm [C/Fe]}$} & 
\colhead{$N_{\rm C}$} & \colhead{[N/Fe]} & \colhead{...} 
}
\startdata
2M13121714+1814178  & M53 & RGB &   4574  &     0.87  &   -2.007   &   0.121  &    \nodata &   0   &   \nodata  &  0   &     \nodata & \\						
2M13122857+1815051  & M53 & RGB &   4202  &    -0.07  &   -1.982   &   0.088  &    \nodata &   0   &   \nodata  &  0   &   0.834 & \\ 
2M13123506+1814286  & M53 & RGB &   4639  &     1.02  &   -1.894   &   0.124  &    \nodata &   0   &   \nodata  &  0   &     \nodata & \\ 
2M13123617+1807320  & M53 & RGB &   4514  &     0.74  &   -1.841   &   0.083  &    \nodata &   0   &   \nodata  &  0   &     \nodata & \\ 
2M13123617+1827323  & M53 & RGB &   4652  &     1.05  &   -1.928   &   0.119  &    \nodata &   0   &   \nodata  &  0   &     \nodata & 
\enddata
\tablecomments{This table is available in its entirety in machine-readable form in the online journal. A portion 
is shown here, with reduced number of columns, for guidance regarding its form and content. 
Star identification from \citet{carretta03} was added in the last column.}
\tablenotetext{a}{The number of lines used in the abundances analysis from BACCHUS \citep{masseron02}.}
\end{deluxetable*}

\begin{deluxetable*}{llrrrrrrrrrrr}
\tabletypesize{\scriptsize}
\tablewidth{0pt}
\tablecaption{Abundance Averages and Scatter}
\tablehead{
\colhead{ID} & \colhead{Name} & 
\colhead{[Fe/H]} & \colhead{[Fe/H]} &
\colhead{Mass} & \colhead{V$_{\rm ABS}$} & 
\colhead{Age} & \colhead{[Fe/H]} & 
\colhead{[Fe/H]} & \colhead{[Fe/H]\tablenotemark{a}} & 
\colhead{[Al/Fe]} & \colhead{[Al/Fe]} \\
\colhead{} & \colhead{} & 
\colhead{Carretta} & \colhead{Pancino} &
\colhead{10$^3$ M$_{\rm \odot}$} & \colhead{} & 
\colhead{} & \colhead{Average} & 
\colhead{Scatter} & \colhead{Error} & 
\colhead{Average} & \colhead{Scatter}
}
\startdata
NGC 104  & 47 Tuc &	-0.768 &	-0.71 &  779	& -9.42	 & 12.8	&	-0.626 & 0.107 & 0.082	& 	0.583 & 0.129		\\
NGC 288	 & 		&	-1.305 &  \nodata	  &  116	& -6.75  & 12.2	&	-1.184 & 0.114 & 0.059	& 	0.368 & 0.175		\\
NGC 362	 & 		&	\nodata		&	-1.12 &  345	& -8.43  & 10.0	&	-1.025 & 0.080 & 0.056	& 	0.241 & 0.240		\\
NGC 1851 & 		&	\nodata		&	-1.07 &  302	& -8.33  & \nodata	&	-1.033 & 0.082 & 0.077	& 	0.192 & 0.251		\\
NGC 1904 & M79	&	-1.579 &	-1.51 &  169	& -7.86  & 12.0 	&	-1.468 & 0.092 & 0.062	& 	0.449 & 0.530		\\
NGC 2808 & 		&	-1.151 &	-1.03 &  742	& -9.39  & 11.2	&	-0.925 & 0.101 & 0.070	& 	0.328 & 0.446		\\
NGC 3201 & 		&	-1.512 &  \nodata     &	 149	& -7.45  & 11.1	&	-1.241 & 0.102 & 0.061	& 	0.099 & 0.345		\\
NGC 4590 & M68	&	-2.265 &  \nodata     &	 123	& -7.37  & 12.7	&	-2.161 & 0.100 & 0.108	& 	0.302 & 0.419		\\
NGC 5024 & M53  &	  \nodata  &   \nodata    &	 380	& -8.71  & 12.7	&	-1.888 & 0.101 & 0.108	& 	0.346 & 0.507		\\
NGC 5053 & 		& 	\nodata    & \nodata      &	  56.6	& -6.76  & 12.3	&	-2.057 & 0.095 & 0.108	& 	0.397 & 0.447		\\
NGC 5139 & \ocen	&	\nodata    &  \nodata	  &	3550	& -10.26 & \nodata	&	-1.511 & 0.205 & 0.077	& 	0.586 & 0.533		\\
NGC 5272 & M3	&	 \nodata   &   \nodata	  &	 394	& -8.88  & 11.4	&	-1.388 & 0.127 & 0.068	& 	0.249 & 0.425		\\
NGC 5466 &   	&	  \nodata  &   \nodata    &	  45.6	& -6.98  & 13.6	&	-1.827 & 0.070 & 0.105	& 	0.246 & 0.663		\\
NGC 5904 & M5	&	-1.340 &  \nodata	  &  372	& -8.81  & 11.5	&	-1.178 & 0.102 & 0.062	& 	0.297 & 0.346		\\
NGC 6121 & M4	&	-1.168 &  \nodata     &	  96.9	& -7.19  & 13.1	&	-1.020 & 0.086 & 0.042	& 	0.708 & 0.121		\\
NGC 6171 & M107 &	-1.033 &  \nodata     &	  87	& -7.12  & 13.4	&	-0.852 & 0.106 & 0.076	& 	0.538 & 0.118		\\
NGC 6205 & M13  &	 \nodata   &   \nodata	  &	 453	& -8.55  & 11.7	&	-1.432 & 0.129 & 0.078	& 	0.536 & 0.517		\\
NGC 6218 & M12	&	-1.310 &  \nodata	  &   86.5	& -7.31  & 13.4	&	-1.169 & 0.094 & 0.073	& 	0.279 & 0.164		\\
NGC 6229 & 		&	\nodata 	& \nodata	  &	 291	& -8.06	 & \nodata	&	-1.214 & 0.127 & 0.038	& 	0.189 & 0.276		\\
NGC 6254 & M10	&	-1.575 &  \nodata     &	 184	& -7.48  & 12.4	&	-1.345 & 0.102 & 0.074	& 	0.451 & 0.549		\\
NGC 6341 & M92  &	  \nodata  &   \nodata    &	 268	& -8.21  & 13.2	&	-2.227 & 0.096 & 0.133	& 	0.562 & 0.414		\\
NGC 6388 & 		&	-0.441 &  \nodata     &	1060	& -9.41  & 11.7	&	-0.438 & 0.074 & 0.152	& 	0.341 & 0.078		\\
NGC 6397 & 		&	-1.988 &  \nodata     &	  88.9	& -6.64  & 13.4	&	-1.887 & 0.092 & 0.088	& 	0.451 & 0.408		\\
NGC 6656 & M22	&	\nodata 	& \nodata	  &	 416	& -8.50	 & 12.7	&	-1.524 & 0.112 & 0.092	& 	0.461 & 0.407		\\
NGC 6715 & M54	&	\nodata 	& \nodata	  &	1410	& -9.98	 & 10.8	&	-1.353 & 0.039 & 0.059	& 	0.189 & 0.499		\\
NGC 6752 & 		&	-1.555 &	-1.48 &  239 	& -7.73  & 13.8	&	-1.458 & 0.076 & 0.052	& 	0.634 & 0.455		\\
NGC 6809 & M55	&	-1.934 &  \nodata     &	 188	& -7.57  & 13.8	&	-1.757 & 0.080 & 0.067	& 	0.358 & 0.454		\\
NGC 6838 & M71  &	-0.832 &  \nodata     &	  49.1	& -5.61  & 12.7	&	-0.530 & 0.112 & 0.088	& 	0.463 & 0.099		\\
NGC 7078 & M15  &	-2.320 &	\nodata	  &	 453	& -9.19	 & 13.6	&	-2.218 & 0.121 & 0.136	& 	0.438 & 0.446		\\
NGC 7089 & M2   &	\nodata		&	-1.47 &  582	& -9.03  & 11.8	&	-1.402 & 0.069 & 0.055	& 	0.400 & 0.464		\\
Pal 5	 & 		&	\nodata    &  \nodata	  &	  13.9	& -5.17  & \nodata	&	-1.214 & 0.085 & 0.073	& 	0.053 & 0.130		\\
\cutinhead{}
	 & &	[Al/Fe]	& [Al/Fe] & [Al/Fe]	& 	f$_{\rm enriched}$	&	S1\tablenotemark{b}   &  S1\tablenotemark{b}	  &	  [N/Fe]	& [N/Fe] & S2\tablenotemark{c}	&	S2\tablenotemark{c} \\
	 & &	Average	& Average & Scatter	& 						& Average    &  Scatter  & Average & Scatter & Average & Scatter  	\\
	 & &	$>$0.3dex	& $<$0.3dex & $>$0.3dex	& 						&     &    &  &  &  &  	\\
\cutinhead{}
NGC 104  & 	47 Tuc &	0.586	& 	\nodata	& 	0.128		& 	\nodata	& 	0.393 & 0.074	& 	0.924 & 0.407	& 	0.486 & 0.112		&		\\
NGC 288	 & 			&	0.462 	& 	0.175 	& 	0.121		& 	\nodata	&	0.418 & 0.054	& 	0.832 & 0.341	& 	0.487 & 0.107		&		\\
NGC 362	 & 			&	0.468 	& 	0.049 	& 	0.125		& 	\nodata	&	0.214 & 0.050	& 	1.038 & 0.360	& 	0.306 & 0.112		&		\\
NGC 1851 & 			&	0.495 	& 	0.033 	& 	0.095		& 	\nodata	&	0.251 & 0.056	& 	1.034 & 0.355	& 	0.274 & 0.128		&		\\
NGC 1904 & 	M79		&	0.826 	& 	-0.136	& 	0.288		& 	0.609  	&	0.248 & 0.029	& 	\nodata	  & 	\nodata	& 	\nodata	 & \nodata	&		\\
NGC 2808 & 			&	0.802 	& 	0.025 	& 	0.341		& 	0.391	&	0.203 & 0.056	& 	0.937 & 0.440	& 	0.327 & 0.120		&	\\
NGC 3201 & 			&	0.635 	& 	-0.081	& 	0.198		& 	0.252  	&	0.221 & 0.053	& 	0.789 & 0.351	& 	0.37 & 0.069		&	\\
NGC 4590 & 	M68		&	0.648 	& 	-0.111	& 	0.207		& 	0.545	&	0.323 & 0.093	& 	\nodata	  & 	\nodata	& 	\nodata	  & \nodata	&	\\
NGC 5024 & 	M53  	&	0.917 	& 	-0.061	& 	0.182		& 	0.417 	&	0.444 & 0.101	& 	\nodata	  & 	\nodata	& 	\nodata	  & \nodata	&	\\
NGC 5053 & 			& 	0.772 	& 	-0.029	& 	0.208		& 	\nodata	& 	0.27 & 0.127	& 	\nodata	  & \nodata		& 	\nodata   & \nodata	&	\\
NGC 5139 & 	\ocen	&	0.935	& 	 0.058	& 	 0.389		& 	0.603	&	0.413 & 0.096	& 	1.273 & 0.452	& 	0.642 & 0.177		&	\\
NGC 5272 & 	M3		&	0.809 	& 	-0.027	& 	0.203		& 	0.331	&	0.303 & 0.083	& 	0.861 & 0.297	& 	0.373 & 0.187		&	\\
NGC 5466 & 	  		&	\nodata	&  	-0.161	& 	\nodata		& 	\nodata &	0.258 & 0.058	& 	\nodata	  & 	\nodata	& 	\nodata	  & \nodata	&			\\
NGC 5904 & 	M5		&	0.604 	& 	0.010 	& 	0.196		& 	0.484	&	0.307 & 0.078	& 	1.094 & 0.393	& 	0.359 & 0.154	&	\\
NGC 6121 & 	M4		&	0.709	& 	\nodata		& 	0.121	& 	\nodata	&	0.489 & 0.064	& 	0.894 & 0.269	& 	0.376 & 0.086 	&\\
NGC 6171 & 	M107 	&	0.538	& 	\nodata		& 	0.118	& 	\nodata &	0.429 & 0.087	& 	0.911 & 0.468	& 	0.6 & 0.123 &	\\
NGC 6205 & 	M13  	&	0.860	& 	-0.050	& 	0.325		& 	0.644	&	0.368 & 0.097	& 	1.248 & 0.268	& 	0.471 & 0.116	&	\\
NGC 6218 & 	M12		&	0.444	& 	0.154 	& 	0.088		& 	\nodata	&	0.373 & 0.064	& 	1.028 & 0.347	& 	0.548 & 0.089	&	\\
NGC 6229 & 			&	\nodata		& 	0.057 	& 	\nodata	& 	\nodata	&	0.283 & 0.056	& 	0.571 & 0.052	& 	\nodata	  & \nodata	 & 	\\
NGC 6254 & 	M10		&	0.981	& 	-0.039	& 	0.265		& 	0.481	&	0.317 & 0.066	& 	1.136 & 0.291	& 	0.512 & 0.096	&	\\
NGC 6341 & 	M92  	&	0.770	& 	-0.092	& 	0.197		& 	0.759	&	0.439 & 0.087	& 	\nodata	  & 	\nodata	& 	\nodata	  & \nodata		&			\\
NGC 6388 & 			&	0.381	& 	\nodata		& 	0.045	& 	\nodata	&	0.158 & 0.088	& 	1.020 & 0.323	& 	0.341 & 0.098	&	\\
NGC 6397 & 			&	0.701	& 	-0.094	& 	0.177		& 	0.686	&	0.338 & 0.092	& 	\nodata	  & 	\nodata	& 	\nodata	  & \nodata	&  	\\
NGC 6656 & 	M22		&	0.662	& 	-0.100	& 	0.248		& 	\nodata	&	0.306 & 0.111	& 	\nodata	  & 	\nodata	& 	\nodata	  & \nodata		&  	\\
NGC 6715 & 	M54		&	\nodata		& 	-0.072	& 	\nodata	& 	\nodata	&	0.243 & 0.025	& 	\nodata	  & \nodata		& 	\nodata	  & \nodata			&	\\
NGC 6752 & 			&	0.832	& 	0.004 	& 	0.326		& 	0.761	&	0.365 & 0.053	& 	1.054 & 0.197	& 	0.38 & 0.106	&	\\
NGC 6809 & 	M55		&	0.734	& 	-0.066	& 	0.249		& 	0.531	&	0.378 & 0.051	& 	1.093 & 0.102	& 	\nodata	  & \nodata		&	\\
NGC 6838 & 	M71  	&	0.477	& 	\nodata		& 	0.088	& 	\nodata &	0.318 & 0.080	& 	0.992 & 0.441	& 	0.661 & 0.113	 	& \\
NGC 7078 & 	M15  	&	0.752	& 	-0.056	& 	0.231		& 	0.613  	&	0.417 & 0.097	& 	\nodata	  & \nodata		& 	\nodata   & \nodata	 	&		\\
NGC 7089 & 	M2   	&	0.785	& 	-0.061	& 	0.212		& 	0.545   &	0.313 & 0.048	& 	1.058 & 0.132	& 	0.413 & 0.154			&	\\
Pal 5	 & 			&	\nodata		& 	-0.009	& 	\nodata	& 	\nodata	&	0.229 & 0.044	& 	0.699 & 0.224	& 	0.353 & 0.087			&
\enddata
\tablecomments{This table lists statistics for 31 GCs remaining after our refining procedure described in Section~2 and 3.1. 
Scatter is defined as the standard deviation around the mean. Masses are taken from \citet{baum01}, and we use the ages compiled 
by \citet{krause01}.}
\tablenotetext{a}{The error of [Fe/H] is the average uncertainty for a given cluster.}
\tablenotetext{b}{[(Mg+Al+SI)/Fe].}
\tablenotetext{c}{[(C+N+O)/Fe].}
\end{deluxetable*}

In this paper we discuss 21 new (mostly southern) clusters observed from both LCO and APO by following the same steps of atmospheric 
parameter and abundance determination as \citet{masseron01} and combine them with the 10 northern clusters discussed 
by \citet{masseron01}. Because M12 was observed from both observatories, we use this cluster to check how homogeneous the 
abundances are from APO and LCO. By combining observations from APO and LCO, we are able to discuss the statistics of 
Al-Mg and N-C anticorrelations as a function of main cluster parameters in a much larger sample of clusters than was  
previously possible. Na-O anticorrelation is not included in our study, because Na lines in the H-band are too weak 
to be observable in almost all of our sample of clusters.

There are various labels used in the literature for stars within GCs that are enriched in He, N,
Na, Al and are depleted in O, C and Mg, such as second generation 
stars and chemically enriched stars. We will use the term second generation/population (SG) stars when referring to stars that 
have [Al/Fe]$>$0.3~dex, and first generation/population (FG) when [Al/Fe]$<$0.3~dex (see Section 5.1). While more than 
two populations 
can be identified based on abundances in some clusters, we focus on simplifying the term to refer to all stars that 
satisfy the above criteria, as second/first generation/population stars for easier discussion. On the other hand 
most metal-rich clusters ([Fe/H]$>-$1) are enriched in Al ([Al/Fe]$>$0.3~dex), but appear to host only a single population of stars, 
so they are chemically enriched but any possible SG stars have the same [Al/Fe] content as FG stars within our errors 
(see Section~7.3 for more discussion). We treat these clusters as having one FG star group when looking at MPs based on Al 
abundances.

\section{Membership Analysis}

\begin{figure}[!ht]
\centering
\includegraphics[width=3.45in,angle=0]{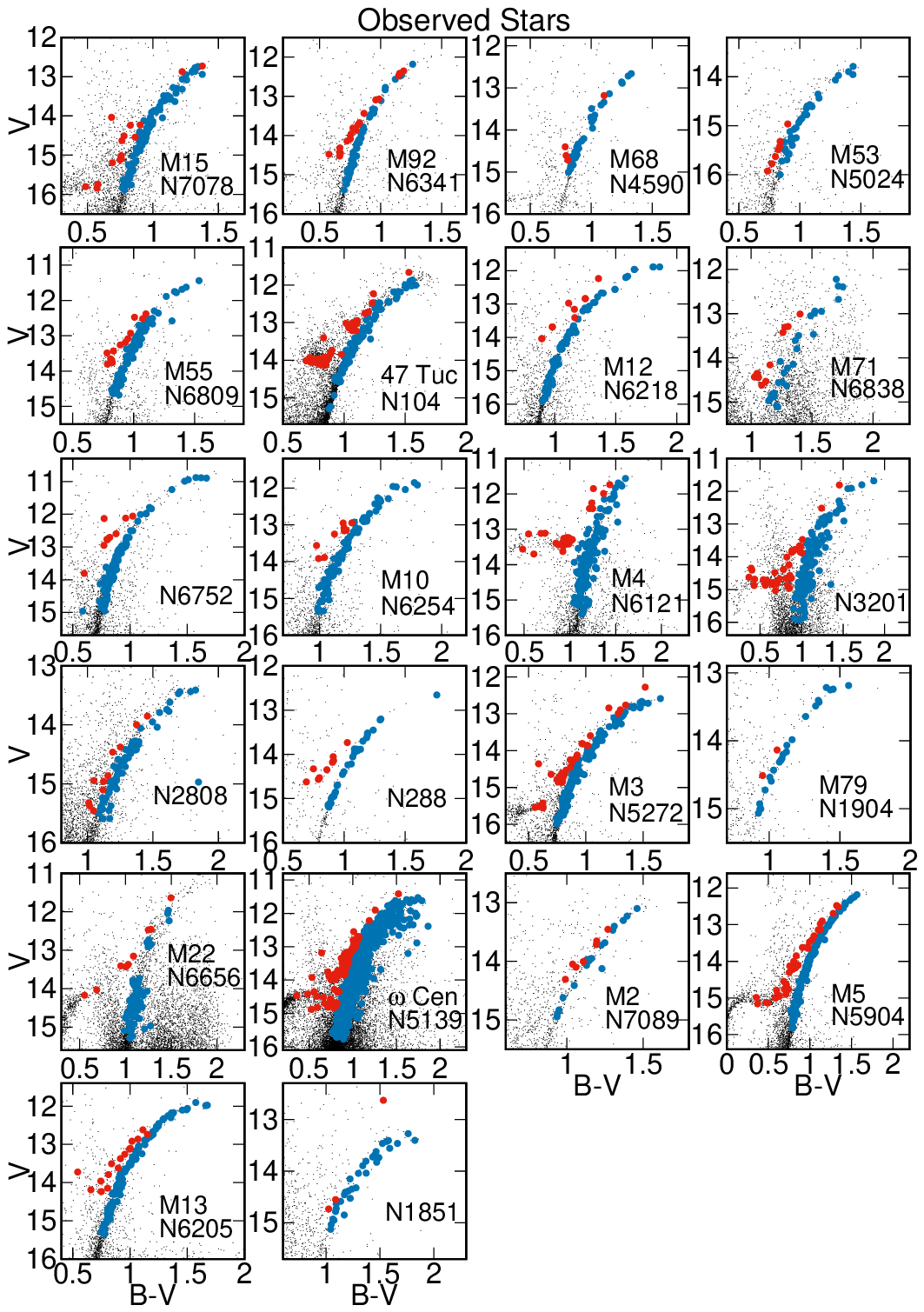}
\caption{The CMD of observed stars by APOGEE in 22 clusters in common with \citet{stetson01}. 
AGB/HB stars are denoted by red dots, the RGB stars are by blue dots.
}
\label{fig:cmd}
\end{figure}

Table~1 lists the globular clusters observed by APOGEE-2, along with the main parameters from 
the Harris catalog \citep{harris01}, Gaia DR2 \citep{baum02} and from \citep{baum01}. A more detailed description of 
the general target selection of APOGEE and APOGEE-2 can be found in \citet{zasowski01} and \citet{zasowski02}, 
respectively. Our target selection follows that of \citet{meszaros03} and  \citet{masseron01}. We select stars based on their 
radial velocity first, their distance from cluster center second, and their metallicity third. In radial velocity, we 
required stars to be within three times the velocity dispersion of the mean cluster velocity, and in distance we 
required stars to be within the tidal radius. The metallicity cut was usually 
set to $\pm$0.5~dex around the cluster average, except for clusters with suspected intrinsic Fe spread for which the 
metallicity cut was skipped, or only obvious field stars were deleted (for example stars with solar-like metallicity in 
otherwise metal-poor clusters). For this paper 
we made important updates by selecting the average cluster radial velocity and its scatter from Gaia DR2 \citep{baum01}
rather than from \citet{harris01}. In addition, we introduced a fourth step that is based 
upon selecting stars that have proper motion within a 1.5$-$2.5 mas yr$^{-1}$ range (depending on the cluster) 
around the cluster average proper motion from the Gaia DR2 catalog \citep{gaia01}. 

These two improvements were not adopted by \citet{masseron01}, but now we refine the list of stars presented 
in that study. While the selected members of those 10 northern clusters have only changed slightly, because some stars 
were added or deleted, we did not 
re-derive atmospheric parameters and abundances for stars that remained members, as our analysis method has not 
changed. It is important to note that only a couple of stars have been deleted from these GCs, and the main science 
results and conclusions presented in \citet{masseron01} remain the same. However, all figures, including data for 
those 10 clusters have been updated for this paper. 

The individual atmospheric parameters and the derived abundances are listed in Table~2, while the 
abundance averages and RMS scatters for each cluster are presented in Table~3. 
Table~2 contains results for all stars and clusters that were analysed, altogether 3382 stars in 44 clusters. 
However, we do not discuss all clusters and stars. We make a quality selection according to the following criteria.
High S/N spectra are essential to determine abundances from atomic and molecular features.  
Most of the tests done by the APOGEE team concluded that abundances become reliable around S/N=70$-$100, however, objects with 
poorer S/N have also been analyzed and included in Table~2. The spectra have been processed by the APOGEE data processing 
pipeline \citep{nidever01}. Another criterion was that a cluster has to have at least 5 members with 
S/N$>$70 to qualify for further analysis. The following clusters did not meet this criterion: NGC~4147, NGC~5634, 
NGC~6316, NGC~6528, NGC~6539, NGC~6760, Pal~6 and Terzan~12. While we do not use these clusters in our analysis, their abundances and 
atmospheric parameters were derived and listed in Table~2 for reference. The remaining 36 clusters were further refined
based on their reddening values as described in the next section. Table~3 contains the clusters remaining 
after our refining procedure.


\section{Atmospheric Parameters and Abundances}

\subsection{BACCHUS description}

Since the method of deriving atmospheric parameters and abundances is identical to that of 
\citet{masseron01}, we only give a short overview of it in this paper. We use the Brussels Automatic Code for 
Characterizing High accUracy Spectra (BACCHUS) \citep{masseron02} to determine the metallicity and abundances, but not  
effective temperatures and surface gravities. Microturbulent velocities were computed from the surface gravities using 
the following equation: 
$$v_{\rm micro} = 2.488 - 0.8665 \cdot {\rm log}~g + 0.1567 \cdot {\rm log}~g \cdot {\rm log}~g$$ 

This relation was originally determined from the Gaia-ESO survey by cancelling the trend
of abundances against equivalent widths of selected Fe I lines \citep{masseron01}. The validity of this relation in the H-band 
was checked by \citep{masseron01}. 
Due to problems with ASPCAP \citep{perez01} effective temperatures at low 
metallicities, [M/H]$<-$0.7~dex \citep[detailed by][]{meszaros03, jonsson01, masseron01, nataf01, nidever02}, these were computed 
from 2MASS colors using the equations from \citet{gonzalez01}. Surface gravities were derived from isochrones 
\citep{bertelli01, bertelli02, marigo01} by taking into account their evolutionary state. The log~g was determined by taking 
the photometric effective temperature and reading the log~g, by interpolating through surface gravities, corresponding to that 
effective temperature from the isochrone. AGB and RGB stars were selected by combining our list of stars with the 
ground-based photometric catalog compiled by \citet{stetson01} for 22 clusters in common with our sample. Our selection 
was based on the star's position on the V$-$(B$-$V) color-magnitude diagram \citep[see e.g.,][]{garcia03} shown in 
Figure~\ref{fig:cmd}. For clusters not listed in the Stetson catalog we assumed all stars to be on the RGB. 
For further information on our abundance determination methods, 
comparisons to ASPCAP, and their accuracy and precision (generally below 0.1~dex) we refer the reader to Section~3 
of \citet{masseron01}. The absorption lines selected for abundance determination are the same as used by \citet{masseron01}. 
Random errors were derived from the line-by-line abundance dispersion.

\begin{deluxetable*}{lrrllll}[!ht]
\tabletypesize{\scriptsize}
\tablecaption{Overview of Homogeneous Spectroscopic Surveys of Globular Clusters}
\tablewidth{0pt}
\tablehead{
\colhead{Reference} & 
\colhead{N$_{\rm stars}$} & 
\colhead{N$_{\rm cl}$} & 
\colhead{Element Pairs\tablenotemark{a}} & 
\colhead{Observatory\tablenotemark{b}} & 
\colhead{Survey} &
\colhead{Comments} 
}
\startdata
\citet{carretta02, carretta03, carretta01}	& 	1958	&	19	&	Na-O, Al-Mg	& ESO/VLT	& Carretta	&	UVES/Giraffe combined.	\\
\citet{meszaros03}							&	 428	&	10	&	Al-Mg, N-C	& APO	& APOGEE	&		\\	
\citet{pancino01}							&	 572	&	9	&	Al-Mg		& ESO/VLT	& Gaia-ESO	&		\\
\citet{masseron01}							&	 885	& 	10	&	Al-Mg, N-C	& APO	& APOGEE	& 	Same clusters as \citet{meszaros03}.	\\
\citet{nataf01}								&	1581	&	25	&	Al-Mg, N-C	& APO/LCO	& APOGEE	&	Payne analysis only. \\
							&		&		&		& 	& 	&	\\
This paper									&	2283	&	31	&	Al-Mg, N-C	& APO/LCO	& APOGEE	&	Includes data from \citet{masseron01}. 
\enddata
\tablecomments{Clusters with less than 5 observed members were excluded from the statistics.}
\tablenotetext{a}{The main element pairs used to study multiple populations.} 
\tablenotetext{b}{ESO/VLT: Very Large Telescope at the European Southern Observatory, 
APO: Apache Point Observatory, LCO: Las Campanas Observatory.}
\end{deluxetable*}

The use of photometric temperatures introduces its own set of problems mostly related to high E(B$-$V) values.
The \citet{gonzalez01} relations are very sensitive to small changes in E(B$-$V), which is very important in high reddening 
clusters that may in addition suffer from significant differential reddening inside the cluster. 
For this reason the list of clusters 
was further limited by removing clusters with E(B$-$V)$>$0.4 according to the Harris catalog. 
Our metallicities derived from highly reddened spectra are also significantly larger than what the optical 
studies have found making us believe that either reddening and/or photometric temperatures are not reliable when 
E(B$-$V)$>$0.4. This issue is explored in more detail in Section~4.1. The following five clusters have at least 5 members 
with S/N$>$70, but have E(B$-$V)$>$0.4: NGC~6441, NGC~6522, NGC~6544, NGC~6553 and Terzan~5. 
The final sample after the S/N and reddening cuts includes
2283 stars in 31 clusters, and we use this sample to study statistics of Mg-Al and N-C anticorrelations 
throughout the paper. Previous homogeneous surveys are listed in Table~4 for easy comparison.

\begin{deluxetable}{lrrr}[!ht]
\tabletypesize{\scriptsize}
\tablecaption{Selected Parameter Cuts for Analysis}
\tablewidth{0pt}
\tablehead{
\colhead{Abundance} & 
\colhead{T$_{\rm eff}$} & 
\colhead{[Fe/H]} & 
\colhead{$\sigma_{\rm [X/Fe]}$} \\
\colhead{} & 
\colhead{K} & 
\colhead{dex} & 
\colhead{dex}
}
\startdata
$\rm [C/Fe]$	& $<$4600 & $>-$1.9	& $<$0.2	\\
$\rm [N/Fe]$ 	& $<$4600 & $>-$1.9	& $<$0.2	\\
$\rm [O/Fe]$ 	& $<$4600 & $>-$1.9	& $<$0.2 	\\
$\rm [Mg/Fe]$ 	& $<$5500 & $>-$2.5	& $<$0.2 	\\
$\rm [Al/Fe]$ 	& $<$5500 & $>-$2.5	& $<$0.2 	\\
$\rm [Si/Fe]$ 	& $<$5500 & $>-$2.5	& $<$0.2 	\\
$\rm [K/Fe]$ 	& $<$4600 & $>-$1.5	& $<$0.2 	\\
$\rm [Ca/Fe]$ 	& $<$5500 & $>-$2.5	& $<$0.2 	\\
$\rm [Fe/H]$  	& $<$5500 & $>-$2.5	& $<$0.2 	\\
$\rm [Ce/Fe]$ 	& $<$4400 & $>-$1.8	& $<$0.2 	\\
$\rm [Nd/Fe]$ 	& $<$4400 & $>-$1.8	& $<$0.2 	
\enddata
\tablecomments{A S/N$>$70 cut is also applied. All averages and scatter values were computed using stars that satisfy 
these conditions including the figures shown in the paper.}
\end{deluxetable}

While Table~2 lists all abundances we were able to measure regardless of S/N, we introduced the previously mentioned 
S/N$>$70 cut in all figures and statistics. Upper limits are also listed in Table~2, but not plotted in any of the figures, or 
included when calculating cluster averages and scatters, because 
we made the decision to study the behavior of anticorrelations based on only real measurements. We 
implemented a maximum temperature cut of 4600~K for CNO and K, because for higher temperatures the 
molecular (atomic in case of K) lines become too weak, rendering abundances of these elements unreliable. 
We use 5500~K for the rest of the elements as maximum temperature above which errors 
start to significantly increase. Stars plotted in all figures in Sections~4 to 8 have elemental abundances with internal errors 
smaller than 0.2~dex to reduce contamination from highly unreliable measurements. Stars with abundances 
outside these parameter regions are published in Table~2, but we caution the reader to carefully examine these values before 
drawing scientific conclusions. These limitations 
were set in place when calculating abundance averages and scatter for all clusters and are listed in Table~5. The error 
in the mean [Fe/H] is smaller than the dot used to represent the data in all figures, thus errorbars were not plotted in any of 
the figures. For the abundance$-$abundance plots we only highlighted the average error of each abundance for simplicity, but 
Table~2 lists all individual errors.

\begin{figure}[!ht]
\includegraphics[width=3.45in,angle=0]{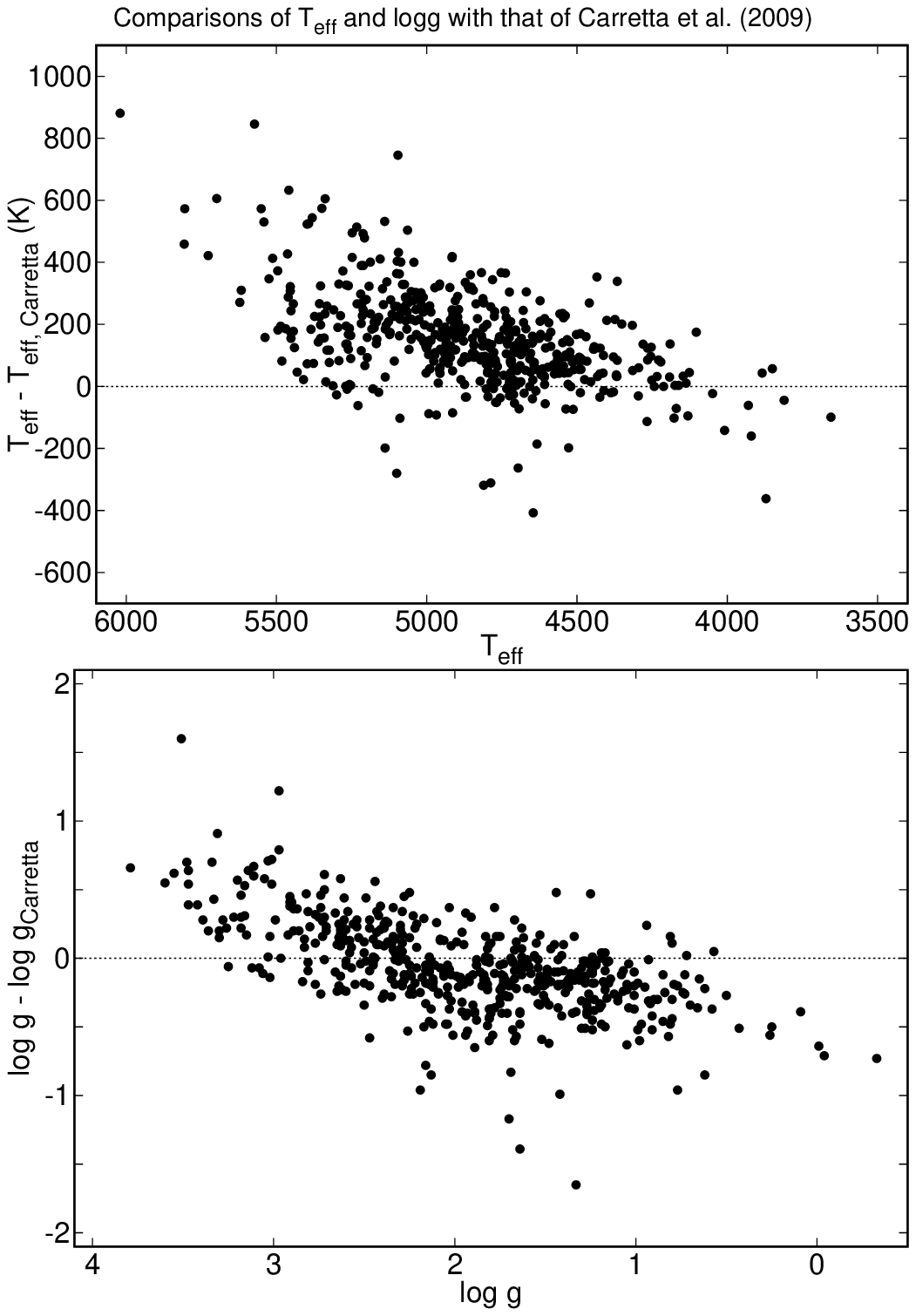}
\caption{Top panel: Comparisons of our T$_{\rm eff}$ scale from \citet{gonzalez01} with \citet{carretta02, carretta03, carretta01}, 
who used \citet{alonso01, alonso02}. Bottom panel: Comparisons of our surface gravities with the same source. 
}
\label{fig:teffcomp}
\end{figure}

\subsection{Comparisons of T$_{\rm eff}$ and log~g Values with the Literature}

We limit our discussion of comparisons of T$_{\rm eff}$ and log~g with literature to that of \citet{carretta02, carretta03, 
carretta01}, since that is the literature source we have the most stars in common with, 514 altogether, out of the list of papers 
in Table~4. Star identification from \citet{carretta03} was added to Table~2 for easy comparison. 

The difference between our parameters and those of \citet{carretta02, carretta03, carretta01} can be seen in 
Figure~\ref{fig:teffcomp}.
The systematic offset seen between the two temperatures are the characteristics of the photometric temperature conversions 
(and differences in colors used to calculate the temperature) of \citet{gonzalez01} and \citet{alonso01, alonso02}, which was 
used by \citet{carretta02, carretta03, carretta01}. The temperature difference is generally between $\pm$300~K, but it increases 
with increasing temperature. 

Similar structure can be seen when comparing surface gravities, because the temperature and log~g have a simple 
linear correlation on the RGB, so any systematic difference seen in the temperature scale will propagate to log~g. These discrepancies 
may also propegate to metallicity, further discussed in Section~4.1, and/or individual abundances, which is expected when 
temperature scales differ from one another.

\subsection{Comparisons of APO and LCO Observations}

As mentioned at the end of the introduction, APOGEE-2 uses two spectrographs identical in design at two observatories, 
APO and LCO to map all parts of the Milky Way. The identical design makes it possible to directly derive  
atmospheric parameters and abundances that are believed to be on the same scale by observing the same stars 
from both observatories. The observing strategy is carefully planned \citep{zasowski02} to observe stars with both telescopes that 
cover the full parameter range ASPCAP operates in so that any differences between the final results can be carefully 
studied and calibrated if necessary. In terms of globular clusters, there is only one that has been observed 
with both the northern and southern telescopes: M12, which limits our comparisons to a small range in metallicity. 

\begin{figure}[!ht]
\includegraphics[width=3.45in,angle=0]{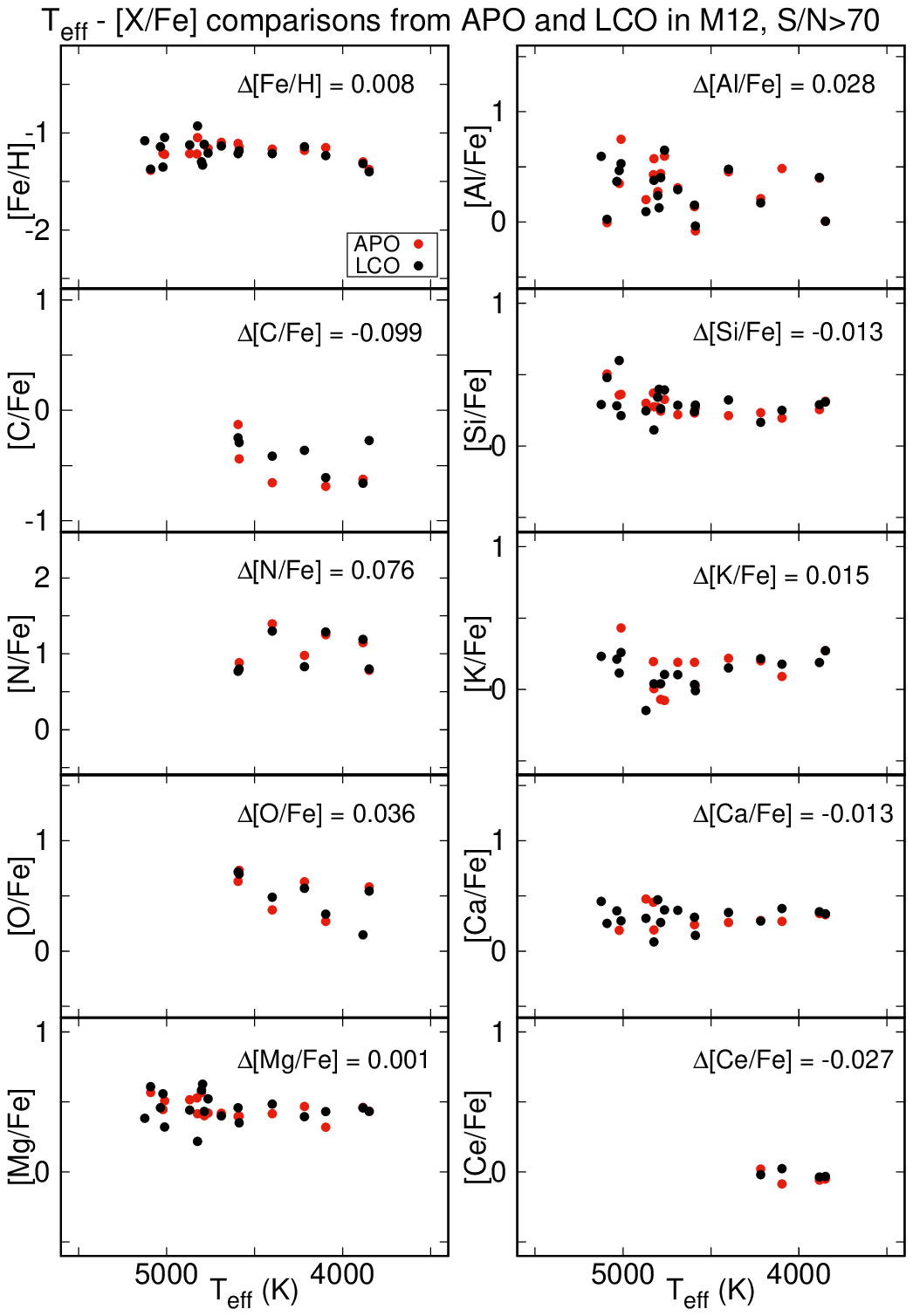}
\caption{Comparison of stars observed from both APO (red dots) and LCO (black dots) in M12. The differences between APO and LCO 
printed in each panel are on the level or smaller than the average internal error of each element.
}
\label{fig:m12}
\end{figure}

\begin{figure*}[!ht]
\centering
\includegraphics[width=4.4in,angle=270]{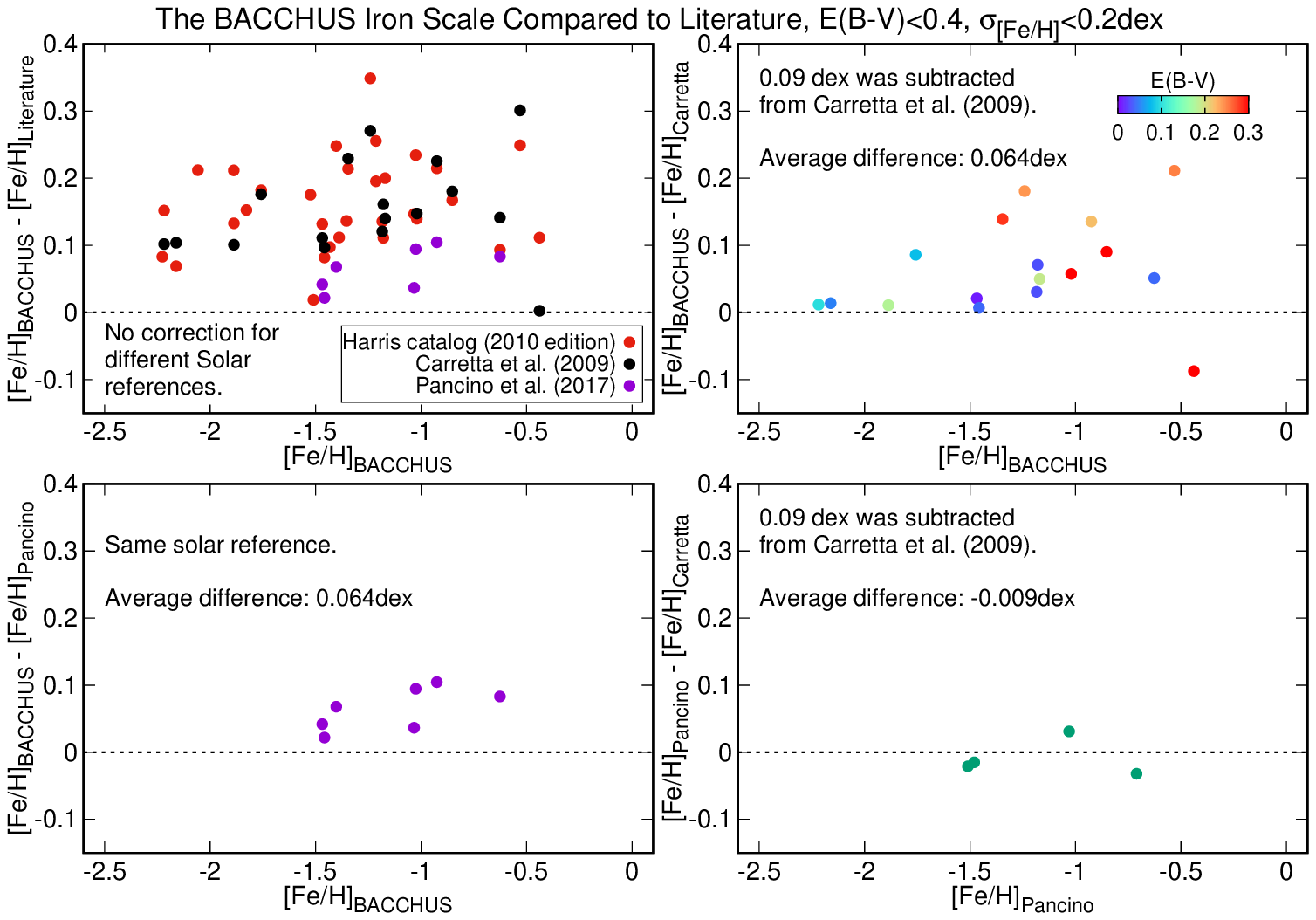}
\caption{Comparison of mean [Fe/H] cluster values from various literature sources. Differences in the solar reference 
Fe abundances was corrected where indicated. The three different Fe scales agree roughly within $\pm$0.1~dex after correction.
}
\label{fig:fehscales}
\end{figure*}

Figure~\ref{fig:m12} shows the BACCHUS derived abundances as a function of effective temperature of the 21 stars 
in M12 that were both observed from APO and LCO. The difference is calculated for each star that was observed 
with both telescopes and then averaged together over all the stars. The differences between the two sets of observations 
range between 0.001~dex for [Mg/Fe] to 0.099~dex for [C/Fe], all of which can be considered as a very good agreement. 
The discrepancy for C, N and O 
are generally larger than for the rest of elements, which is understandable considering that it is more difficult to 
fit these molecular lines than simple unblended atomic absorption lines. All the differences are on par or smaller than 
the average error in M12, and thus we conclude that observations from APO and LCO can be directly compared to each other without
worrying about any possible large systematic errors. While this test is limited to a unique metallicity ([Fe/H]=$-$1.2), similar 
tests on much lager samples of APO-LCO overlapping stars have been done on the ASPCAP-analysis 
of the DR16 data, suggesting that the data from APO and LCO indeed are of similar quality and 
yield very similar stellar parameters  and abundances \citep{hol03}.


\section{The Fe scale}

The amount of iron observed in GCs allows the investigation of the history of stars and intra-cluster medium 
from which the GCs have formed, because Fe is mostly the result of core-collapse supernovae of high and intermediate mass stars. 
Additionally, Fe is traditionally used as the tracer of metallicity - the overall abundance of metals in a star. 
Abundances of iron from homogeneous high-resolution spectroscopic studies are also used to calibrate low-resolution 
spectroscopic and photometric indices. Setting a true and absolute Fe scale is, thus, one of the most important goals 
of high-resolution abundance analysis. 

\subsection{Comparisons with Literature}

We compare our metallicity scale with those of the Harris catalog, \citet{carretta01} and 
Gaia-ESO \citep{pancino01}. The Harris catalog is a compilation of various literature sources and 
all our clusters were selected from it. The largest homogeneous study of iron abundances from high-resolution spectra was 
previously carried out by \citet{carretta01}, 17 of their clusters are in common with our sample, and we have 7 clusters 
that were also observed by Gaia-ESO. We show the four different iron scales on the top left panel of Figure~\ref{fig:fehscales}.
We find that the [Fe/H] metallicities we derive are on average 0.162~dex higher than those from the Harris catalog, 
0.154~dex higher than \citet{carretta01}, 0.064~dex higher than \citet{pancino01}.

\begin{figure*}[!ht]
\centering
\includegraphics[width=4.4in,angle=270]{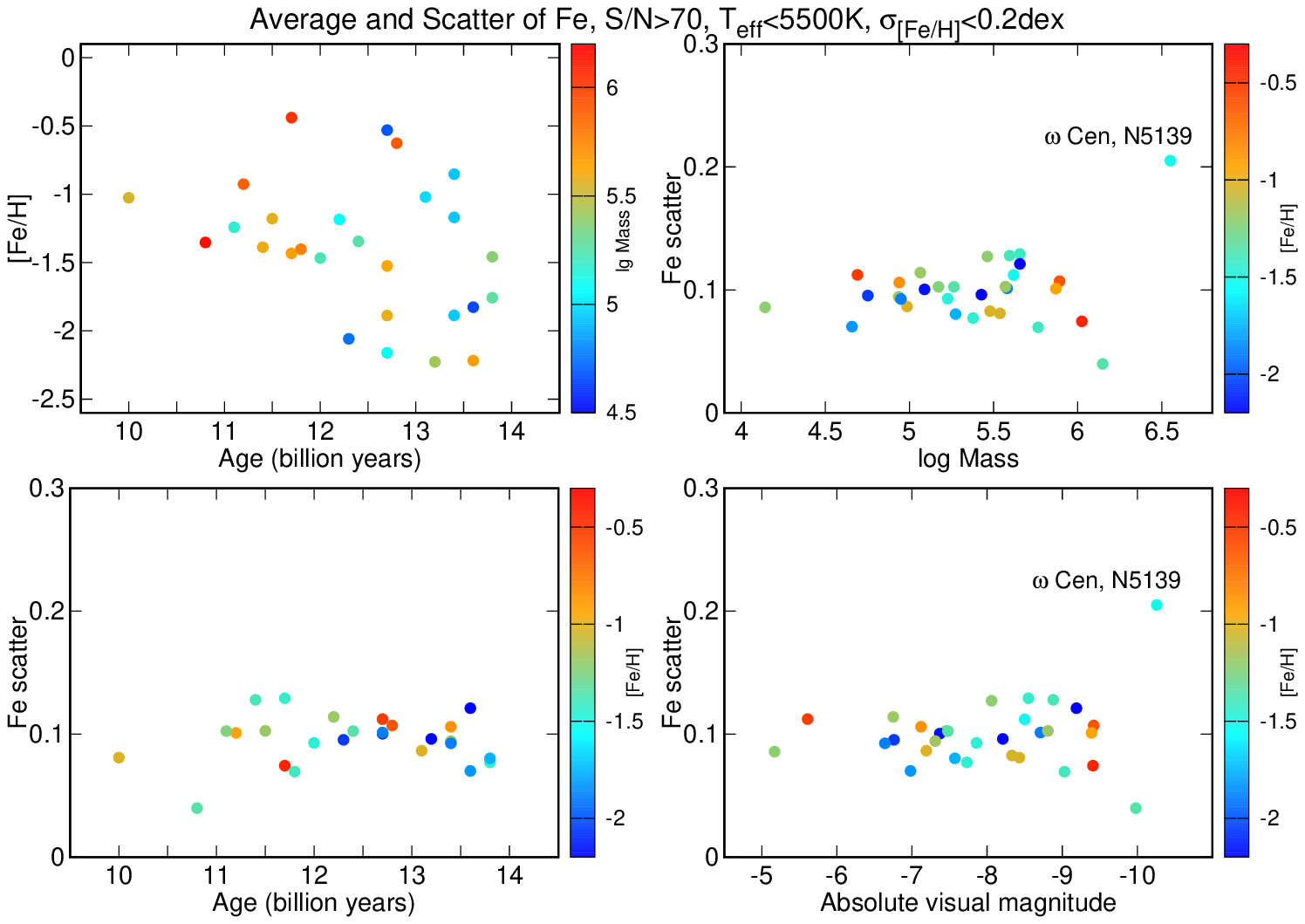}
\caption{The age-metallicity relation and spread of Fe as a function of cluster properties. No significant correlation is observed with mass, 
V$_{\rm Abs}$ and age. \ocen \ is the only cluster with significant Fe variations from our sample. 
}
\label{fig:fehspread}
\end{figure*}

These metallicity differences of GCs have been present in the APOGEE data since the very first data release 
\citep{meszaros02} and remained in place in all subsequent data releases \citep{hol01, hol02, jonsson01}. This was verified by 
\citet{meszaros03} and by \citet{masseron01} using the APOGEE line list, but effective temperatures and surface gravities independent 
of ASPCAP. Interestingly, 
\citet{pancino01} have also found a similar, although slightly smaller, 0.08~dex higher metallicities than \citet{carretta01} 
in the 7 clusters in common with our sample. This latter study was carried out by the Gaia-ESO survey, completely independent of 
APOGEE observations and using optical spectra instead of the H-band. We speculate that the nature of these discrepancies between 
the three different studies can be attributed to three main factors: 

1. Most of the differences can be explained by the choice of the reference solar abundance table.  \citet{carretta01} and some of the 
compilation found in the Harris catalog used the Fe reference value of A(Fe)$_{\odot}$=7.54 derived by \citet{gratton03}, 
while \citet{pancino01} and APOGEE use 7.45 from 
\citet{gre01}. The difference of 0.081~dex between \citet{pancino01} and \citet{carretta01} is on the level of the change 
coming from the different solar Fe references. After applying a correction to account for the different solar reference values, the 
\citet{pancino01} results become almost identical (difference is $-$0.009~dex on average) to the \citet{carretta01} 
results (bottom right panel in Figure~\ref{fig:fehscales}). An important aspect of the \citet{pancino01} study is that it 
used spectroscopic temperatures directly derived from the VLT spectra, while \citet{carretta01} used the \citet{alonso01, alonso02}
conversions. The difference between our study and that of 
\citet{carretta01} reduces to 0.064~dex on average, with a dispersion of 0.073~dex, which is not too different from 
the statistical uncertainties given by \citet{carretta01}.

2. A separate comparison to \citet{carretta01} is shown on the upper right panel of Figure~\ref{fig:fehscales} after both 
metallicities are converted to the same scale by subtracting 0.09~dex from \citet{carretta01}.
A slight correlation can be seen between these two homogeneous studies that is dependent on the E(B$-$V) value of each cluster. 
For most of the clusters with E(B$-$V)$>$0.2 (NGC~2808, NGC~3201, M10, and M71), we find higher metallicities than 
for clusters with low reddening, which are still slightly more metal-rich than \citet{carretta01}. While one cluster with high 
reddening, M4, have an average metallicity closer to that of \citet{carretta01} after the correction, we believe that either the 
photometric calibration does not work at high reddening, the reddening of these clusters is not correct, or this is the result 
of a systematic difference in the temperature scales of \citet{gonzalez01} and that of \citet{alonso01, alonso02}, or 
a combination of any of these. Generally, these photometric temperature conversions are very sensitive to the reddening, so a small 
error in the E(B$-$V) can lead to a large change in temperature, perhaps pushing M4 closer to \citet{carretta01}. 
Also, considering that small errors of E(B$-$V) may result in large errors 
in temperature, and high reddening clusters may have significant differential reddening, we 
exclude the five clusters with E(B$-$V)$>$0.4 listed in Section~3.1 from our study.

\begin{deluxetable*}{lrrrrrl}[!ht]
\tabletypesize{\scriptsize}
\tablecaption{Correlation of Parameters with Cluster Properties}
\tablewidth{0pt}
\tablehead{
\colhead{Parameter} & 
\colhead{N\tablenotemark{a}} & 
\colhead{a} & 
\colhead{b} &
\colhead{r} &
\colhead{p-value} & 
\colhead{Comments} \\
\colhead{Pair} & 
\colhead{} & 
\colhead{} & 
\colhead{} &
\colhead{} & 
\colhead{} & 
\colhead{}  
}
\startdata
Fe average $-$ Age				& 27 & 	\nodata	&  \nodata	&  \nodata 	& \nodata 	&   Non-linear correlation, Section~4.2. \\
Fe scatter $-$ log(Mass)		& 30 & 	-0.002690	&  0.110497 	&  -0.0615 	& 0.7468 	&   No correlation, \ocen \ not fitted.	\\
Fe scatter $-$ V$_{\rm Abs}$	& 30 &   0.001578 	&  0.108669 	&  0.0915 	& 0.6306 	&	No correlation, \ocen \ not fitted.	\\
Fe scatter $-$ Age				& 27 &   0.001722 	&  0.074533 	&  0.0884 	& 0.6610	&	No correlation. 	\\
\hline
\hline
N$_{\rm SG}$/N$_{\rm tot}$ $-$ [Fe/H]			& 16 &  -0.182509	&   0.249953	&  -0.4899 	& 0.0541 	&   Weak/No correlation, Section~5.4.	\\
N$_{\rm SG}$/N$_{\rm tot}$ $-$ log(Mass)		& 16 & 	-0.013256	&   0.613653   	&  -0.0348	& 0.8982	&   No correlation.	\\
N$_{\rm SG}$/N$_{\rm tot}$ $-$ V$_{\rm Abs}$	& 16 &   0.020454 	&   0.710909 	&  0.1328 	& 0.6239 	&	No correlation. 	\\
N$_{\rm SG}$/N$_{\rm tot}$ $-$ Age				& 15 &   0.105273	&  -0.770896 	&  0.6850 	& 0.0048 	&	Strong correlation, Section~5.4. 	\\
\hline
\hline
Al scatter $-$ [Fe/H]			& 31 & \nodata	&  \nodata 	&  \nodata 	& \nodata 	&   Non-linear correlation, Section~6.1.	\\
Al scatter $-$ log(Mass)		& 10 & 	0.071677	&   0.054672  	&  0.7426 	& 0.0139 	&   Moderate correlation, [Fe/H]$<-$1.5, Section~6.2.	\\
Al corrected scatter $-$ log(Mass)		& 9 & 	0.099891	&   -0.310714  	&  0.8134 	& 0.0077 	&   Moderate correlation, [Fe/H]$<-$1.5, Section~6.2.	\\
Al scatter $-$ V$_{\rm Abs}$	& 10 & -0.030952 	&   0.190358 	&  -0.6809 	& 0.0301 	&	Moderate correlation, [Fe/H]$<-$1.5, Section~6.2. 	\\
Al corrected scatter $-$ V$_{\rm Abs}$	& 9 & -0.038668 	&   -0.082526 	&  -0.7091 	& 0.0324 	&	Moderate correlation, [Fe/H]$<-$1.5, Section~6.2. 	\\
Al scatter $-$ Age				& 27 &  \nodata	&  \nodata 	&  \nodata 	& \nodata 	&	Non-linear correlation, Section~6.3. 	\\
\hline
\hline
Mg+Al+Si average $-$ [Fe/H]			& 31 & 	-0.050382	&   0.254633 	&  -0.2851 	& 0.1200 	&   No correlation.	\\
Mg+Al+Si average $-$ log(Mass)		& 31 & 	-0.000689	&   0.327513  	&  -0.0041	& 0.9825	&   No correlation.	\\
Mg+Al+Si average $-$ V$_{\rm Abs}$	& 31 &   0.002073	&   0.340408 	&  0.0299	& 0.8731 	&	No correlation. 	\\
Mg+Al+Si average $-$ Age			& 27 &   0.053967	&   -0.341424 	&  0.6468 	& 0.0003 	&	Strong correlation, Section~6.4. 	\\
\hline
\hline
Mg+Al+Si scatter $-$ [Fe/H]			& 31 & 	-0.016603	&   0.049214  	&  -0.3227 	& 0.0767 	&   Weak/No correlation, Section~6.4.	\\
Mg+Al+Si scatter $-$ log(Mass)		& 31 & 	 0.003140	&   0.055162  	&  0.0641	& 0.7323	&   No correlation.	\\
Mg+Al+Si scatter $-$ V$_{\rm Abs}$	& 31 &  -0.001116	&   0.063077 	&  -0.0553	& 0.7689 	&	No correlation. 	\\
Mg+Al+Si scatter $-$ Age			& 27 &   0.006675	&  -0.009527  &  0.2737 	& 0.1671 	&	No correlation. 	\\
\hline
\hline
N scatter $-$ [Fe/H]			& 21 &  0.187394	&   0.52432 	& 0.5341	& 0.0126 	&   Moderate correlation, Section~7.2.	\\
N scatter $-$ log(Mass)			& 21 &  0.030739	&   0.144789 	& 0.1367 	& 0.5546 	&   No correlation.	\\
N scatter $-$ V$_{\rm Abs}$		& 21 & -0.012339 	&   0.211891	& -0.1345 	& 0.5611	&	No correlation. 	\\
N scatter $-$ Age				& 17 & -0.024625	&   0.621128 	& -0.2533 	& 0.3266 	&	No correlation. 	\\
\hline
\hline
C+N+O average $-$ [Fe/H]		& 19 & 	0.015709	&   0.45331 	&  0.0431 	& 0.8609 	&   No correlation.	\\
C+N+O average $-$ log(Mass)		& 19 & -0.013718	&   0.50998  	&  -0.0650	& 0.7915 	&   No correlation.	\\
C+N+O average $-$ V$_{\rm Abs}$	& 19 &  0.015851	&   0.56303 	&  0.1841	& 0.4506 	&	No correlation. \\
C+N+O average $-$ Age			& 16 &  0.060981	&  -0.30185		&  0.6011 	& 0.0138 	&	Moderate correlation, Section~7.4.	\\
\hline
\hline
C+N+O scatter $-$ [Fe/H]		& 19 & 	-0.029856	&   0.085138  	&  -0.3024 	& 0.2083 	&   No correlation.	\\
C+N+O scatter $-$ log(Mass)		& 18 & 	 0.024809	&  -0.017488    &  0.4207	& 0.0821	&   No correlation, \ocen \ not fitted, Section~7.4.\\
C+N+O scatter $-$ V$_{\rm Abs}$	& 18 &  -0.010535	&   0.031638	&  -0.4656	& 0.0515 	&	No correlation, \ocen \ not fitted, Section~7.4.\\
C+N+O scatter $-$ Age			& 16 &  -0.006833	&   0.198537  	&  -0.24 	& 0.3706 	&	No correlation. 	
\enddata
\tablecomments{The correlation is determined by fitting the f(x)=a*x+b equation. The P-value expresses the probability of getting a significant correlation if only numeric 
fluctuations were present and no signal.}
\tablenotetext{a}{The number of clusters included in the statistics.}
\end{deluxetable*}

3. An important source of systematic error can be seen in the bottom left panel of Figure~\ref{fig:fehscales}, when comparing our 
metallicities with those of \citet{pancino01}. This discrepancy is similar to what can be seen in the top right panel 
when comparing the low-reddening clusters with \citet{carretta01}. 
In this case, all the clusters in common have low reddenings, thus errors
from the wrong estimate of E(B$-$V), or a possible error of the \citet{gonzalez01} conversion at high reddenings is minimal. 
This offset could also be due to how differences in spectroscopic temperature from Gaia-ESO and photometric ones 
from \citet{gonzalez01} affect the metallicities. Another possibility is that this 0.064~dex constant offset is the result of 
NLTE and/or 3D effects which are currently not modeled when fitting the APOGEE \citep{masseron01} or the Gaia-ESO spectra.

While our sample is larger than that of \citet{carretta01} and it naturally gives the opportunity to update the iron scale, 
the choice to do so is tainted by the fact that three independent survey analysis only agree within roughly $\pm$0.1~dex 
across the clusters in common. Also, different photometric temperatures from \citet{gonzalez01} and from \citet{alonso01, alonso02} 
might also introduce a systematic offset when the reddening is too high, possibly both affecting \citet{carretta01} 
and the results presented in this paper. On top of this, we suspect that a combination of NLTE and 3D effects introduce 
another systematic offset compared to optical studies. As of writing this paper, NLTE/3D corrections of 
iron lines are not available for the H-band, and any future study using APOGEE data when updating the cosmic iron 
scale from that of \citet{carretta01} must account for NLTE (and/or 3D) effects of iron lines, which may be as high as 
$~$0.06~dex \citep{masseron01}. For these reasons we estimate that the current absolute accuracy of the 
iron scale is roughly $\pm$0.1~dex based on these three independent studies. Overall, we conclude that after the 
correction for different solar Fe abundances, our values are still 0.064~dex higher on average than the 
optical Carretta scale.

\subsection{Intrinsic iron spread in clusters}

We defined the RMS scatter (R$_{\rm X}$, where X is the particular
element) of each element or sum of elements as the standard deviation around the mean value in each
cluster using the restrictions listed in Table~5. 
Detection of an intrinsic Fe abundance spread requires an accurate knowledge of the abundance measurement error within the sample.
The rms and cluster average iron errors are listed in Table~3. The errors are underestimated 
when [Fe/H]$<$-1.6, and overestimated for most of the more metal-rich clusters, which is probably the result of over and 
underestimating the effect of some sources of error, like dependence on effective temperature, S/N etc. 
There are two obvious outliers when comparing errors to the internal spread: 
\ocen \ and NGC~6229. \ocen \ is well known to host multiple populations with an Fe spread \citep{johnson02, gratton04}. 
We believe that our errors on the metallicity for NGC~6229 are significantly underestimated because 
the spread of Fe is 0.128~dex, while the error is 0.038~dex, the lowest in our sample. 

The RMS scatter in relation to the cluster age can be seen in the bottom left panel of Figure~\ref{fig:fehspread}. We 
use a recent compilation of ages by \citet{krause01}, which omits \ocen \ from its sample. 
Alternatively one can use the ages from \citet{marin01}, but results presented in this paper are not affected by the 
difference between these two ages. 
Table~6 contains the statistics of correlations between metallicity and cluster parameters. The age-metallicity relationship 
shown in Figure~\ref{fig:fehspread} is very similar to those of \citet{marin01} and \citet{krause01}. We refer the reader to these 
papers to provide a detailed discussion on this topic.

The measured RMS as a function of the main cluster parameters, mass, absolute visual magnitude can be seen in 
Figure~\ref{fig:fehspread}. \citet{carretta01} has reported that the iron spread is correlated with absolute visual 
magnitude and mass. From Figure~\ref{fig:fehspread} we are not able to confirm this; we find that the spread 
of Fe does not depend on either the mass, absolute visual magnitude or the age of the clusters (see Table~6 for 
statistical analysis). The lack of 
confirmation of the correlation may be due to our errors being slightly larger than that of \citet{carretta01}, although 
we believe our precision should be high enough to confirm such a correlation if it existed.

The iron spread in most clusters spans from 0.040~dex to 0.129~dex, with the exception of one  
cluster with 0.205~dex: \ocen. Not counting \ocen, the average spread of iron in 30 clusters is 0.096~dex. 
The true intrinsic iron spread can be computed by subtracting the effect of random error of the average value 
in quadrature from the measured cluster Fe RMS. As mentioned before, our estimated errors can technically be somewhat smaller 
or larger than the measured scatter, here, the average level of the error is assumed to be equal to the scatter for simplicity. 
With this 
simplification the true real iron spread is around 0.068~dex on average across 30 clusters. \citet{carretta01} reported an average 
iron spread value of 0.048~dex based on 19 GCs, and our value is 0.065~dex for the 17 clusters in common with that sample. 
Our study is lower resolution than \citet{carretta01}, which is the most likely 
source of our slightly higher internal errors.

While it is widely known that \ocen \ has a significant spread in iron that is of astrophysical origin, e.g. \citep{johnson02}, 
other clusters have been reported to have a significant spread, but this does not appear in our measurements.   
From the overview of \citet{costa01} these clusters are: NGC~1851, \ocen, NGC~362, NGC~5286, NGC~5824, M19, M22, M54, M75 and M2. 
Our data set includes five of those clusters. The iron spread in M22, M2, NGC~362, and NGC~1851 has been  
debated later, in particular because they can be introduced artificially by how atmospheric parameters were derived for those 
studies \citep{mucci01, lardo02}. In $\omega$ Cen, which has a wide metallicity distribution, we find an Fe scatter of 0.2 dex, 
clearly above our errors. M2 has a range in metallicity  \citep{yong06, lardo02}, with a high-metallicity population at 
[Fe/H]$\sim -1.0$ that comprises only 1\% of the cluster. Our measured Fe scatter in 
M2 is 0.06 dex, which is consistent with having observed entirely stars from the dominant 
population. All four of these clusters (M22, M2, NGC~362, NGC~1851) show Fe spreads expected from our 
internal errors (see Table~3 for individual values), and while our measurements do not disagree with the literature, 
we cannot make strong statements about the intrinsic Fe scatter in these four clusters. 
APOGEE observed only 7 stars with S/N$>$70 in M54, a 
sample not large enough to confirm or deny the broad metallicity distribution reported by \citet{carretta04}. Terzan~5 was also 
reported to have a multimodal metallicity distribution \citep{massari01}, but this cluster was excluded from our analysis 
due to large uncertainties in T$_{\rm eff}$ coming from its very high reddening.


\

\section{Multiple Populations Based on Al and Mg}

\subsection{The Al-Mg anticorrelation}

\begin{figure*}[!ht]
\centering
\includegraphics[width=6.2in,angle=0]{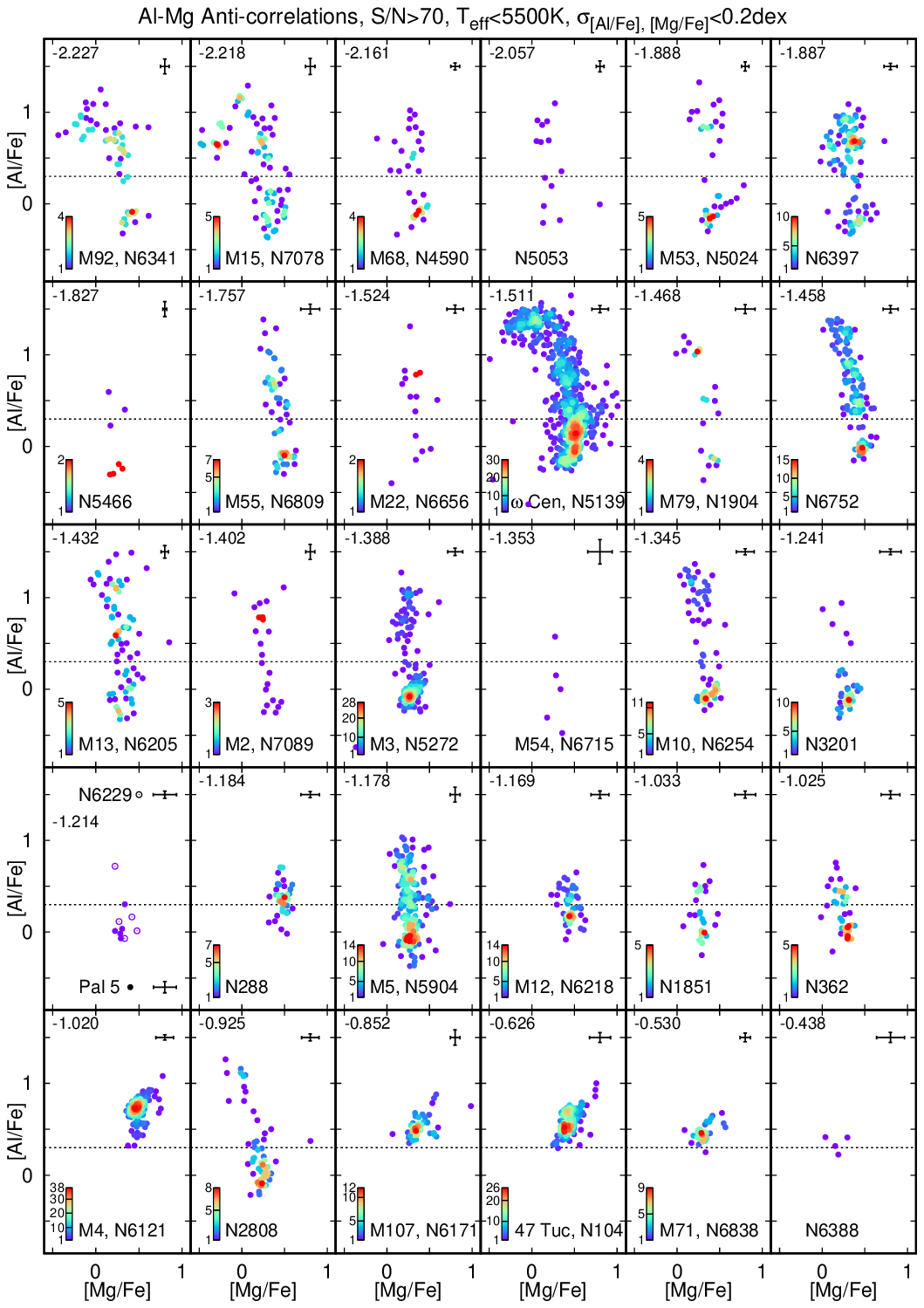}
\caption{Al-Mg anticorrelations in 31 clusters, NGC~6229 and Pal~5 are plotted in the same panel for simplicity. Each panel 
is color-coded linearly by the density of points calculated in a $\pm$0.05~dex range around each point. The color legend shows 
the density range for each cluster. The dotted line drawn at [Al/Fe]=0.3~dex denotes a generalized separation of classic first 
and second generation stars. Clusters are ordered by decreasing average metallicity, which is indicated in the top left corner 
in each panel.
}
\label{fig:almg}
\end{figure*}

\begin{figure*}
\centering
\includegraphics[width=6.2in,angle=0]{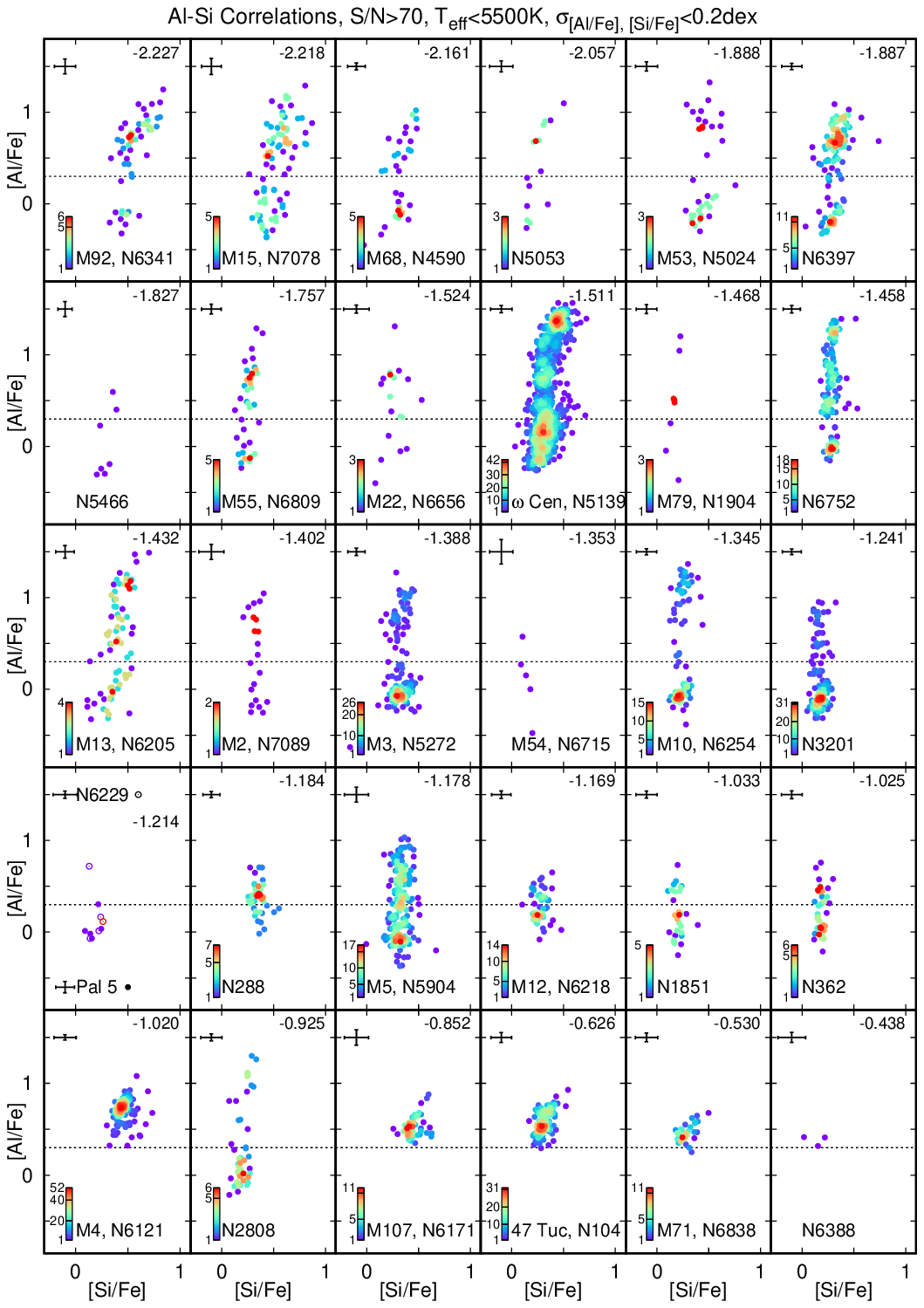}
\caption{Al-Si correlations in 31 clusters. The meaning of color legends and the line drawn at [Al/Fe]=0.3~dex are the same 
as in Figure~\ref{fig:almg}. Clusters are ordered by decreasing average metallicity, which is indicated in the top right corner 
in each panel.
}
\label{fig:alsi}
\end{figure*}

\begin{figure*}
\centering
\includegraphics[width=6.2in,angle=0]{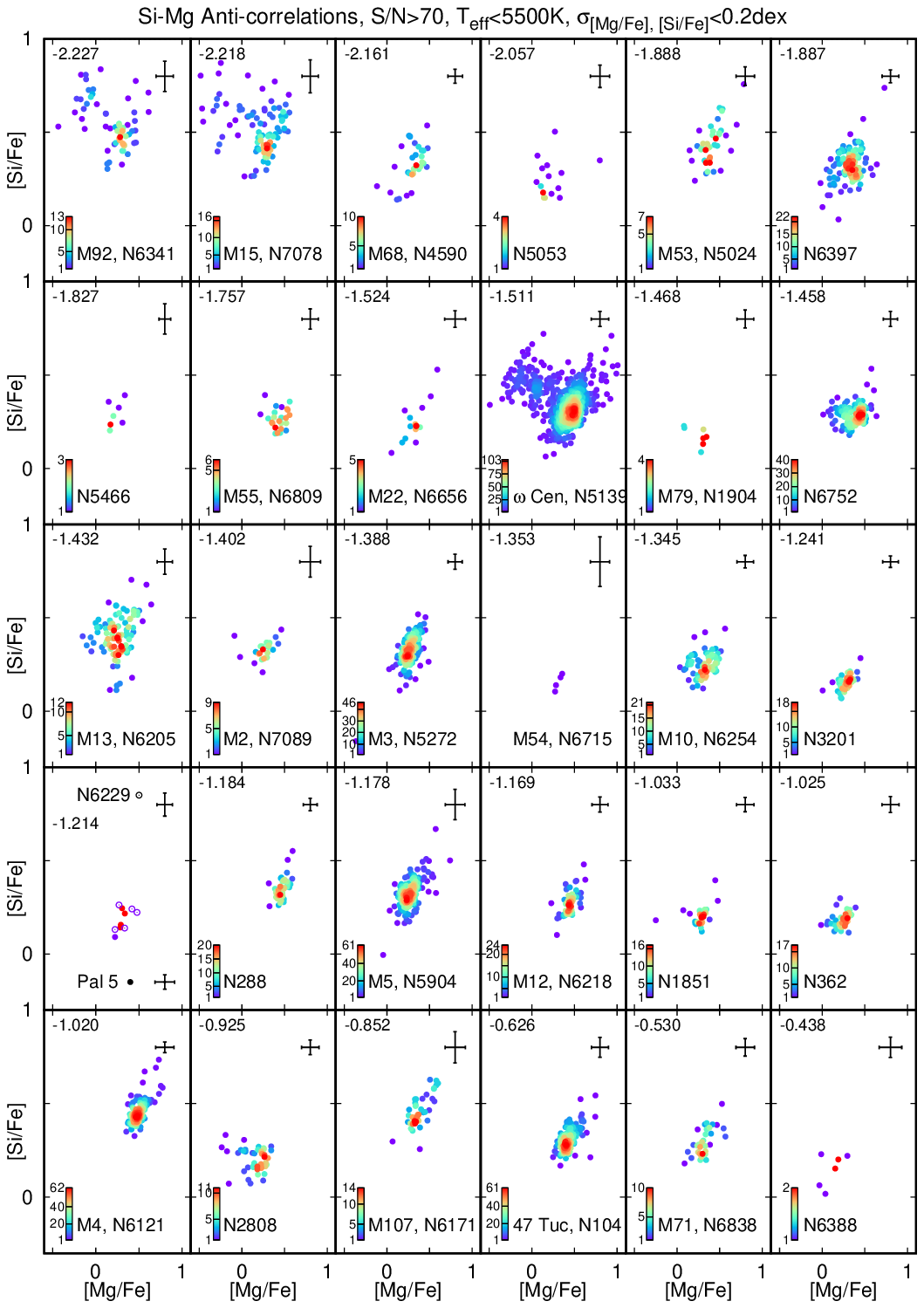}
\caption{Mg-Si anticorrelations in 31 clusters. Clear anticorrelation can be seen in only three clusters: M92, M15 and \ocen. 
Clusters are ordered by decreasing average metallicity, which is indicated in the top left corner 
in each panel.
}
\label{fig:mgsi}
\end{figure*}

It has been shown by several groups  \citep[e.g.,][]{carretta02, carretta03, gratton02} that variations in C, N, O, and Na can be seen in all 
observed GCs, but this is not 
the case for Al and Mg. The Mg-Al cycle needs large temperatures ($>$70 million Kelvin) to operate, temperatures that only the 
core of low metallicity polluters are capable of reaching. This is reinforced by the observation that 
some metal-rich clusters can be described by one single [Al/Fe] value, while in others the variation in the Al content spans a 
large range, as first reported by \citet{shetrone01}. We are able to discuss the dependence of the shape of Al-Mg on 
cluster parameters in more detail than it was possible before due to the increased number of observed clusters 
with a sufficient number of stars.

In this paper we discuss the largest sample of the Al-Mg anticorrelation and Al-Si correlation to date in 31 clusters, 
plotted in Figures~\ref{fig:almg}, \ref{fig:alsi}, and \ref{fig:mgsi}, in which the clusters are ordered by decreasing metallicity. 
An anticorrelation between Al and Mg is weakly present in most clusters, with a typical Mg range is 0.2 $<$ [Mg/Fe] $<$ 0.6~dex, 
much smaller than that of Al. RGB and AGB stars do not appear to follow different paths, or group separately in any of 
the abundance-abundance figures presented in the paper. There are two clusters shown here that have had no Al-Mg 
anticorrelation investigated before: NGC~6229 and Pal~5. While we have only five members in each of the three clusters that 
make our parameter cuts, it is enough to observe elevated Al abundances showing the signs of the Mg-Al cycle. It is clear that the 
extended distribution of Al, which is much larger than the typical errors of [Al/Fe] and [Mg/Fe], is present in most metal-poor
clusters, and clearly shows the past presence of the Mg-Al cycle. \citet{meszaros03} used an extreme-deconvolution method 
to identify population groups based on Mg, Al, Ca, and Si abundances. They found that it was Al abundances that drive the separation between 
stars, and northern clusters (except M107 and M71) presented in that paper could be divided into only two populations corresponding 
to first and second generation stars. Because an initial separation of FG and SG stars can be simply done by setting the [Al/Fe] 
limit at around 0.3~dex, we opted against doing a detailed population analysis again based on the same method, and instead use 
density maps and histograms of Al to explore MPs in all clusters. This is further motivated by the fact that most of the 
clusters show bimodal or continuous 
distributions in Al, in the latter selecting groups will always be difficult. While it is certainly possible that more than two 
populations are present in all of these clusters, their effect in the distribution of Al can be blurred out by any 
bias in target selection, and/or any measurement error we have, even if those are smaller than 0.1~dex in most cases. 
We set a limit of [Al/Fe]=0.3~dex to act as a guide to quickly and easily separate FG and SG stars.  
This limit is drawn in both Figures~\ref{fig:almg} and \ref{fig:alsi}.

\begin{figure}[!ht]
\centering
\includegraphics[width=3.45in,angle=0]{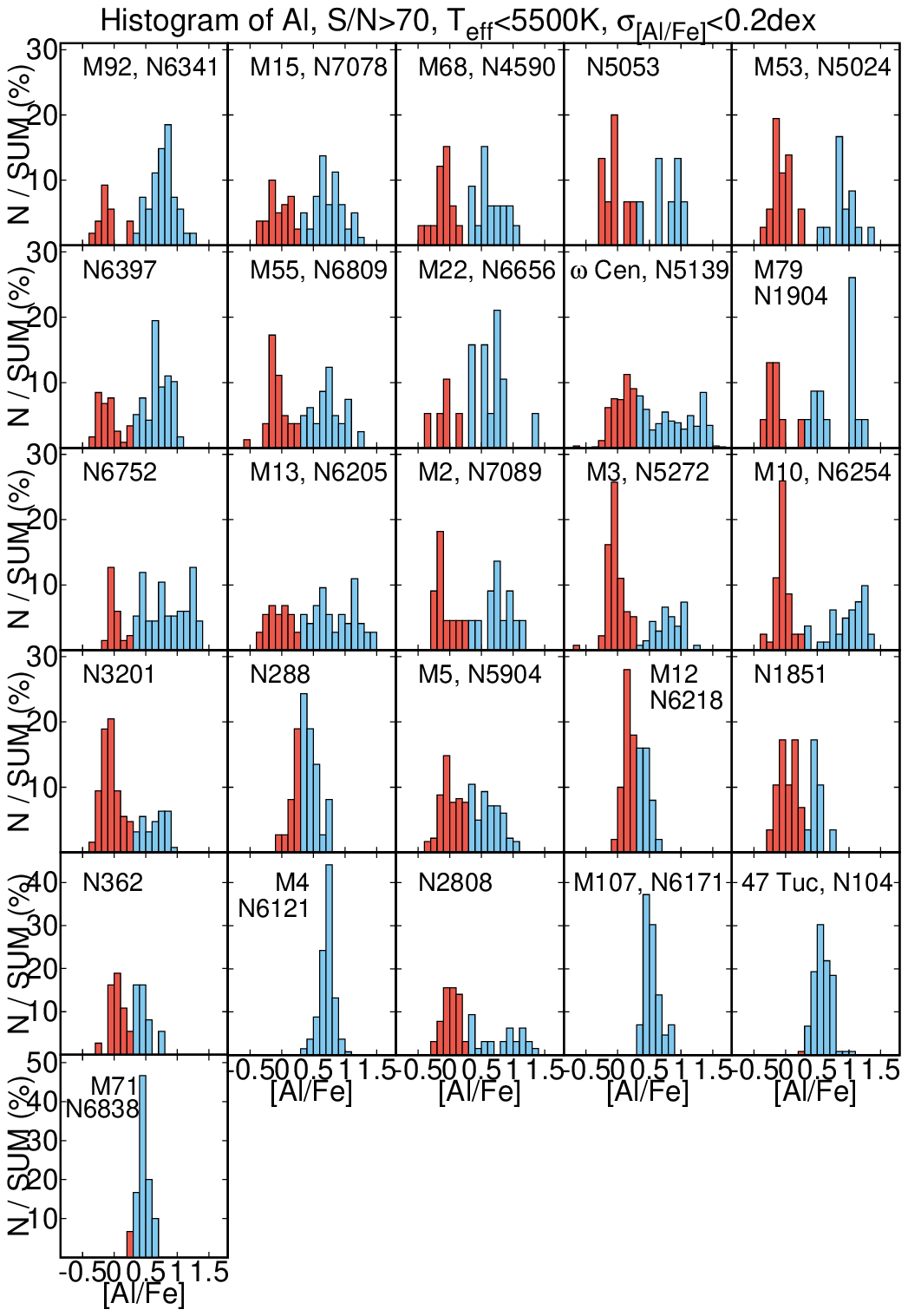}
\caption{The histogram of Al distribution in 0.1~dex bins. Stars with [Al/Fe]$<$0.3~dex are denoted by red, stars 
with [Al/Fe]$>$0.3~dex with blue to indicate classic FG/SG separation. 
}
\label{fig:alhist}
\end{figure}

Instead of coloring the Al-Mg and Al-Si plane according to population, we color them by their respective density calculated in a 
$\pm$0.05~dex range around each star. While this coloring method does not provide significant information if the number of 
stars in a cluster is small (NGC~5466, M54, Pal 5, NGC~6229, NGC~6388), our sample is large enough in most clusters to use this as a 
tool of analyzing multiple populations in GCs instead of the previously mentioned extreme-deconvolution method. This is further 
motivated by the findings of \citet{carretta06} in NGC~6752 for which multiple populations manifested themselves in the 
enhancement and depletion around discrete abundance values 
in an otherwise rather continuous distribution. In clusters that are considered to have one population with enriched [Al/Fe] 
values, or where the scatter of Al is smaller than 0.2~dex, the density profile also shows that most stars are concentrated 
around a single value of [Al/Fe]. These are the metal-rich clusters M4, M107, 47~Tuc, and M71. 

On the other hand, clusters with scatter of Al larger than 0.4~dex 
have vastly different density profiles even compared to each other. In some GCs the FG stars are concentrated 
around one single value of Al, like in \ocen, NGC~6752, M3, M10, M5, NGC~3201, NGC~2808, and perhaps M68. Other GCs with extended 
Al distribution do not show this behavior so clearly: M15, M92, M79, M2, M13, M55, and M53. The reason behind this varies from 
cluster to cluster. In M15 and M92, the [Mg/Fe] distribution of the FG stars is more sparse than in the more metal-rich clusters 
smoothing out any obvious density peaks. M79 and M2 may have too few stars observed in them to make a definite conclusion. 
M13 has a clear continuous 
distribution of Al abundances with no density peak in its FG stars, while M55 and M53 show only a small density peak below 
[Al/Fe]$<$0.3~dex. It appears that there is no clear correlation between the cluster metallicity, or the shape of the Al 
distribution and the existence of a density peak inside the FG stars.

\begin{deluxetable}{llrl}[!ht]
\tabletypesize{\scriptsize}
\tablecaption{The Description of MPs based on the Al Distribution.}
\tablewidth{0pt}
\tablehead{
\colhead{ID} & 
\colhead{Name} & 
\colhead{N$_{P}$} & 
\colhead{Description} 
}
\startdata
NGC 104  & 47 Tuc &	1	&		no Al spread \\
NGC 288	 & 		&	2?	&		small Al spread \\
NGC 362	 & 		&	2	&		small Al spread \\
NGC 1851 & 		&	2	&		small Al spread \\
NGC 1904 & M79	&	3	&		trimodal, but need more data \\
NGC 2808 & 		&	2 	&		continuous \\
NGC 3201 & 		&	2	&		bimodal/continuous? \\
NGC 4590 & M68	&	2	&		bimodal	\\
NGC 5024 & M53  &	2	&		bimodal	\\
NGC 5053 & 		&   2?	&		bimodal, but need more data \\
NGC 5139 & \ocen &	3	&		continuous with density peaks \\
NGC 5272 & M3	&	2	&		bimodal \\
NGC 5466 &   	& \nodata	&	not enough data \\
NGC 5904 & M5	&	2	&		continuous \\
NGC 6121 & M4	&	1	&		no Al spread \\
NGC 6171 & M107 &	1	&		no Al spread \\
NGC 6205 & M13  &	2	&		continuous with density peaks \\
NGC 6218 & M12	&	2?	&		small Al spread \\
NGC 6229 & 		&  \nodata	&	not enough data \\
NGC 6254 & M10	&	2	&		bimodal \\
NGC 6341 & M92  &	2	&		continuous with gap and Al turnover	 \\
NGC 6388 & 		&	1	&		no Al spread, but need more data \\
NGC 6397 & 		&	2	&		bimodal \\
NGC 6656 & M22	&	\nodata	&	not enough data \\
NGC 6715 & M54	&	\nodata	&	not enough data \\
NGC 6752 & 		&	4	&		continuous with gap and density peaks \\
NGC 6809 & M55	&	2	&		continuous with gap  \\
NGC 6838 & M71  &	1	&		no Al spread \\
NGC 7078 & M15  &	2	&		continuous with Al turnover	\\
NGC 7089 & M2   &	2	&		continuous, but need more data \\
Pal 5	 & 		&	\nodata	&	not enough data 
\enddata
\tablecomments{The number of populations were determined using the distribution of Al abundances only. 
The most metal-rich clusters have no Al spread, but still have large N variations proving the existence of MPs. 
See Sections~5.1 and 6.2 for more discussion.}
\end{deluxetable}

The histogram of Al can corroborate the findings from the density maps by integrating any spreads coming from Mg and Si together. This 
histogram is plotted in Figure~\ref{fig:alhist} using $\delta$[Al/Fe]=0.1~dex bins for clusters with significant number of stars observed. 
The histogram is normalized in each panel to the total number of stars in each cluster. While the density plots give more detail, 
the advantage of the histogram is that it can give a more complete picture if we have reliable [Al/Fe] abundances, but 
Mg or Si measurements are missing, like in NGC~3201 in which several Al rich stars do not have measurements of [Mg/Fe] or [Si/Fe]. 
When analyzing the number of populations in each cluster we use the density 
maps in Figures~\ref{fig:almg} and \ref{fig:alsi} and the histogram in Figure~\ref{fig:alhist} in a complementary fashion. 
Table~7 summarizes how many populations were identified in each cluster based on these methods  
and provides a short description of the Al distribution.

Both \citet{carretta02} and \citet{meszaros03} have reported observing bimodal and continuous distributions of Al. 
Our extended sample of stars 
and clusters paint a more complex picture on the distribution of Al by smoothing out the differences between bimodality and 
continuousness. For example, \citet{meszaros03} observed a clear bimodality in M3 and M53, but the distribution in the 
current (larger) sample is more continuous. On the other hand, 
M53 is still clearly bimodal. The classical bimodality/continuous distinction is further complicated by the fact that there 
are several clusters with continuous distribution of Al, but with well defined Al-Mg density peaks: \ocen, NGC~6752, and perhaps M13. 
Another interesting observation in both bimodal and continuous clusters is the existence of a gap with no or very few stars 
between [Al/Fe]=0.1~dex and 0.3~dex. The following clusters meet this criterion: M10, M3, NGC~6752, M55, NGC~6397, M53, M68, M92. 
NGC~6752 is particularly interesting, because it exhibits multiple of these properties, it has an extended and continuous SG 
distribution with four density peaks that is separated from the FG stars by an 0.2~dex wide almost empty gap. This also confirms and 
adds one more population to the results by \citet{carretta06}, who has found three populations using [Al/Fe] in an otherwise 
continuous Al distribution. Based on these observations it is clear that it is hard to generalize MPs 
from the properties of Al-Mg, and in reality every cluster has its own specific pattern of MPs showing a high degree of variety.

\subsection{The presence of Si-Mg anticorrelation}

Weak Si-Mg anticorrelations were observed in a small number of massive and 
metal-poor GCs before: NGC~6752 \citep{yong05}, NGC 2808, and M15 \citep{carretta02}. This implies leakage from the 
MgAl chain into Si production through the $^{26}\rm{Al}$(p,$\gamma$$^{27}\rm{Si}$(e-,$\nu$)$^{27}\rm{Al}$(p,$\gamma$)$^{28}\rm{Si}$ reactions at high temperature. 
Without this leakage, we would expect a simple correlation between Mg and Si since they are both alpha elements. 

\begin{figure}
\centering
\includegraphics[width=3.45in,angle=0]{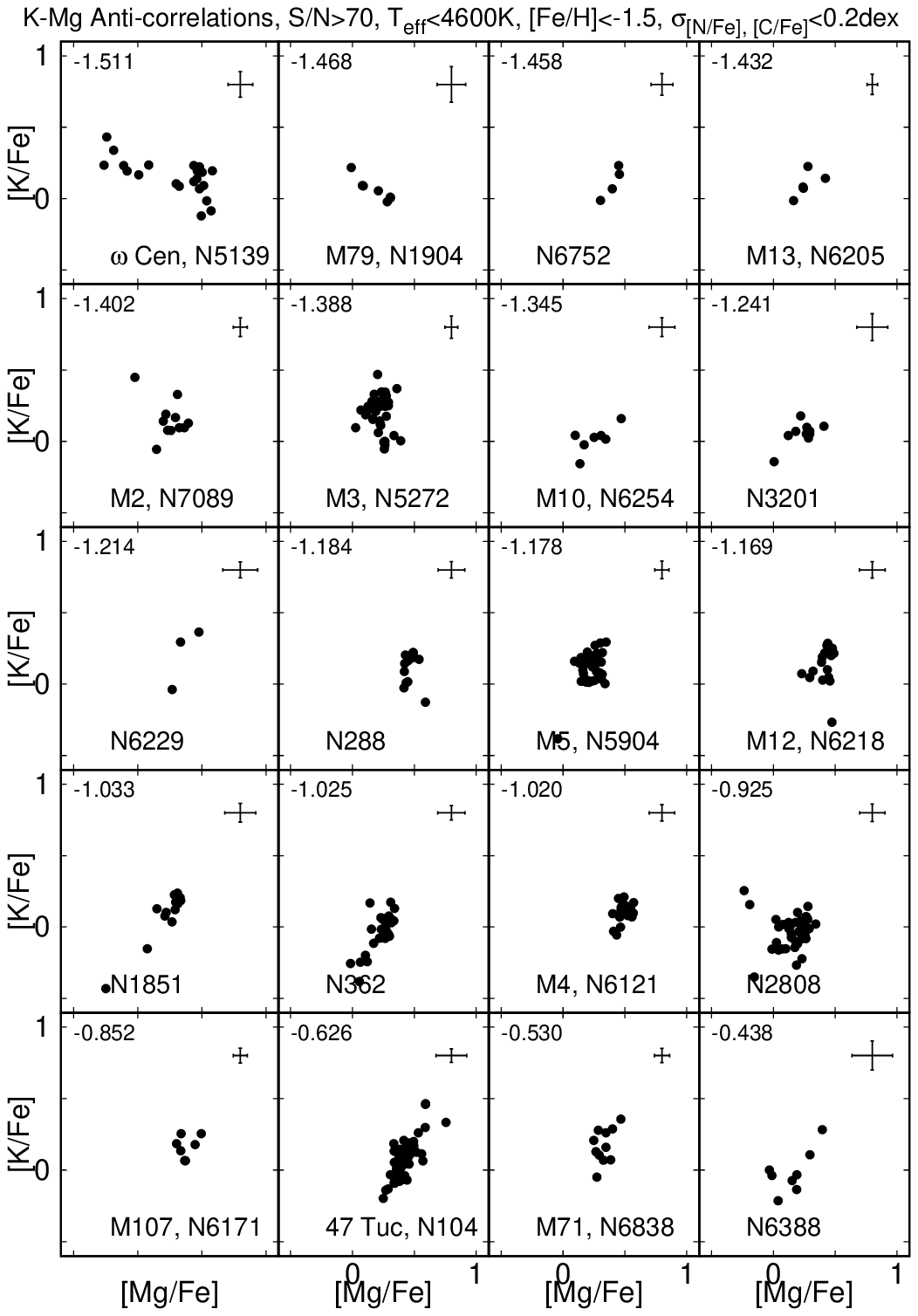}
\caption{Mg-K anticorrelations in 20 clusters. An anticorrelation might be observed in three clusters only: M79, NGC~2808 and \ocen.
Clusters are ordered by decreasing average metallicity, which is indicated in the top left corner 
in each panel.
}
\label{fig:mgk}
\end{figure}

From Figure~\ref{fig:mgsi} we are able to confirm the Si-Mg anticorrelation observed in NGC~2808 by \citet{carretta02}, but 
the case of NGC~6752 (as observed by \citet{yong05}) is less convincing. Although some stars seem to have lower Mg abundances, 
[Mg/Fe]$<$0.2, than where the most part of the cluster lies at [Mg/Fe]$>$0.3, these stars do not show larger Si abundances 
than their Mg rich counterparts. Thus, our data do not confirm the occurrence of hot proton burning in the early populations 
of NGC~6752.

\begin{figure*}[!ht]
\centering
\includegraphics[width=4.4in,angle=270]{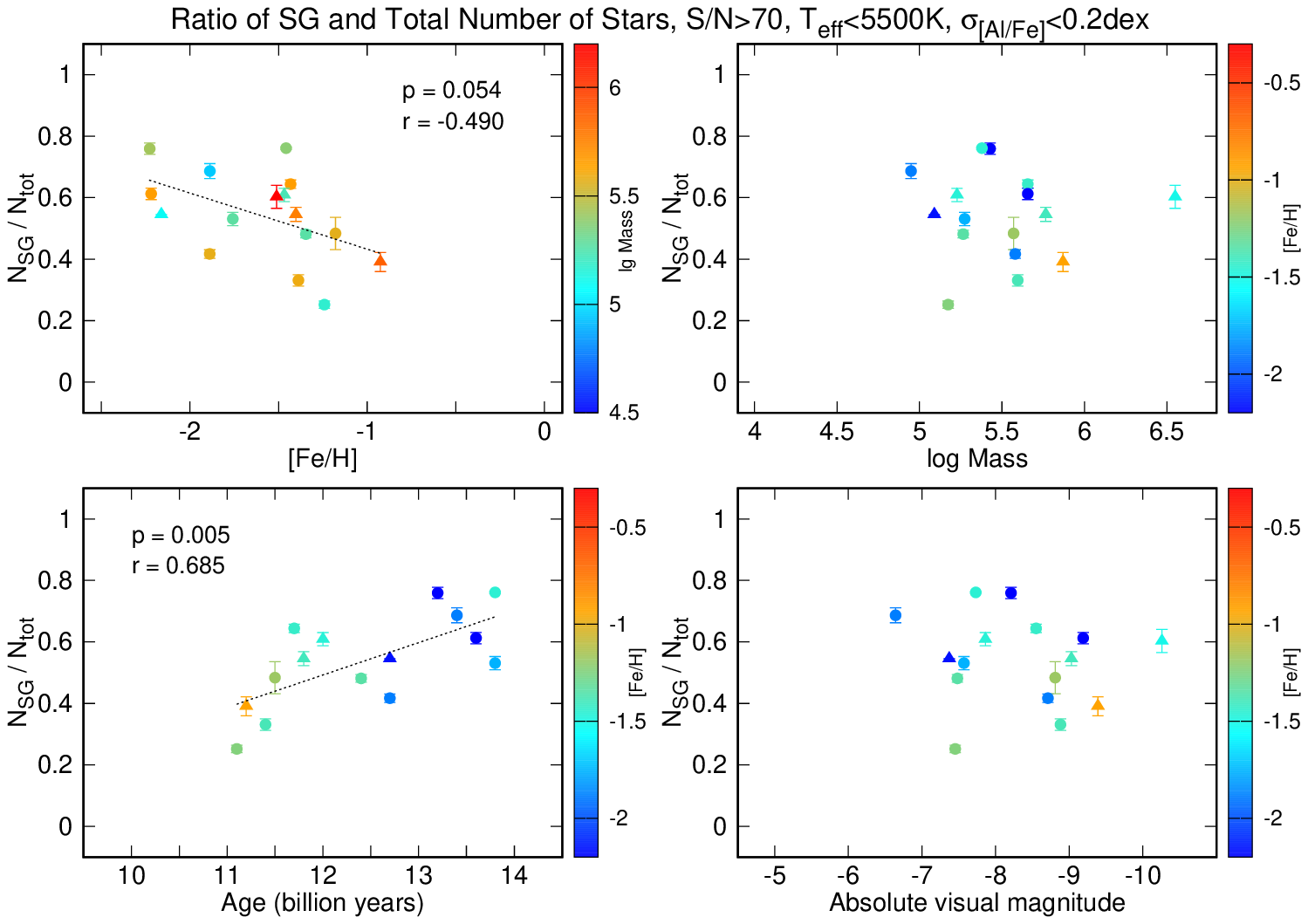}
\caption{The ratio of the number of FG and SG stars as a function of cluster parameters. Accreted clusters are denoted by triangles, 
in situ clusters by solid circles. The most metal-rich GCs are not included because those exhibit one single population based on 
Al alone. 
}
\label{fig:pop}
\end{figure*}

An Al-Si correlation in M15 and M92 was also observed by \citet{meszaros03}, but it was \citet{masseron01} who has 
discovered more stars in M15 and M92 that show an extreme Mg depletion with some 
Si enhancement while at the same time Al depleted relative to the most Al rich stars in these clusters, displaying an unexpected 
turnover in the Mg-Al diagram. 

In this paper we present the same type of behavior of the Al-Mg anticorrelation in \ocen \ shown in 
comparison with M15 and M92 in Figure~\ref{fig:almg}. It can be clearly seen that the most extreme Mg-poor stars in \ocen \ have 
lower Al content than 
what is expected from the traditional shape of the Al-Mg anticorrelation, while they are also the most Si rich stars in the 
cluster. The Si-Mg anticorrelation is clear in all of these three clusters (Figure~\ref{fig:mgsi}). 
\citet{masseron01} explained the shape of Al-Mg by suggesting that Al has been partially 
depleted in their progenitors by very hot proton-capture nucleosynthetic processes occurring above 80~MK temperatures. While 
M15 and M92 are two of the most metal-poor clusters, \ocen \ is significantly more metal-rich, showing that the observation 
of the turnover of the 
Al-Mg diagram at different metallicities may be the result of multiple mechanisms. 
Because this paper focuses on the overall characteristics of globular clusters, the detailed discussion of \ocen \ is out the 
scope of this study. We will present the detailed analysis of \ocen \ in the third 
part of our series.

\subsection{The presence of K-Mg anticorrelations?}

Stars showing a large range of K abundances were first discovered by \citet{mucci03} in NGC~2419. Later, \citet{mucci02} 
observed a large K enhancements in four stars with very low Mg abundances in NGC~2808. The enhancement of K 
is currently not understood. \citet{ventura03} attempted to explain the origin of a Mg-K anticorrelation by suggesting 
that this population might have directly formed from super-AGB ejecta. NGC~2419 is not in our sample so we can only 
examine the existence of K rich stars in NGC~2808, as shown in Figure~\ref{fig:mgk}. Our confirmation is based on two 
stars with very low, [Mg/Fe]$<$0.0~dex, Mg abundances that are slightly enhanced in K compared to the more Mg rich, 
mostly FG stars.

Interestingly, \ocen \ contains seven stars with [Mg/Fe]$<$0.0~dex, 
previously discussed in Section~5.2, that are also slightly enriched in K compared to the classical FG stars, drawing a 
weak anticorrelation between K and Mg in Figure~\ref{fig:mgk}. However, the two K lines found in the H-band are fairly 
weak at low metallicities and high temperatures (they are also often blended), thus it is necessary to implement a 
strict cut on these two parameters (Table~5) to cut out upper limits, even when BACCHUS reports real detection. 
Considering these issues, our conclusion 
is that the discovery of K enhancement of the Mg-poor stars in \ocen \ cannot be convincingly claimed from our spectra, the 
anticorrelation is weak and we need independent confirmation from 
optical spectra before the extent of the enhancement can be reliably discussed. 

There is another cluster in our sample, M79, which shows a weak K-Mg anticorrelation shown in Figure~\ref{fig:mgk}, although 
the extent of the Mg abundances in the M79 are on the level of the reported uncertainties. In such a case the 
observed anticorrelation is more likely the result of correlated errors, and not of an astrophysical origin. 

The weak correlation between Mg and K observed in NGC~1851, NGC~362 and 47~Tuc exists because there are 
2-3 outlier stars with very high or very low [K/Fe] abundances in each of these clusters, which are most likely bad 
abundance determinations.

\begin{figure*}[!ht]
\centering
\includegraphics[width=4.4in,angle=270]{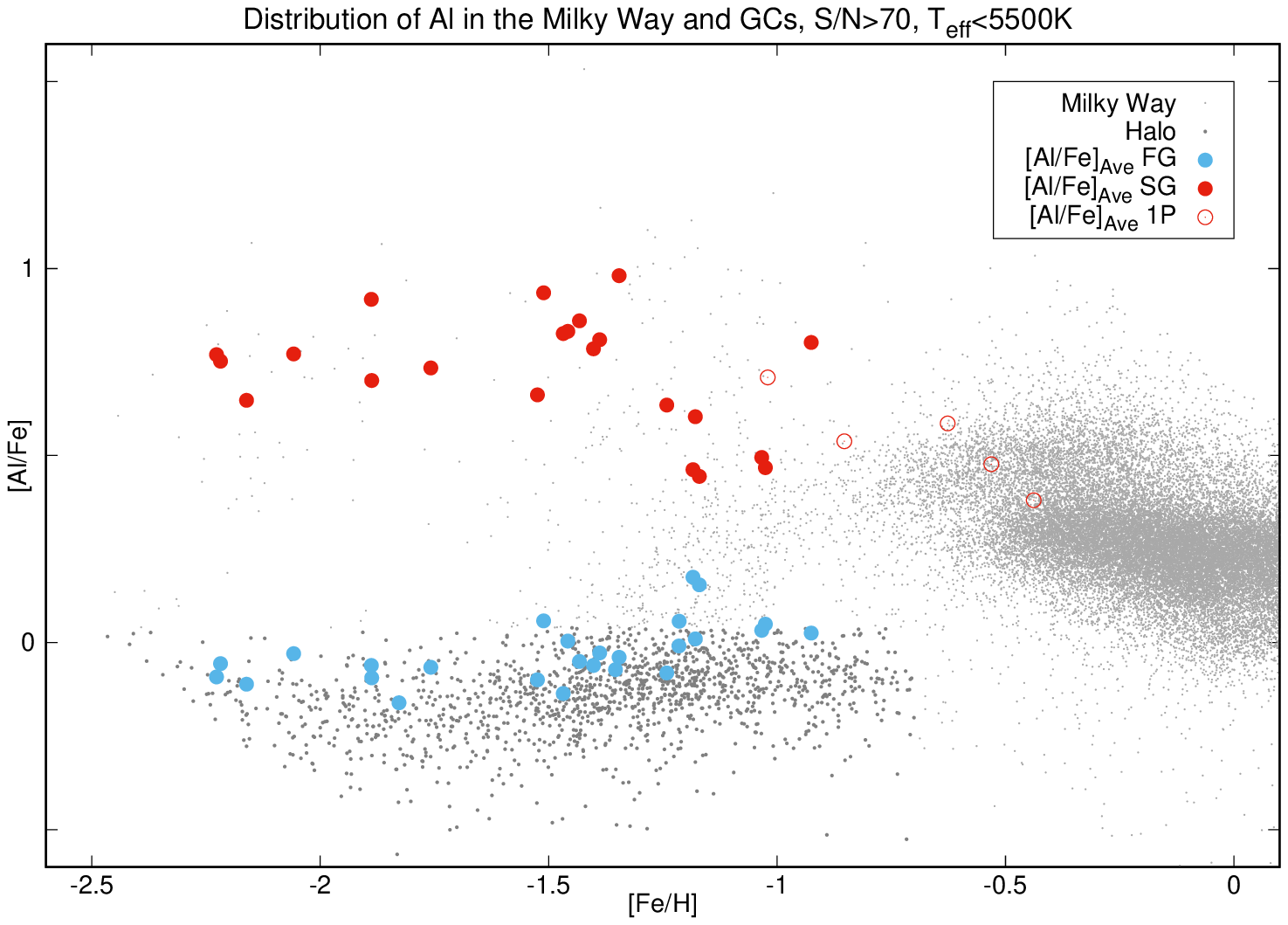}
\caption{Chemical evolution of Al in the Milky Way. Small grey dots are standard Milky Way stars, dark grey dots are stars 
mostly from the Galactic Halo. Blue dots denote the average [Al/Fe] of the FG stars with [Al/Fe]$<$0.3~dex, red dots denote 
the average [Al/Fe] of the SG stars with [Al/Fe]$>$0.3~dex. Open red dots show the clusters that do not show signs of Al 
enrichment due to pollution.
}
\label{fig:milky}
\end{figure*}

\subsection{The ratio of FG and SG}

The discussion of the ratio of SG vs. FG stars is generally difficult because there needs to be a significant number of stars 
observed in each cluster. 
For this reason we limit our discussion to clusters that have at least 20 stars observed. As in 
photometric studies such as \citet{milone02}, we use the definition 
f$_{\rm enriched}$ = N$_{\rm SG}$/N$_{\rm tot}$ to examine the extent of enrichment.
The computed f$_{\rm enriched}$ ratios can be found in Table~3. When calculating the number of FG and SG stars 
we used the limit of [Al/Fe]=0.3 to separate FG and SG stars. 
However, error bars were computed by varying the limit from 0.25 to 0.35~dex and the ratio recalculated. The f$_{\rm enriched}$ 
ratio is plotted against the cluster properties in Figure~\ref{fig:pop}. The resulting error bars are 
generally small and do not affect any conclusion on how the ratio depends on cluster parameters.

We listed the statistics of the f$_{\rm enriched}$ correlation with cluster properties in Table~6. A very weak, statistically 
barely significant with p=0.0541, linear correlation was found against metallicity, in which more metal-poor clusters exhibit more 
SG stars than FG stars. Considering that the ratio can be improved by observing more stars, this correlation may move closer to or 
farther from statistically significant, but in this paper we do not explicitly conclude that this correlation exists. The 
correlation is similar to what \citet{bastian02} have found using spectroscopic results collected from the literature, 
which has been confirmed by \citet{milone02} with the HST Legacy Survey \citep{soto01, piotto02}. The only clear and 
statistically significant correlation (p=0.0048) is with cluster age, with the younger clusters exhibiting lower 
f$_{\rm enriched}$ than the older ones. 

In terms of absolute values of f$_{\rm enriched}$ we have a good agreement with 
\citet{milone02} as both studies measured f$_{\rm enriched}$ between 0.4 and 0.8 for most clusters. However, a correlation 
with mass and absolute visual magnitude is non-existent in our data, which is in sharp contrast with what \citet{milone02} observed. 
They observed that more massive clusters have more SG stars than less massive ones. 
Our study is biased towards the outer cluster regions, because the fiber-collision radius does not allow 
the APOGEE instrument to properly sample the inner regions. The HST data samples the inner 2 arcminutes of the clusters, thus 
there are very few stars which 
overlap between APOGEE and HST observations \citep{meszaros04}. The significant difference between our correlations 
with f$_{\rm enriched}$ and that of \citet{milone02} may arise from 
cluster properties which depend on distance from the cluster core.

\section{The spread of Al abundances}

\begin{figure*}
\centering
\includegraphics[width=6.2in,angle=0]{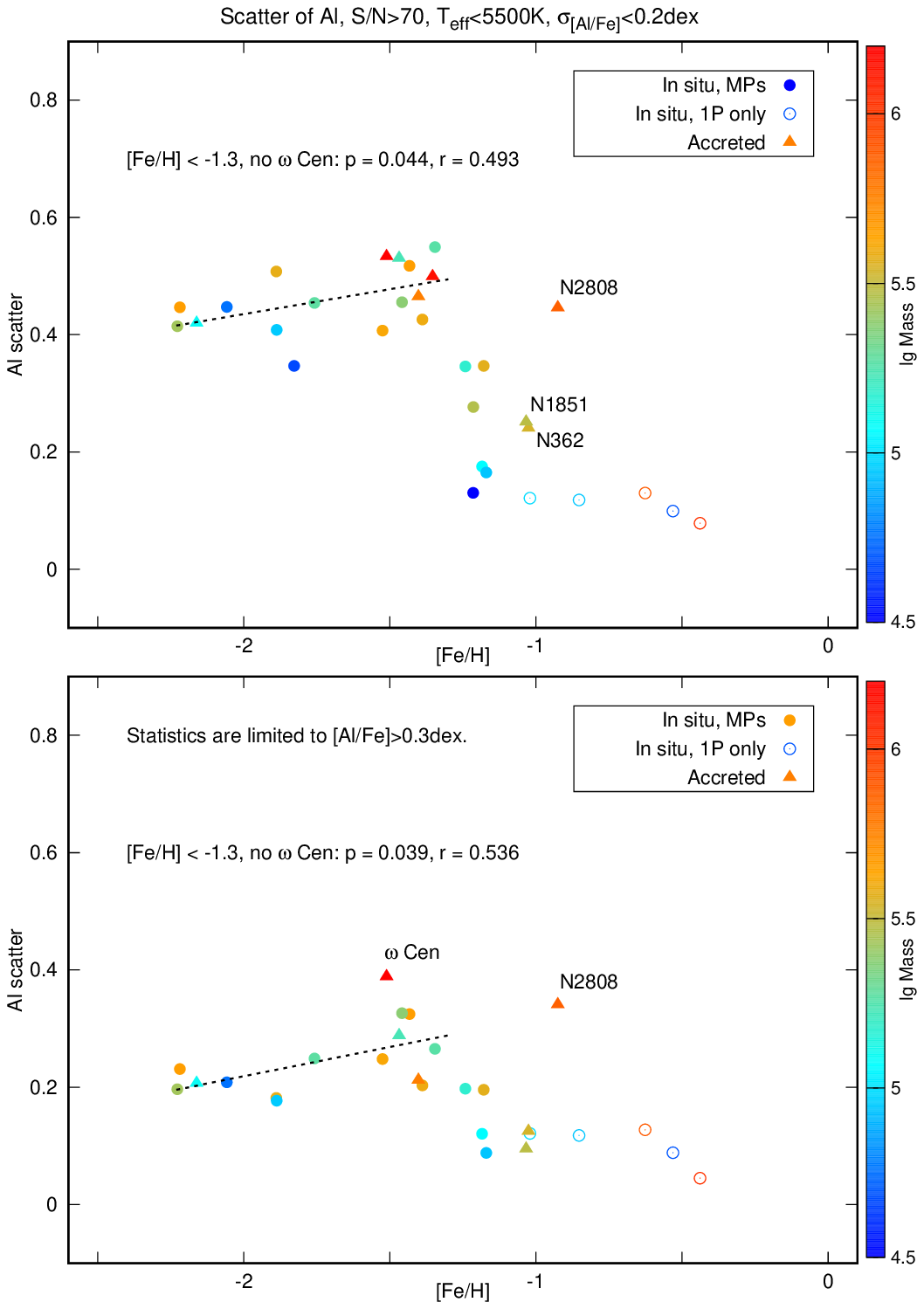}
\caption{Scatter of Al as a function of average cluster [Fe/H] color coded by mass for clusters with at least three members. 
Accreted clusters are denoted by triangles, in situ clusters by solid circles. Top panel shows R$_{\rm Al}$ as directly observed, 
the bottom panel shows the scatter of Al after excluding the FG stars with [Al/Fe]$<$0.3~dex from the sample. See Section~5.1 for 
discussion.
}
\label{fig:alsc1}
\end{figure*}

\begin{figure*}
\centering
\includegraphics[width=6.2in,angle=0]{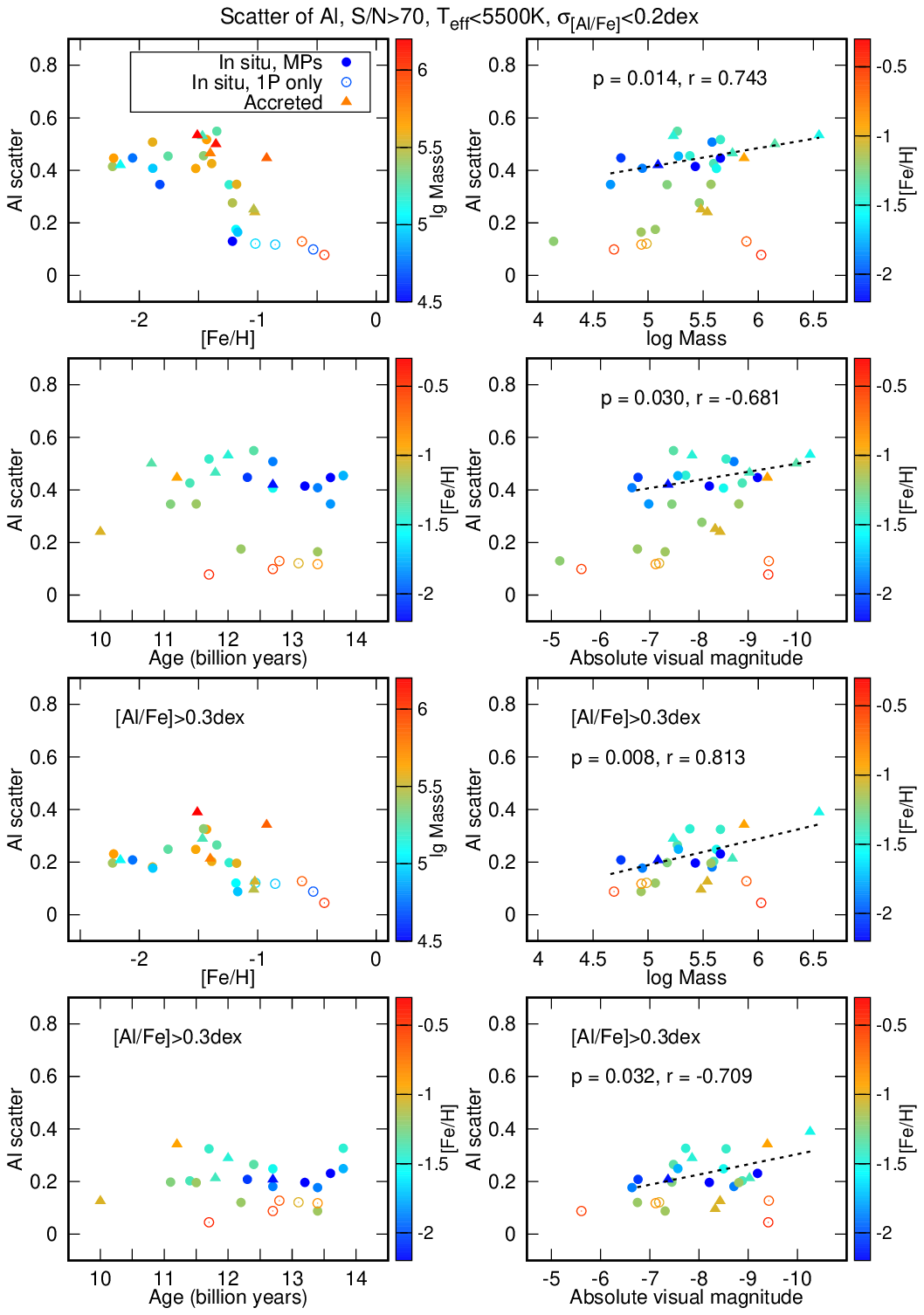}
\caption{Scatter of Al as a function of cluster parameters for clusters with at least three members. 
Accreted clusters are denoted by triangles, in situ clusters by solid circles. The top four panels show R$_{\rm Al}$ as directly 
observed, the bottom four panels show the scatter of Al after excluding the FG stars with [Al/Fe]$<$0.3~dex from the sample. 
See Section~5.1 for discussion.
}
\label{fig:alsc2}
\end{figure*}

In order to properly discuss the Al scatter as a function of cluster parameters, we need to distinguish between clusters that 
form in situ 
with the Milky Way and those that were accreted by the Milky Way. In the most recent models, the haloes of galaxies similar 
to the Milky Way are believed to be formed from the accretion of smaller galaxies \citep{bullock01, abadi01, font01}. 
These small dwarf galaxies are disrupted and incorporated into the larger galaxy, only very dense components like 
globular clusters will survive intact 
\citep{pen01}. The GCs that are formed from this accretion process are then added to the rest of the clusters
formed in situ within the Milky Way. There have been several efforts made to identify these accreted clusters 
(Gaia sausage clusters, CMa and Sag clusters, Sequoia clusters), but the following are in common with our sample: 
NGC~1851, NGC~1904, NGC~2808, NGC~362, NGC~7089 \citep{forbes01, myeong01}, 
NGC~4590 \citep{forbes01}, and \ocen \ \citep{bekki01}. 

It is also important to take the standard chemical evolution of the Milky Way into account, which was explored by \citet{hayes01}
using APOGEE DR13 data. This is illustrated in Figure~\ref{fig:milky}, in which we plotted the average [Al/Fe] of FG and SG stars 
as defined in Sections~5.1 for all clusters with at least three members in each population on top of stars observed in the 
Galaxy. Stars from the Milky Way were selected by applying the criteria defined by \citet{hayes01} to the DR14 data. 
The average [Al/Fe] of FG stars (blue dots) agree well with the Al abundances observed in the Galactic halo, denoted by dark 
grey points, while the average [Al/Fe] of SG stars is elevated (red dots). The slight, roughly 0.1~dex systematic offset between 
the average [Al/Fe] of FG stars and the [Al/Fe] of Galactic halo stars is most likely due to systematics between BACCHUS and ASPCAP, 
the latter used by \citet{hayes01}. As metallicity increases, the two averages get 
closer to each other. The metal-rich clusters (red open dots) that only show a single Al population with an average [Al/Fe]  
close to what is observed in the Galactic thick and thin disc. The fact that the average [Al/Fe] of the FG is lower at low 
metallicities may introduce a bias to how the scatter of Al depends on cluster parameters. This is because metal-rich 
clusters formed in parts of the Galaxy where more Al was present to begin with. 

We defined the RMS scatter of Al (R$_{\rm Al}$) as the 
standard deviation around the mean value of [Al/Fe] in each cluster. Another measure that can be introduced is the difference between 
the maximum and minimum value of an abundance inside a cluster. This measure is less robust as it is more sensitive to any biases 
in target selection and less accurate when only a small number of stars are observed. As a test, we carried out 
the same statistical analysis of correlations by using both the scatter and the max$-$min of [Al/Fe] and found that the main 
conclusions are the same in both cases, but the relationships when using the max$-$min of [Al/Fe] are less defined and more noisy. 
For this reason we limit our discussion in Section~6 to that of R$_{\rm X}$ only.

\subsection{Metallicity}

As of now only a handful of studies have examined the behavior of Al spread as a function of cluster parameters. 
As previously mentioned in Section 5.1, \citet{carretta02, meszaros03, pancino01, masseron01, nataf01} have reported that the 
extent of the Al distribution linearly depends on cluster metallicity, but all of those studies were carried out using 
only a handful of clusters or spanned a relatively small metallicity range, and did not take the evolution of Al in the Milky 
Way \citep{hayes01} into account. 
Here, we are able to significantly increase the sample size to 31 clusters, and also cover a large metallicity range  
between [Fe/H]=$-$2.23 and $-$0.44~dex. Figure~\ref{fig:alsc1} shows the measure RMS scatter of Al (R$_{\rm Al}$) as a function 
of cluster average metallicity. The observed distribution of Al scatter is more complex than previously found, but it is 
also biased because low metallicity halo stars have lower [Al/Fe] content than high metallicity disc stars \citep{hayes01}. 

\begin{figure}[!ht]
\centering
\includegraphics[width=3.45in,angle=0]{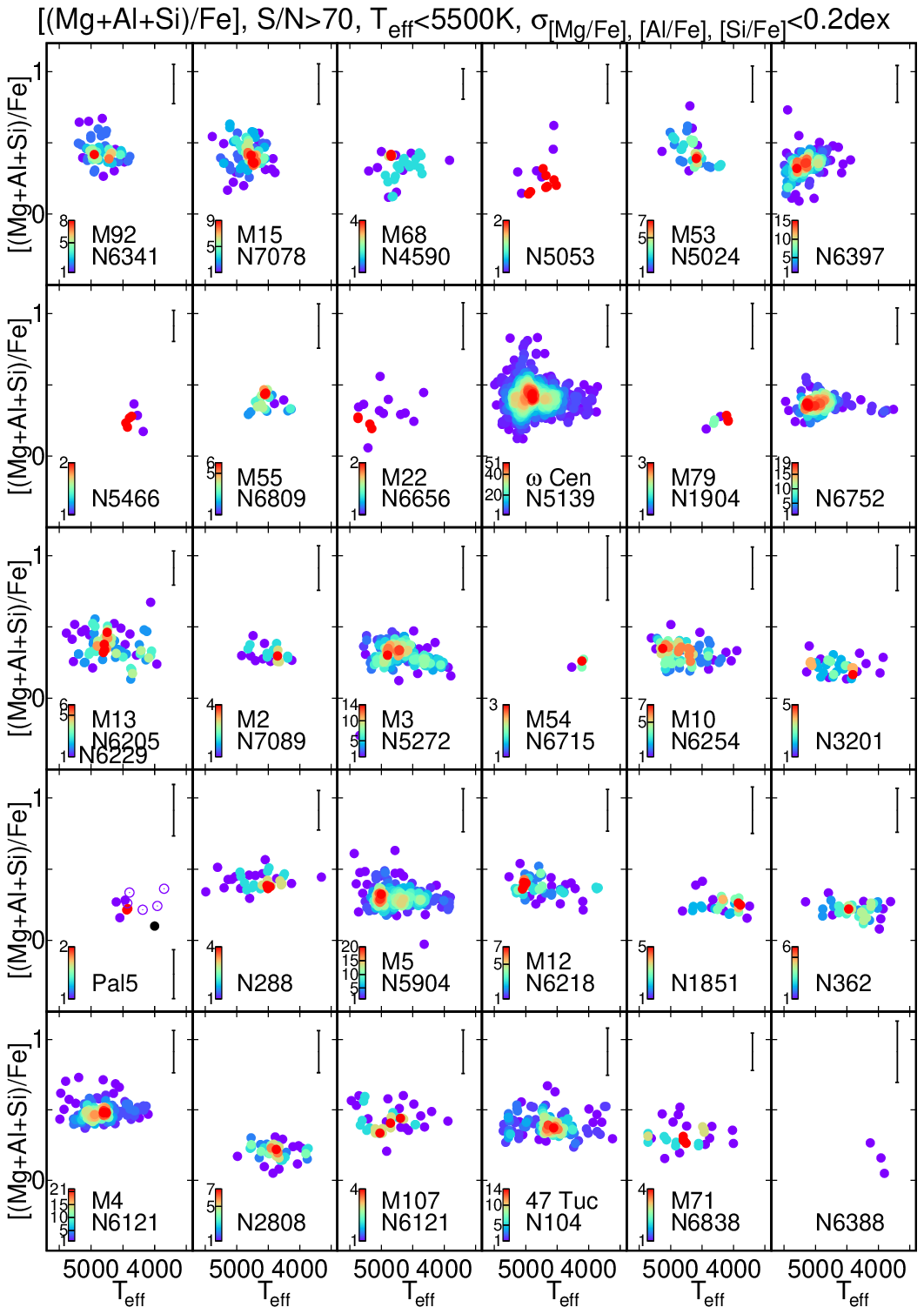}
\caption{Mg+Al+Si as a function of effective temperature. Each cluster exhibits the same [(Mg+Al+Si)/Fe] value across MPs.
}
\label{fig:mgalsi}
\end{figure}

Based on the top panel of Figure~\ref{fig:alsc1} there are three main groups that can be identified: 
\begin{itemize}
\item{[Fe/H]$<$-1.3:} The scatter of Al in all clusters is larger than 0.35~dex. These clusters show clear Al-O, Al-N 
(anti)correlations (see Section 7.2). In this metal-poor region the correlation between [Fe/H] and R$_{\rm Al}$ is 
statistically significant but weak. Accreted clusters have very similar R$_{\rm Al}$ to that of those formed in situ. 

\begin{figure*}[!ht]
\centering
\includegraphics[width=4.4in,angle=270]{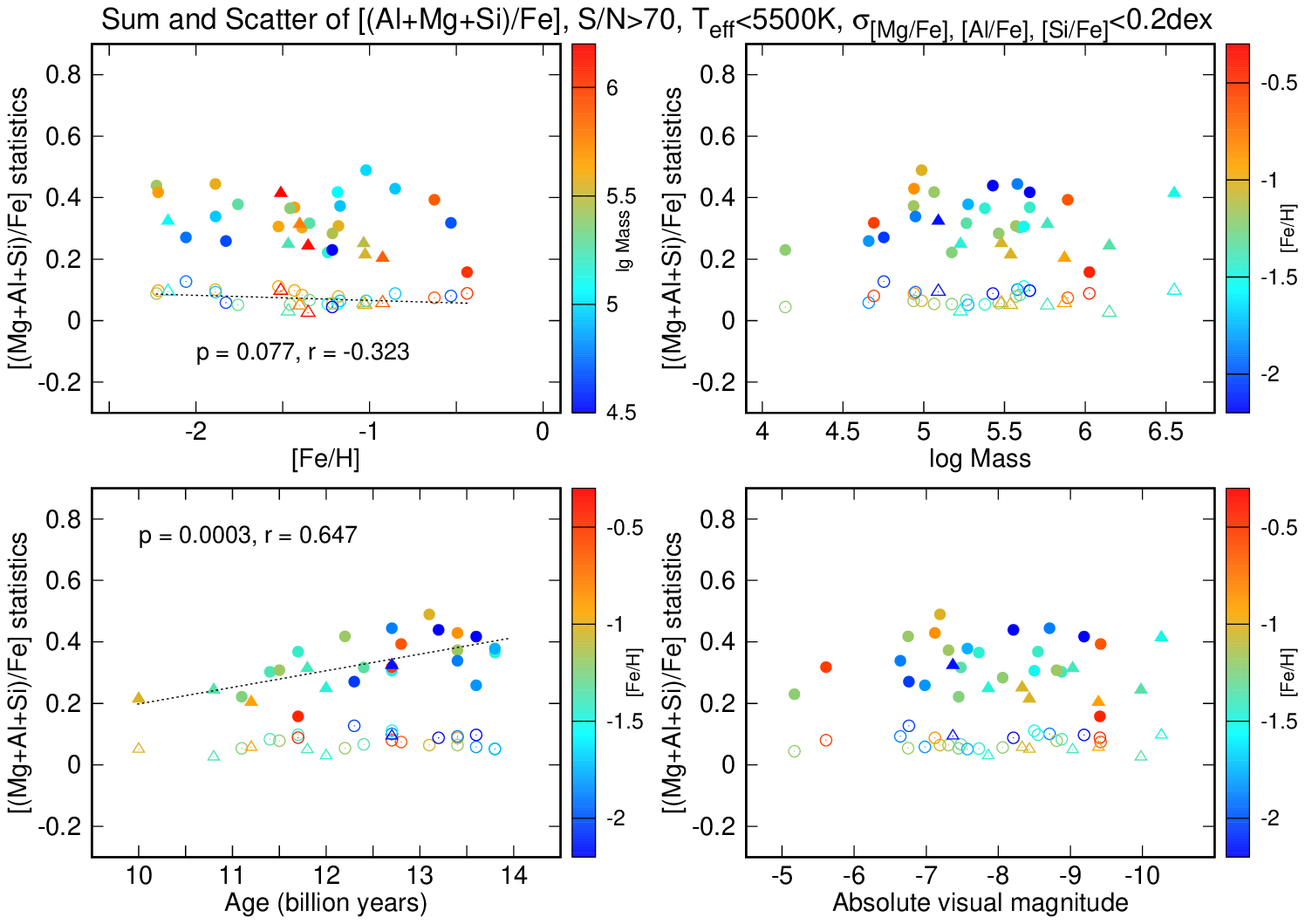}
\caption{Statistics of Mg+Al+Si as a function of cluster [Fe/H] and mass. Filled symbols represent the average of Mg+Al+Si, 
open symbols represent the scatter of Mg+Al+Si. Triangles are accreted clusters. 
}
\label{fig:mgalsi2}
\end{figure*}

\item{-1.3$>$[Fe/H]$<$-1.0:} In this transition region there is a sharp drop of R$_{\rm Al}$ from about 0.5~dex to 0.18~dex. 
Here, accreted clusters are not present in our selection. These clusters also show clear Al-O, Al-N 
(anti)correlations. 
\item{[Fe/H]$>$-1.0:} The R$_{\rm Al}$ is constant as a function of [Fe/H], and remains lower than 0.18~dex. 
However, there is a significant difference between the accreted GC NCG~2808 and other clusters. NCG~2808 has 
significantly higher R$_{\rm Al}$ than its in situ and other accreted counterparts with similar metallicity. Other than 
NGC~2808, none of these clusters have any Al-O, Al-N (anti)correlations. 
\end{itemize}

The average error of [Al/Fe] spans a range from 0.03 to 0.09~dex, except for M54 for which $\sigma_{\rm [Al/Fe]}=0.13$, which is 
roughly half of the R$_{\rm Al}$ measured when [Fe/H]$>$-1.0. Considering that the calculated R$_{\rm Al}$ is the quadratic sum of 
the intrinsic Al spread and the error, the logical conclusion would be that these clusters do not bear the signs of past Al-Mg cycles 
in the progenitors. We believe this is not the case, because abundance is measured on a logarithmic scale, a larger absolute 
enrichment is required to see the same change in [Al/Fe] in metal-rich clusters than in metal-poor clusters.
We provide more discussion on this topic in Section 7.3. It is important to note that the three 
accreted clusters (NGC~362, NGC~1851, NGC~2808) are the among the most metal-poor ones in this third group of otherwise metal-rich 
GCs, meaning they may more naturally belong to the transition metallicity zone where R$_{\rm Al}$ drops suddenly. 
Nevertheless, accreted clusters with Al spreads close to the estimated errors are not observed.

As mentioned before, this picture may be biased because metal-rich clusters have
an initial composition more Al-rich than metal-poor clusters due to chemical evolution in the Galaxy seen in 
Figure~\ref{fig:milky}. What we want to know is how the extent of the enrichment of GC stars depends on cluster 
parameters if we remove the effect of Galactic chemical evolution on the FG Al abundance.

In order to compensate for the chemical 
enrichment and to compute the scatter of Al more objectively, we exclude all stars from the sample that 
have [Al/Fe]$<$0.3, this is shown in the bottom panel of Figure~\ref{fig:alsc1}. This is possible taking into 
account that the Al production is more sentive to the abundance of Mg available for the proton capture 
channel $^{25}Mg(p,\gamma)^{26}Al$ than to the initial Al abundance. After removing the bias introduced by 
the standard chemical evolution, the correlation of the [Fe/H]$<$-1.3 
region remains the same and barely statistically significant, but the difference in R$_{\rm Al}$ between metal-rich 
and metal-poor clusters decreases significantly. The overall trend over the full metallicity range still shows that 
low metallicity clusters have higher Al scatter than the metal-rich ones. 

However, there are two outliers after the correction, \ocen \ and NGC~2808, that lie above other GCs at similar metallicities, 
showing larger Al enrichment than expected. We know that NGC~2808 and Omega Cen are among
the most massive clusters and to properly discuss their behavior one has to look at the mass dependence first.

\subsection{Mass and V$_{\rm ABS}$}

\citet{carretta07} used 19 GCs to look for correlations between the extent of the Na-O anticorrelation and  
cluster properties. The strongest relation found was with cluster mass, with higher mass clusters showing larger Na-O 
abundance spreads. A similar positive correlation between He spread and mass was found by \citet{milone03} in which higher mass 
cluster exhibit larger He spreads. Looking at Al-Mg, both \citet{carretta02, carretta03} and \citet{pancino01} found that 
massive metal-poor clusters tend to have larger Al-Mg anticorrelations than their lighter counterparts. \citet{nataf01} 
used APOGEE data to show that the slope of the [Al/Fe] vs. [N/Fe] relation depends on both metallicity and mass. 
Without dark matter, a globular cluster's ability to gather or retain material for star formation is 
tied directly to its stellar mass, and so (in a two-generation scenario) one might expect higher-mass clusters to show 
more abundance variation, agreeing with the observations.

The corrected and uncorrected R$_{\rm Al}$ as a function of mass and absolute visual magnitude are plotted in the right panels 
of Figure~\ref{fig:alsc2}. 
As significant Al spread was observed only in metal-poor clusters ([Fe/H]$<$-1.3), we explored the correlation between mass and 
R$_{\rm Al}$ for these metal-poor GCs separately from the metal-rich clusters. The correlation found, although moderate, is 
statistically significant both with mass 
and V$_{\rm ABS}$, with p=0.0139 and 0.0301 respectively. The appearance of correlation in both mass and V$_{\rm ABS}$ is trivial 
to understand because more massive clusters have higher luminosities. These results confirm previous literature findings for 
metal-poor GCs, however, high metallicity clusters with [Fe/H]$>$-1.3 do not have an obvious R$_{\rm Al}-$mass correlation.

When looking at the R$_{\rm Al}$ dependence on mass and V$_{\rm abs}$ after the correction 
(bottom right panels in Figure~\ref{fig:alsc2}) we find that the correlation appears more clearly, because the clusters are now 
not polluted by low Al FG stars. We therefore conclude that the extent of the enrichment of GCs stars is a function of both 
the cluster mass and metallicity, 
the correlation with mass becomes stronger when [Fe/H]$<-$1.3, while the correction removes most of the step from the metallicity 
dependence that is introduced because FG stars in low metallicity clusters have significantly lower [Al/Fe] than metal-rich 
clusters. This is in contrast with previous findings because the bias due to chemical evolution was not taken into account.

NGC~2808 and \ocen \ are among the most massive accreted clusters in our sample. \ocen \ does not exhibit larger Al 
enrichment than the rest of the clusters if the enrichment is 
plotted against the metallicity without the correction, which shows the importance of the correction. 
They separate more from the rest of the clusters when plotted against metallicity after the correction, because 
the extent of the Al enrichment in the SG stars is larger in more massive clusters.

\subsection{Age}

R$_{\rm Al}$ as a function of age \citep{krause01} is plotted in Figure~\ref{fig:alsc2}. 
There are two distinctive groups visible in the R$_{\rm Al}-$age diagram without the correction (upper panels), one 
is the metal-poor group ([Fe/H]$<$-1.3) for which there is no correlation between R$_{\rm Al}$ and age in the first 
three billion years, but that is only because our sample of GCs does not contain young, metal-rich clusters. 
The other one is the metal-rich group 
([Fe/H]$>$-1.1) with low R$_{\rm Al}$, and all of these clusters are older than 
11.5 billion years. Again there seems to be no correlation with age in this group either. 
The lower left panel shows the R$_{\rm Al}-$age diagram
after the correction, in which the metal-poor group only has slightly larger scatter than the metal-rich ones. This
is similar to what is shown in the metallicity panels.

\subsection{Mg+Al+Si}

\begin{figure}[!ht]
\centering
\includegraphics[width=3.45in,angle=0]{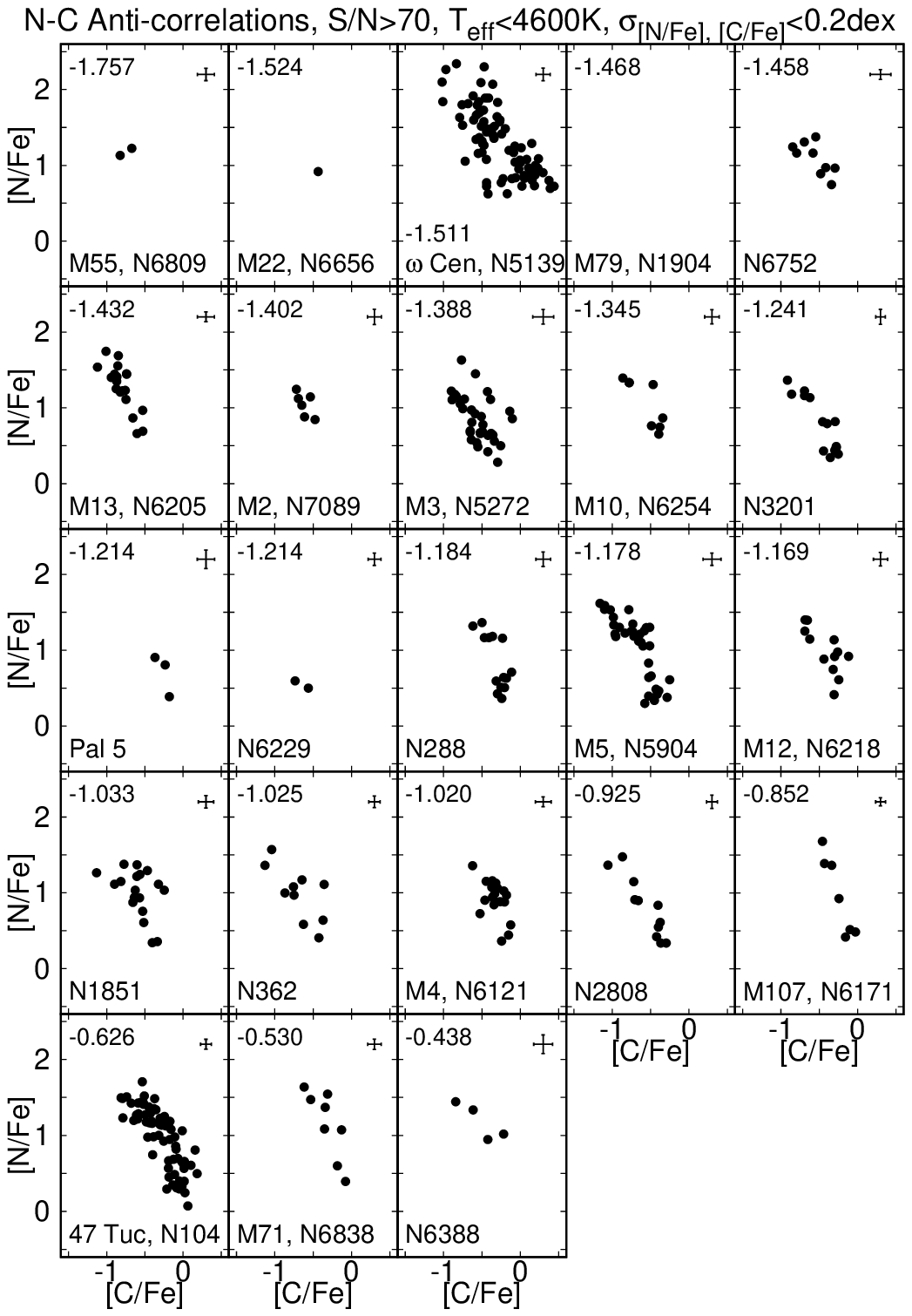}
\caption{N-C anticorrelations. Most clusters show continuous distributions, only M10 and NGC~288 exhibit clear bimodality. 
Clusters are ordered by decreasing average metallicity, which is indicated in the top left corner 
in each panel.
}
\label{fig:cn}
\end{figure}

The summed abundance of [(Mg+Al+Si)/Fe] is expected to be constant as a function of T$_{\rm eff}$ 
and that is what our results show in Figure~\ref{fig:mgalsi}.
Both FG and SG stars have the same [(Mg+Al+Si)/Fe] within the errors, and there are no density peaks observed in any of the 
clusters. The R$_{\rm Mg+Al+Si}$ is very similar, or slightly larger than 
the average error of [(Mg+Al+Si)/Fe], which is what needs to be observed if the Mg-Al cycle operates normally. There 
seems to be no difference in [(Mg+Al+Si)/Fe] between in situ and accreted clusters.

As in previous sections, we explore the statistical significance of the correlation between the cluster properties 
and the sum and scatter of [(Mg+Al+Si)/Fe], shown in Figure~\ref{fig:mgalsi2}. There is a very minimal trend (p=0.0767) between 
R$_{\rm Mg+Al+Si}$ and metallicity, that is interpreted as errors of individual line fitting piling up with decreasing 
metallicity. Small correlations can appear on the level of the average error and usually are the result of correlated 
errors when the measured scatter is on the same level. When looking at the sum of [(Mg+Al+Si)/Fe] there are no 
such correlations present with metallicity, mass or absolute visual magnitude, as expected. But there is a significant 
(p=0.0003) correlation with age. This is due to standard chemical evolution, which we confirmed by only looking 
at the statistics of FG stars that have halo-like chemical composition. This trend is dominated by [Mg/Fe] and [Si/Fe], 
which decreases as metallicity increases. This standard chemical evolution and structure of the Milky Way was recently overviewed by 
\citet{hayden01} and \citet{weinber01} based on APOGEE data.


\section{Multiple Populations Based on N and C}

\begin{figure}
\centering
\includegraphics[width=3.45in,angle=0]{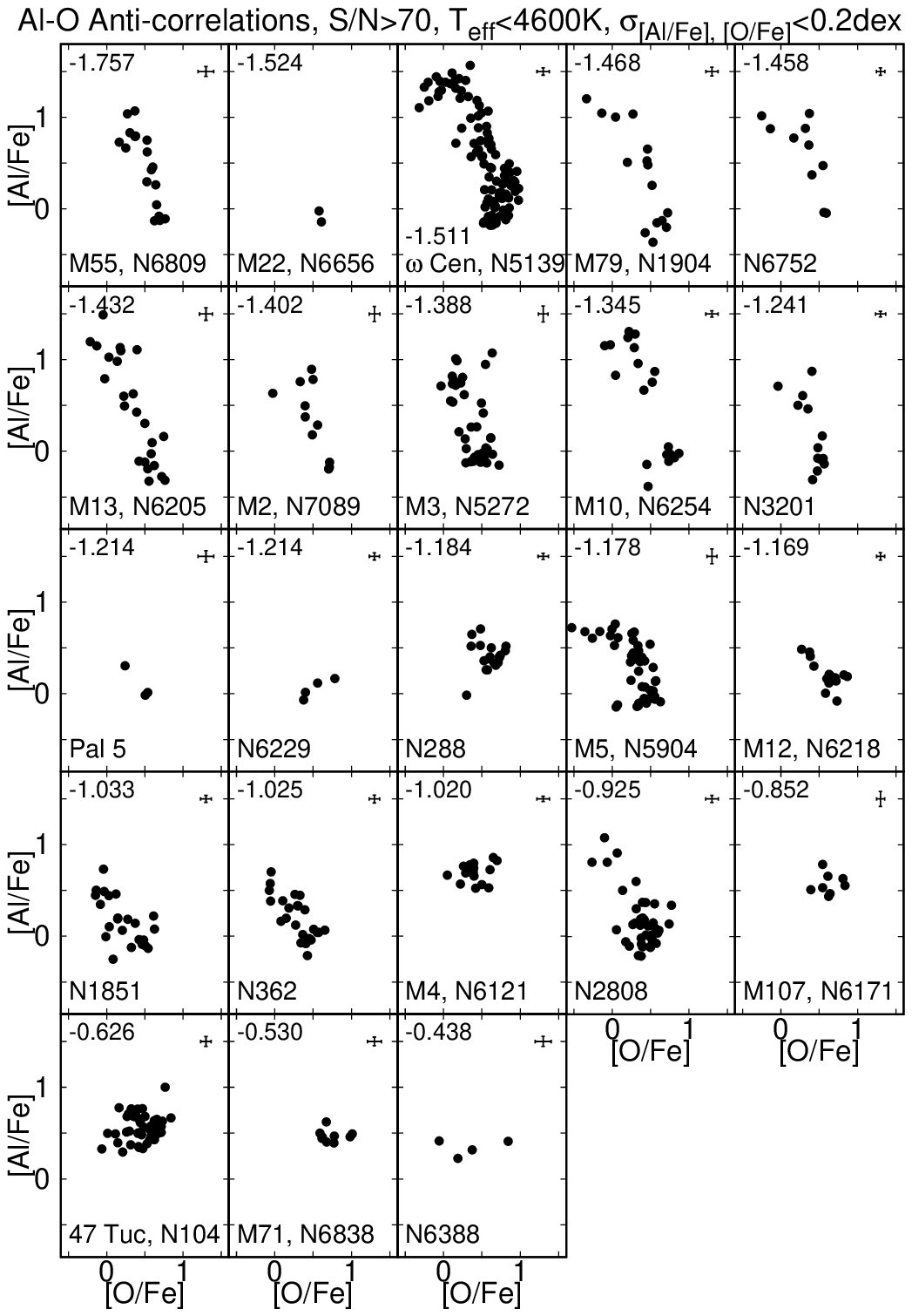}
\caption{Al-O anticorrelations. All cluster show clear Al-O anticorrelations, except 47~Tuc, M4, M107, NGC 6388 and M71, 
cluster with no significant Al spread. Clusters are ordered by decreasing average metallicity, which is indicated in the top left corner 
in each panel.
}
\label{fig:alo}
\end{figure}

\begin{figure}
\centering
\includegraphics[width=3.45in,angle=0]{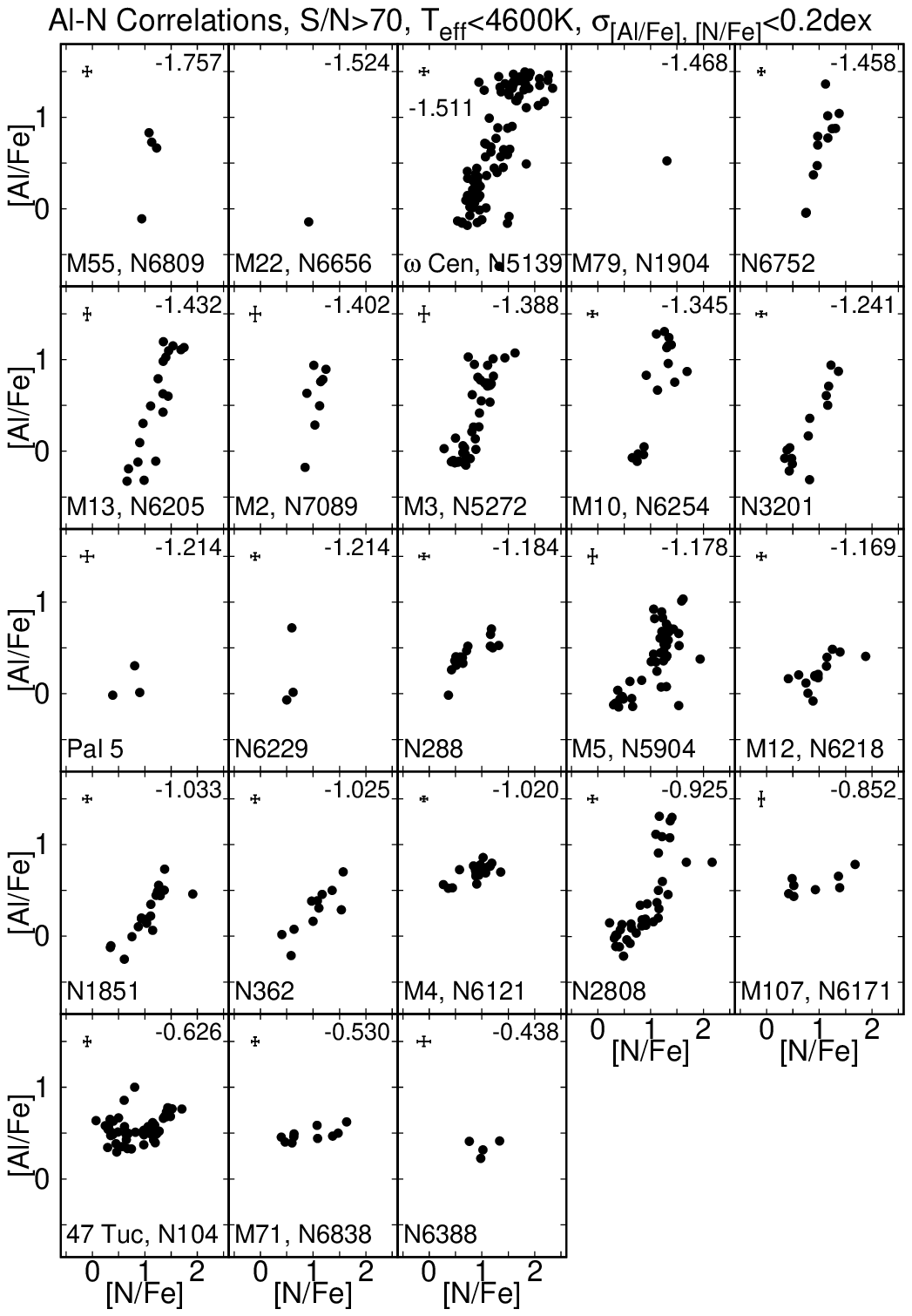}
\caption{Al-N correlations. All cluster show clear Al-N correlations, except 47~Tuc, M4, M107, NGC 6388 and M71, 
cluster with no significant Al spread. Clusters are ordered by decreasing average metallicity, which is indicated in the top right corner 
in each panel.
}
\label{fig:aln}
\end{figure}

\begin{figure*}[!ht]
\centering
\includegraphics[width=4.4in,angle=270]{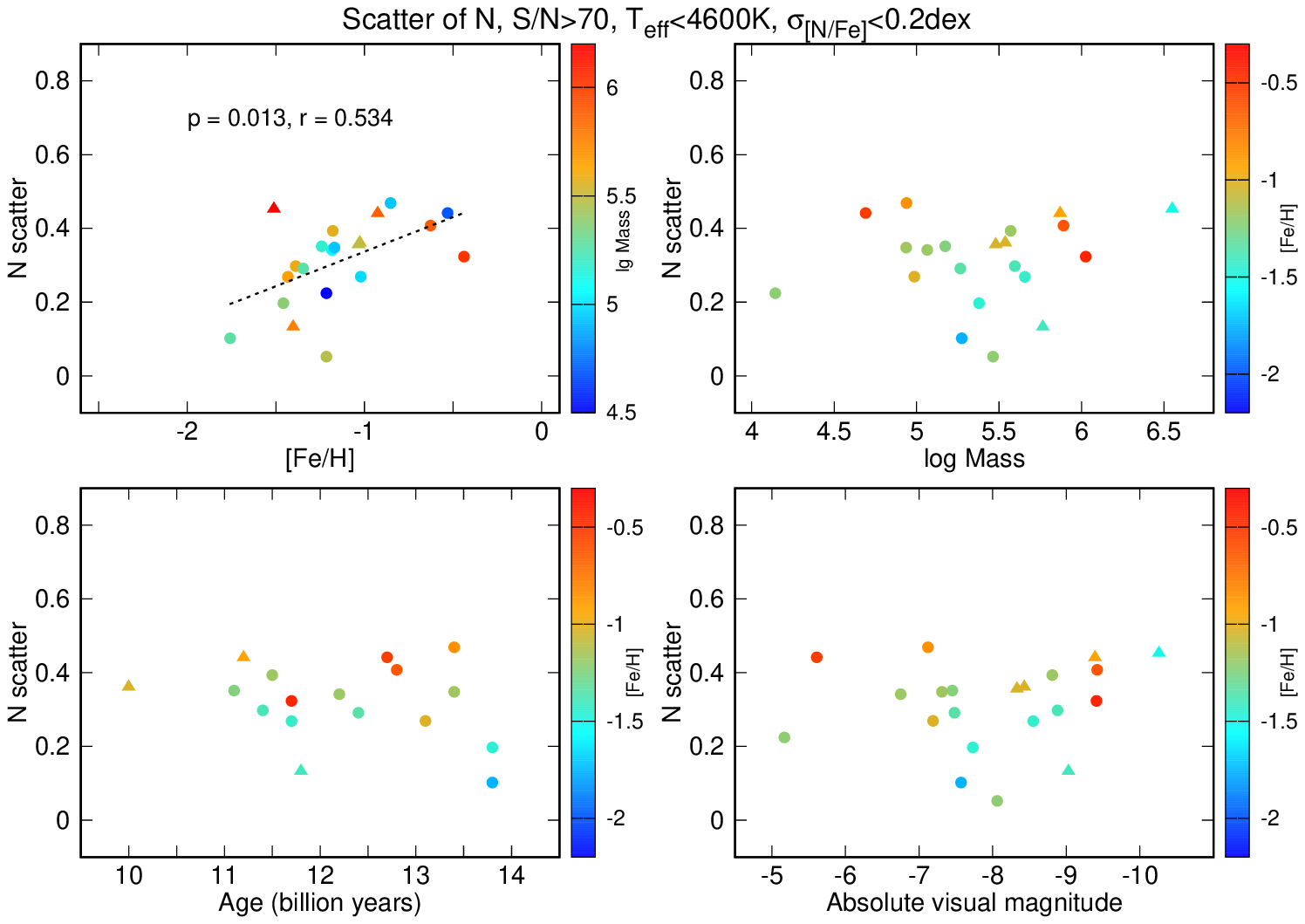}
\caption{Scatter of N as a function of average cluster [Fe/H] color coded by mass for clusters with at least three members. 
Accreted clusters are denoted by triangles, in situ clusters by solid circles. 
}
\label{fig:scn}
\end{figure*}


\subsection{The N-C anticorrelation}

C and N abundances are affected by two different astrophysical processes: 1. deep mixing occurring on the RGB, and 2. pollution 
from FG stars which is similar to the O-Na and Al-Mg patterns. Generally, N is anticorrelated with C in all observed clusters, 
but the slope of the anticorrelation is the combination of these two effects. The N-C and Al-O anticorrelations are shown in 
Figures~\ref{fig:cn} and \ref{fig:alo} respectively. The Al-N correlation is plotted in Figure~\ref{fig:aln}, 
upper limits are omitted from the figures. Clusters with [Fe/H]$<-1.8~$dex are not plotted because the CO and CN lines 
in these stars are too weak to derive reliable abundances. The observed 
slopes were not corrected for deep mixing, but [C/Fe] is strongly correlated with temperature, a clear evidence of 
mixing occurring in every cluster. 

The extended variations in N and C are observed in all GCs in our sample, in accordance with the earliest optical 
CH, CN, and NH observations \citep{norris01}. There are several clusters in our sample that had no 
N-C anticorrelation published before: NGC~2808, M12, NGC~6229, M10, NGC~6388 and Pal~5, all of these exhibit 
clear N enhancements. Previous literature sources have reported both bimodal and continuous distributions 
of N abundances in GCs. Interestingly, multimodality cannot be easily identified in our data. This is because the number of stars 
with CN abundances is less than those with Mg and Al. There are two clusters in which 
bimodality can be convincingly determined: NGC~288 and M10. To a lesser extent, M5, NGC~3201 and M107 appear to  
have two distinctive populations based on N, but their existence is up to interpretation. 
All other clusters exhibit clear continuous distributions, which of course does not mean that multiple density peaks in the 
N-C plane, similar to that of Al-Mg, do not exist, but this can only be proved with more precise measurements of even more stars. 
M3 is an interesting case because it appears to have continuous N, but rather bimodal Al distribution (see Section~5.1), as 
reported by \citet{meszaros03}.

\subsection{The spread of N abundances}

The scatter of N (R$_{\rm N}$) as a function of [Fe/H] paints a very different picture from the scatter of Al, seen in 
Figure~\ref{fig:scn}. A correction to the Galactic evolution of N is not necessary, because N did not go through the same 
chemical evolution as Al \citep{hayes01}. Here, we observe
a slight positive correlation (p=0.0126) with metallicity. The number of stars for which the derivation of [N/Fe] is possible 
quickly decreases as metallicity decreases, because more and more stars are warmer and reach our determination limit of 4600~K
and are the spectroscopic features also intrinsically weaker at lower metallicity. 
We required clusters to have at least 3 stars with [N/Fe] values to be included in this part of the analysis. 
This is somewhat offset by the 
expected increased errors at low metallicities, thus it is hard to judge how much these two systematics affect the correlation. 
The correlation remains even if we exclude the two most metal-poor GCs, thus focusing on the [Fe/H]$>-$1.5 region, in which 
these two sources of error are small. R$_{\rm N}$ does not appear to be correlated with either mass, V$_{\rm obs}$, nor age. 
Also, it seems that both in situ and accreted clusters show similar R$_{\rm N}$ at 
the same metallicity. 

As previously reported in the literature \citep{norris01} all metal-rich GCs have extended N distributions, even the ones with no 
significant Al scatter. This is the case for 8 clusters in our sample: 47~Tuc, NGC~288, M4, M107, M12, NGC~6388, M71, and Pal~5.
All these clusters have high metallicities, in which the Mg-Al cycle cannot start due to the polluting stars not reaching the 
necessary high temperatures in the stellar interiors. This can also be seen in the Al-N correlations and Al-O anticorrelations. 

Al is expected to correlate with other elements produced during the proton-capture process, like Na and N, and anticorrelate 
with O and C. The Al-N and Al-O relationships (Figures~\ref{fig:aln} and \ref{fig:alo}) also help to identify 
whether pollution from the Mg-Al cycle occurred in the clusters with relatively low Al scatter, NGC~1851 and NGC~362. While their 
slightly increased R$_{\rm Al}$ values, 0.25 and 0.24, respectively, suggest some Al enhancement, only the Al-N and Al-O diagram 
can give convincing results by showing a clear (anti-)correlation between these abundances. NGC~288 and M12, two clusters with 
even lower R$_{\rm Al}$ values (0.18 and 0.16), that we assigned only one population to based on Al in Table~7, also seem to 
exhibit some Al-N anticorrelation, but only one of them, M12 has an Al-O anticorrelation. While NGC~288 and M12 are 
less certain to show Al pollution, the Al-Mg anticorrelation in these four clusters will need to be studied in a larger 
sample to reach more conclusive results on the parameter space in which the Mg-Al cycle contributes. 

\subsection{N spread in Clusters with no Al spread}

\begin{figure*}[!ht]
\centering
\includegraphics[width=4.4in,angle=270]{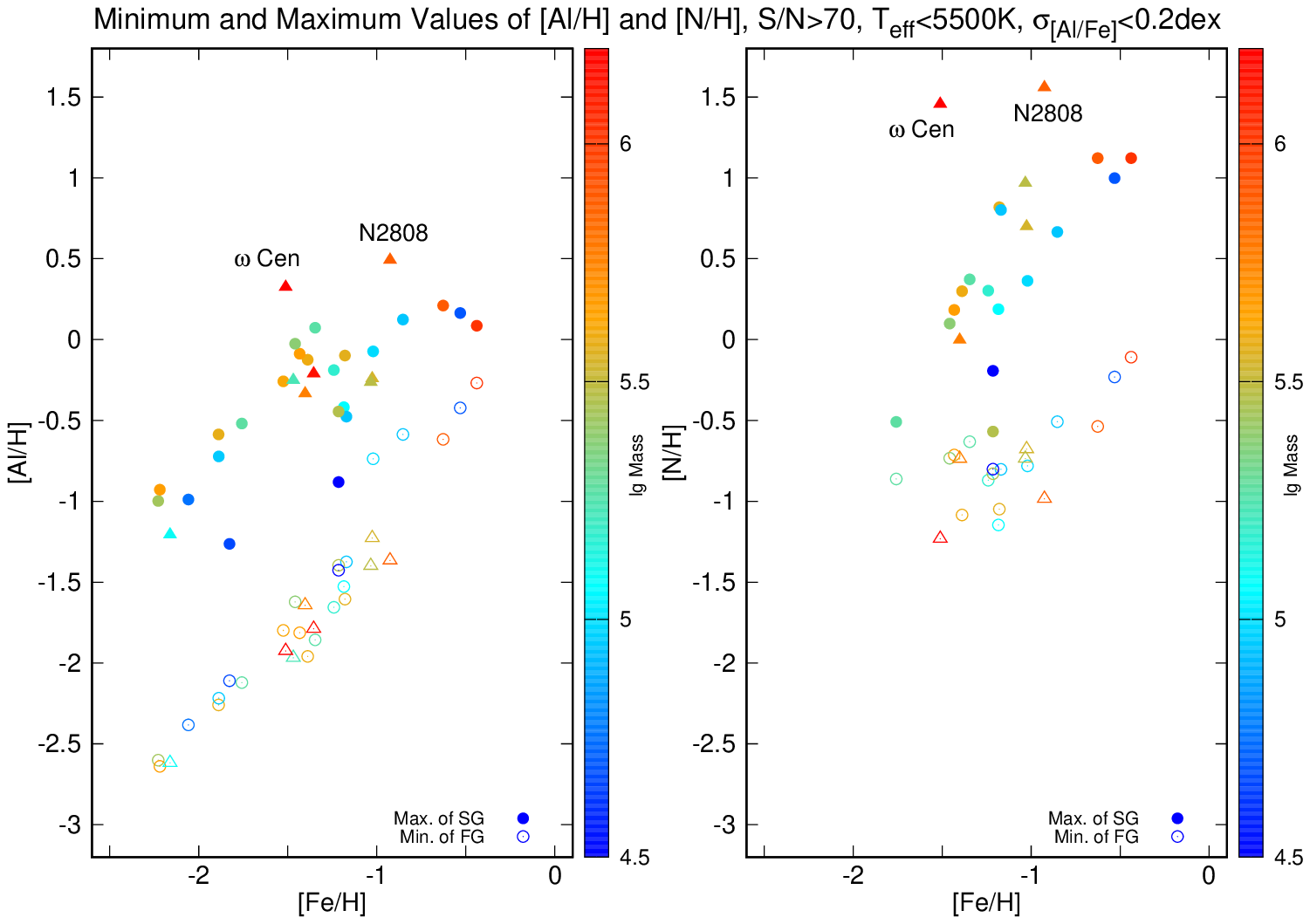}
\caption{The maximum (solid symbols) and minimum (open symbols) values of [Al/H] and [N/H] in each cluster. 
Accreted clusters are denoted by triangles, in situ clusters by circles. The Al and N enrichment is strongest in 
the two massive accreted clusters \ocen \ and NGC~2808.
}
\label{fig:alminmax}
\end{figure*}

In Section~6 we concluded that there are five metal-rich clusters in our sample that do not exhibit large Al-spread. These clusters 
are the following (from Table~7): 47~Tuc, M4, M107, NGC~6388 and M71. From Figures~\ref{fig:alo} and \ref{fig:aln} we can conclude 
that these clusters have no clear Al-N correlation, or Al-O anticorrelations either, but clearly exhibit N-C anticorrelations 
(Figure~\ref{fig:cn}) and large N spreads (Figure~\ref{fig:scn}). M71 is particularly interesting because \citet{ram01} 
have observed a weak Al-Na correlation, and \citet{yong01} showed slightly non-standard isotope ratios suggesting some 
Mg-Al processing may have taken place. These clusters clearly have MPs based on their N abundances 
despite appearing to have single populations in the Al abundance. This is illustrated in Figure~\ref{fig:alminmax}, 
in which we plotted the minimum value of the [Al/H] and [N/H] in the FG stars by comparing it with the maximum value of 
[Al/H] and [N/H] in the SG stars as a function of metallicity. The extent of enrichment of Al is clearly the 
largest at the lowest metallicities and slowly decreases as metallicity increases, while the enrichment of N 
increases with increasing metallicty. The enrichment of Al and N is the largest in the two massive accreted 
clusters, \ocen and NGC~2808, as previously found in Sections~6.1 and 7.2. We explore two different possible explanations of this 
observation. 

The first explanation is as follows: because of the chemical evolution of Al, the FG stars in metal-rich clusters have already 
elevated [Al/Fe] abundances. This is not the case in the [N/Fe] dimension, since chemical evolution of N is not as 
steep as Al \citep{hayes01}. As mentioned in Section~6.1, when the [Al/Fe] of the FG stars are elevated, significantly more Al 
production is needed to be observable in the logarithmic abundance scale. Because of this it is entirely possible that Al 
production existed in these clusters (independent from the nature of the polluters), but did not reach the observable level, 
because the FG stars are mixed up with SG stars in the [Al/Fe] dimension. Both \citet{schiavon01} and \citet{tang01} have 
observed large Al spread in NGC~6553 using ASPCAP data, which is one of the most metal-rich GCs with [Fe/H]=$-$0.15~dex 
\citep{tang01}. NGC~6553 is also in our sample, but we excluded it from our analysis, because its reddening (E(B$-$V)=0.63) is 
too high to derive reliable metallicities using photometric temperatures (see Section~4.). These observations in M71 and 
NGC~6553 strongly supports this theory. 

In the second case, the Mg-Al cycle is modest when [Fe/H]$>-$1~dex. If the GC polluters are massive AGB stars, we  
would expect a small Al production in the metal-rich clusters, because hydrogen burning in AGB stars operates at a higher 
temperature in lower-metallicity stars, and so one might expect higher-metallicity clusters 
to show less variation in elements that participate in the MgAl chain. Thus, in massive metal-rich AGB stars N variations 
are expected without, or very little, variation in Al, meaning that 
N is the best generation indicator for those metal-rich clusters. Indeed, this Al production dependence with 
metallicity has been used before to favor the massive AGB hypothesis \citep{ventura04}. This is supported 
by the dependence of Al on metallicity, which remains even after correcting for the standard chemical evolution discussed in 
Section~6. 

The case of NGC~2808 is peculiar since it is a massive cluster with a large spread of Al and also has similar metallicity 
to these five clusters (47~Tuc, M4, M107, NGC~6388 and M71). One possible explanation is that NGC~2808 has not been formed 
in the Milky Way, such that Al was not high at the time of the formation and FG stars had lower Al than other clusters. 
At the same time other discussions regarding any pollution scenarios need detailed computations and analysis, which are 
far from the scope of the present investigation.

\subsection{C+N+O}

\begin{figure}[!ht]
\centering
\includegraphics[width=3.45in,angle=0]{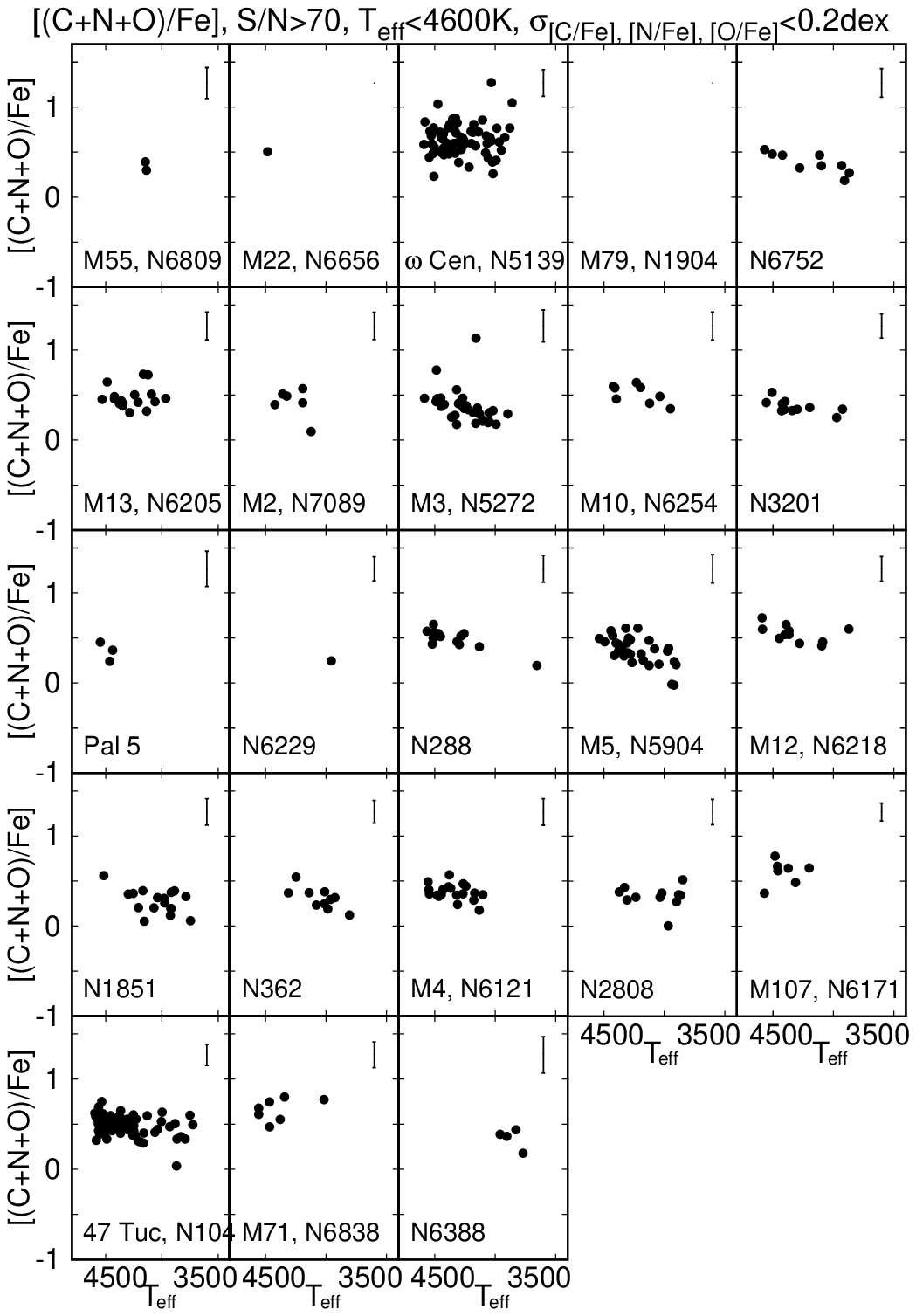}
\caption{The C+N+O in each cluster is constant. Clusters are ordered by decreasing metallicity from left to right and top to bottom.
}
\label{fig:cno}
\end{figure}

While deep mixing affects the C-N diagrams, the C+N+O should be remain constant in each cluster as material is fully 
processed during the CNO cycle \citep{dickens01}. This is what we observe in all clusters, plotted as a function of 
T$_{\rm eff}$ in Figure~\ref{fig:cno}. Some slight correlations on the level of the average error can be seen in some 
clusters. However, these are most likely not of astrophysical origin, but the result of correlated errors between T$_{\rm eff}$ and
CNO abundances. As previously mentioned, as temperature rises, the CN, CO, and OH lines become weaker and harder to measure.

\begin{figure*}
\centering
\includegraphics[width=4.4in,angle=270]{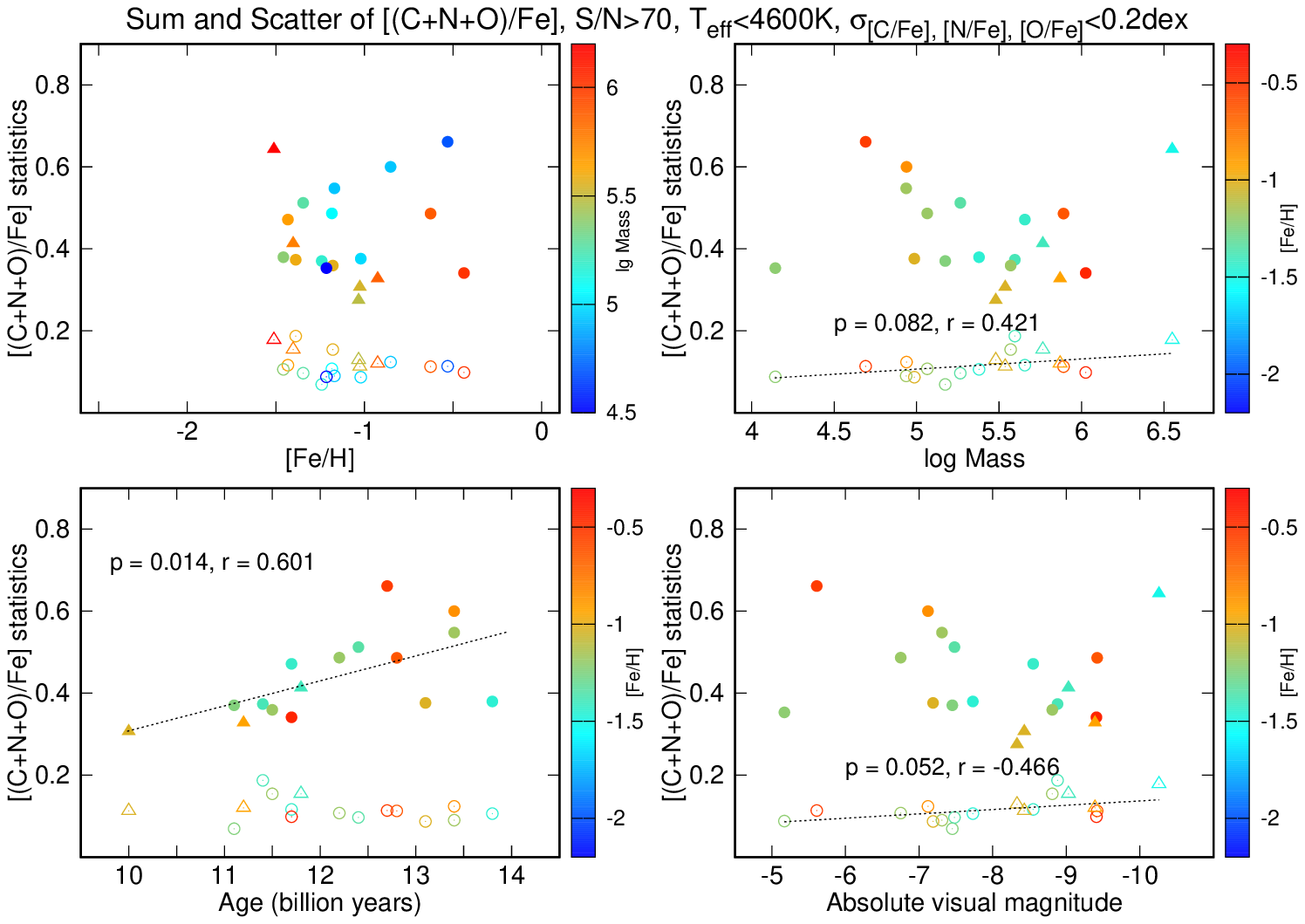}
\caption{Statistics of C+N+O as a function of cluster [Fe/H] and mass. Filled symbols represent the average of C+N+O, open
symbols represent the scatter of C+N+O. Triangles are accreted clusters.
}
\label{fig:sccno}
\end{figure*}

The C+N+O cluster average is consistent with that observed in field stars at similar metallicities \citep{gra06}. 
By looking at Figure~\ref{fig:sccno}, no correlation with metallicity, mass or V$_{\rm Abs}$ can be seen. In situ clusters 
do not have smaller or larger [(C+N+O)/Fe] than those captured via accretion. As with 
[(Mg+Al+Si)/Fe], the significant correlation between age and [(C+N+O)/Fe] is the result of standard chemical 
evolution and is dominated by [O/Fe]. 

An increase in the sum and also the scatter of CNO as a function of metallicity 
has been observed by \citet{johnson02} and by \citet{marino03}. The increased scatter 
was the result of a dependence of C+N+O on [Fe/H]. While \ocen \ will be discussed in detail in the third part of our 
series, we can briefly report that C+N+O is indeed larger than in other clusters ([(C+N+O)/Fe]=0.64~dex). There is 
another cluster, M71, which has an even more elevated CNO sum, [(C+N+O)/Fe]=0.66~dex. This is significantly higher 
than the typical value of [(C+N+O)/Fe], that varies from 0.3 to 0.6~dex in all but three clusters. The third is M107, 
[(C+N+O)/Fe]=0.6~dex, but both clusters differ from \ocen \ in that they are mono-metallic. M107 and M71 are part of those 
five clusters that do not have significant Al spread and their chemical evolution is more like that of 
the thick disc than the traditional halo (Figure~\ref{fig:milky}). The other three clusters (47~Tuc, M4, NGC~6388) 
have [(C+N+O)/Fe]$<$0.49~dex, so the sum of C+N+O does not become elevated for all metal-rich clusters.

The scatter of C+N+O (R$_{\rm CNO}$) shows a correlation with mass and V$_{\rm Abs}$. These 
are moderate correlations, with p=0.013 and p=0.0115, respectively. The average error of C+N+O spans a similar range 
to R$_{\rm CNO}$. Most of the correlation is the result of the increased C+N+O scatter of \ocen, which is the most 
massive cluster in our sample. If \ocen \ is not included in the fit, the statistical significance as a function of mass 
drops down dramatically to p=0.0821 erasing most of the correlation. Thus, our conclusion is that there is no clear 
correlation between R$_{\rm CNO}$ and mass or V$_{\rm Abs}$.

\section{Other Elements}

\subsection{Ca}

All clusters are expected to have uniform and constant [Ca/Fe], because Ca is not affected by H-burning process as it is mostly 
produced by supernovae. This is what we see in our whole sample, Ca is constant in all clusters and its scatter 
is on the level of errors. 

The only GC with a reported Ca spread is M22 \citep{marino04}, which 
was later disputed by \citet{mucci01} explaining the Ca spread with the presence of NLTE effects, similar to that of Fe discussed 
in Section~4.1. M22 is in our sample, however we were able to measure [Ca/Fe] in only a handful of stars, because  
the S/N of the M22 observations are low and Ca lines are generally weak at low metallicities. There are only three 
stars in our sample that satisfy the criteria set in Section~3.1 for analysis, and those three stars span a range of 
[Ca/Fe]=0.35~dex to 0.5~dex, but two of those have errors of $\sigma_{\rm [Ca/Fe]}$=0.19~dex. This data is not sufficient  
to confirm or reject the findings of \citet{marino04}.

\subsection{Ce and Nd}

\begin{figure*}[!ht]
\centering
\includegraphics[width=4.4in,angle=270]{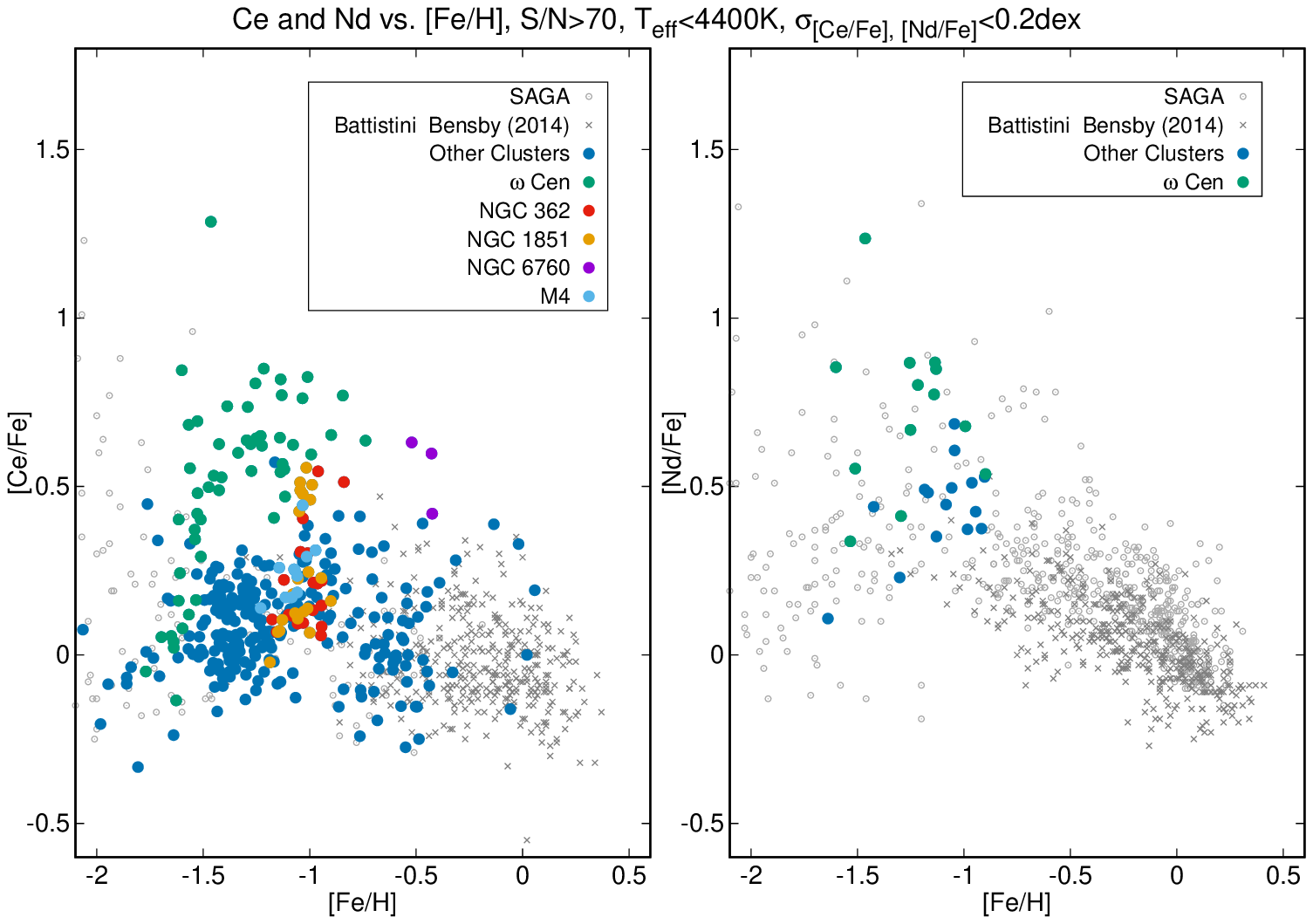}
\caption{[Ce/Fe] and [Nd/Fe] as a function of metallicity for the sample
stars. Background grey circles and crosses are field stars abundances
extracted from the SAGA database \citep{suda01} and
\citet{batt01}. 
}
\label{fig:cend}
\end{figure*}

S-process element enhancements are usually rare and are reported for only a few clusters \citep[M~22, M~15, M~92, 
M4, NGC~362 and NGC~1851]{marino04, sobeck01, roederer01, shingles01, marino05, carretta05}. 
Nd II and Ce II lines have been discovered in the APOGEE spectral region by \citet{hass01, cunha01}. 
In Figure~\ref{fig:cend}, we show Ce and Nd 
abundances obtained from our sample. While there are only few constraining Nd measurements, we consider here stars with
s-process enhancements such that [Ce/Fe]$>$0.4 based on the
comparison with field stars measurement and our typical uncertainties.
We can confirm s-process enhancement in all above-mentioned clusters
except M22, for which our temperature cut off do not leave any stars to
be analyzed. There is, however, several clusters in our sample with clear s-process enhancement: \ocen \, NGC~362, NGC~1851, 
NGC~6760 and M4. \ocen \ shows a clear increase of the Ce abundance as metallicity increases, confirming the early
findings of \citet{norris02} and supporting the pollution of this cluster by low mass AGB stars. 
In addition, we could identify one new cluster with s-process enhancement: NGC~6760 in which all 
three members show enhanced Ce.

\citet{masseron01} discussed the case of M~15 and M~92 where they observe
star-to-star variations of Ce compatible with the halo scatter. Consequently, they interpret that the Ce
enhancement was inherited form the initial gas composition of the
clusters. However, the other clusters with some s-process enhancement are more metal-rich than M15 and M92. At such
metallicities, the Ce scatter in the field is much lower and the initial
composition of the cluster gas can certainly be considered as
homogeneous. Therefore, the s-process enhanced stars observed in M4,
NGC~362, NGC~1851 and NGC~6760 (as well as \ocen) have probably been
polluted in Ce after the clusters have formed. Nevertheless, the
presence of s-process rich stars is not correlated with the Al
enhancement, nor it is with the cluster  metallicity or the cluster
mass. Thus, we believe that the s-process enrichment has been produced
by a different source than the progenitor of the Mg-Al and Na-O
anticorrelations, possibly by low-mass AGBs.


\section{Summary}

In this paper we investigated the Fe, Mg, Al, C, N and O abundances of 2283 red giant stars in 31 GCs 
from high-resolution spectra obtained by the SDSS-IV APOGEE-2 survey. We reported on the properties of MPs
based on their Al-Mg, and C-N anticorrelations and also explored the dependence of the abundance spread of Fe, Al and N on 
cluster properties. To summarize our results, we find the following:

1. The scatter of Fe does not depend on mass, V$_{\rm Abs}$ or age. The uncertainty coming from possible 
3D/NLTE and reddening through 
photometric temperatures does not allow us to further refine the metallicity scale from the literature. By comparing three 
independent metallicity scales we determine that the metallicities of GCs derived from the H-band are 0.064~dex higher 
on average (in absolute terms) than the optical Fe scale.

2. Other than the well-known Fe spread in \ocen, we do not observe significant Fe variations in any of the clusters from 
our sample even though we have the precision to do so. This includes clusters with previously reported Fe spreads: 
M22, NGC~1851 and M54. 
While in M22 and NGC~1851 we have more than enough stars to sample multiple Fe populations, in 
M54 we only observed 7 stars with S/N$>$70. We most likely have not sampled enough stars with different Fe abundances, 
possibly due to limitations of the APOGEE fiber collision constraints which limit sampling the inner cluster regions.

3. By using density maps of the Al-Mg anticorrelations we were able to identify multimodality in several clusters, including 
M79, \ocen, and NGC~6752. While \ocen \ and NGC~6752 were previously known to host more than two populations based on Al 
from the literature, M79 has not been previously reported on.

4. In \ocen, we observe a turnover in Al abundances 
for the most Mg-poor stars, similar to that of M15 and M92. Some of these Mg-poor stars are also slightly K enriched compared 
to standard FG stars drawing a weak K-Mg anticorrelation. However, the weak and blended K lines do not allow us to 
present a firm discovery of this K enrichment. Followup observations are needed to confirm or to contradict our findings.

5. We are able to confirm the Si-Mg anticorrelation observed in NGC 2808 by \citet{carretta02}, but the case 
of NGC 6752, as observed by \citet{yong05} is less convincing in our data.

6. The ratio of the number of FG/SG stars depends on metallicity and age, but not on mass, which contradicts the findings
of \citet{milone02}. This may be explained by a sample bias created by selecting stars from the outer regions of the clusters 
which affects f$_{\rm enriched}$ compared to HST studies which sampled the inner 2 arcminutes of the clusters. 

7. We find a complex relationship between the spread of Al and cluster average metallicity and mass. We identified three distinctive
groups in Al scatter - [Fe/H] diagram: a.) clusters with [Fe/H]$<$-1.3 have a near constant high R$_{\rm Al}$ value above 0.4~dex;  
b.) clusters between $-$1.0$<$[Fe/H]$<-$1.3 show a wide variety of Al spread; c.) the more metal-rich GCs have a small Al spread,
comparable in size to the errors. This picture is changed when a correction for the chemical evolution of Al in the Milky Way is 
introduced. After the correction, the scatter of Al decreases and most of the large step between metal-poor and 
metal-rich clusters is removed, but the complex nature of the correlation with metallicity remains. The dependence 
of R$_{\rm Al}$ with cluster mass is increased suggesting that the extent of Al enrichment as a function of mass 
was suppressed before the correction.

8. Metal-rich accreted clusters, NGC~2808 and \ocen \ show significantly higher R$_{\rm Al}$ than their 
counterparts formed in situ. The rest of the accreted GCs appear to have similar Al spreads to the in situ clusters. 

9. The measured N-C anticorrelation is generally continuous with the exception of NGC~288 and M10, which show 
clear bimodality. This is in contrast with previous literature observations which generally found bimodal distributions. 

10. We measure constant Mg+Al+Si and C+N+O within all clusters. The sum does not depend on 
metallicity, mass, V$_{\rm Abs}$, but on age, which is the result of standard chemical evolution. The scatter of Mg+Al+Si increases 
with decreasing metallicity which is most likely the result of accumulated errors at low metallicities. The scatter of 
C+N+O in \ocen \ is larger than in other clusters, agreeing with previous literature finds. 

11. The five clusters (47~Tuc, M4, M107, NGC 6388 and M71) that have large variations in N, but Al scatter close to our 
uncertainties, appear to not show the signs of the Mg-Al cycle because their FG stars have elevated [Al/Fe] similar to thick 
disc stars. Considering that it is necessary to produce significantly more Al to reach the observational limit in the 
logarithmic abundance scale in metal-rich clusters than in metal-poor clusters, and the observations of Al 
rich stars in NGC~6553 by \citet{schiavon01, tang01}, we conclude that our observations of low Al scatter in these five clusters 
do not rule out the existence of the Mg-Al cycle.

12. \ocen \ shows a clear increase of the Ce abundance as metallicity increases, confirming the early
findings of \citet{norris02} and supporting the pollution of this cluster by low mass AGB stars. We identified a 
new cluster, NGC~6760, with clear Ce enhancement.


\acknowledgements{SzM has been supported by the Premium Postdoctoral
Research Program and J{\'a}nos Bolyai Research Scholarship of the Hungarian Academy of
Sciences, by the Hungarian NKFI Grants K-119517 and GINOP-2.3.2-15-2016-00003 of the Hungarian National
Research, Development and Innovation Office. DAGH, TM, OZ, and FDA acknowledge support 
from the State Research Agency (AEI) of the Spanish Ministry of Science, Innovation and 
Universities (MCIU) and the European Regional Development Fund (FEDER) under grant AYA2017-88254-P. 
J.G.F-T is supported by FONDECYT No. 3180210. D.G. gratefully acknowledges support from the 
Chilean Centro de Excelencia en Astrof{\'{\i}}sicay Tecnolog{\'{\i}}as Afines (CATA) BASAL grant AFB-170002.
D.G. also acknowledges financial support from the Dirección de Investigación y Desarrollo de
la Universidad de La Serena through the Programa de Incentivo a la Investigación de
Académicos (PIA-DIDULS). T.C.B. acknowledges partial support from grant PHY 14-30152; Physics
Frontier Center/JINA Center for the Evolution of the Elements (JINA-CEE), awarded by the US National Science Foundation, and
from the Leverhulme Trust (UK), during his visiting professorship at the University Of Hull. 
SLM acknowledges the support of the Australian Research Council through Discovery Project grant DP180101791.

Funding for the Sloan Digital Sky Survey IV has been provided by the Alfred P. Sloan Foundation, 
the U.S. Department of Energy Office of Science, and the Participating Institutions. SDSS-IV acknowledges
support and resources from the Center for High-Performance Computing at
the University of Utah. The SDSS web site is www.sdss.org.

SDSS-IV is managed by the Astrophysical Research Consortium for the 
Participating Institutions of the SDSS Collaboration including the 
Brazilian Participation Group, the Carnegie Institution for Science, 
Carnegie Mellon University, the Chilean Participation Group, the French Participation Group, 
Harvard-Smithsonian Center for Astrophysics, 
Instituto de Astrof\'isica de Canarias, The Johns Hopkins University, 
Kavli Institute for the Physics and Mathematics of the Universe (IPMU) / 
University of Tokyo, Lawrence Berkeley National Laboratory, 
Leibniz Institut f\"ur Astrophysik Potsdam (AIP),  
Max-Planck-Institut f\"ur Astronomie (MPIA Heidelberg), 
Max-Planck-Institut f\"ur Astrophysik (MPA Garching), 
Max-Planck-Institut f\"ur Extraterrestrische Physik (MPE), 
National Astronomical Observatories of China, New Mexico State University, 
New York University, University of Notre Dame, 
Observat\'ario Nacional / MCTI, The Ohio State University, 
Pennsylvania State University, Shanghai Astronomical Observatory, 
United Kingdom Participation Group,
Universidad Nacional Aut\'onoma de M\'exico, University of Arizona, 
University of Colorado Boulder, University of Oxford, University of Portsmouth, 
University of Utah, University of Virginia, University of Washington, University of Wisconsin, 
Vanderbilt University, and Yale University.
}


\end{document}